\documentclass[iop, twocolappendix, revtex4]{emulateapj}
\usepackage{amsmath,amsfonts,amssymb,amsthm,pdfpages,graphicx,mathtools,subfigure,bold-extra,natbib,courier,array,tabulary}
\usepackage[draft]{hyperref}

\shorttitle{Systematic Effects in the $M_\star-Z-\mathrm{SFR}$ Relation}
\shortauthors{Telford et al.}

\begin{document}

\title{Exploring Systematic Effects in the Relation Between Stellar Mass, Gas Phase Metallicity, and Star Formation Rate}

\author{O. Grace Telford\altaffilmark{1} and Julianne J. Dalcanton}
\affil{University of Washington Astronomy Department, Box 351580, Seattle, WA 98195-1580, USA}

\author{Evan D. Skillman}
\affil{Minnesota Institute for Astrophysics, School of Physics and Astronomy, 116 Church Street, S.E., University of Minnesota, Minneapolis, MN 55455, USA}

\author{Charlie Conroy}
\affil{Harvard-Smithsonian Center for Astrophysics, 60 Garden Street, Cambridge, MA 02138, USA}

\altaffiltext{1}{otelford@uw.edu} 

\begin{abstract}
There is evidence that the well-established mass-metallicity relation in galaxies is correlated with a third parameter: star formation rate (SFR). The strength of this correlation may be used to disentangle the relative importance of different physical processes (e.g., infall of pristine gas, metal-enriched outflows) in governing chemical evolution. However, all three parameters are susceptible to biases that might affect the observed strength of the relation between them. We analyze possible sources of systematic error, including sample bias, application of signal-to-noise ratio cuts on emission lines, choice of metallicity calibration, uncertainty in stellar mass determination, aperture effects, and dust. We present the first analysis of the relation between stellar mass, gas phase metallicity, and SFR using strong line abundance diagnostics from \citet{dopita13} for  $\sim 130,000$ star-forming galaxies in the Sloan Digital Sky Survey and provide a detailed comparison of these diagnostics in an appendix. Using these new abundance diagnostics yields a $30-55\%$ weaker anti-correlation between metallicity and SFR at fixed stellar mass than that reported by \citet{mannucci10}. We find that, for all abundance diagnostics, the anti-correlation with SFR is stronger for the relatively few galaxies whose current SFRs are elevated above their past average SFRs. This is also true for the new abundance diagnostic of \citet{dopita16}, which gives anti-correlation between $Z$ and SFR only in the high specific star formation rate (sSFR) regime, in contrast to the recent results of \citet{kashino16b}. The poorly constrained strength of the relation between stellar mass, metallicity, and SFR must be carefully accounted for in theoretical studies of chemical evolution. 
\end{abstract}


\section{Introduction}

The stellar mass of a galaxy ($M_\star$) is correlated with the gas phase metallicity ($Z$, the oxygen abundance relative to hydrogen) such that galaxies with higher $M_\star$ have larger oxygen abundance. This well-studied correlation is known as the mass-metallicity relation (MZR; e.g., \citealt{lequeux79, tremonti04}). However, the measurement of a galaxy's characteristic gas-phase metallicity is the subject of ongoing debate. Different techniques for measuring metallicity disagree by up to $0.7 \, \mathrm{dex}$ and produce MZRs with different slopes and normalizations \citep{ke08}. The largest discrepancies are between metallicities measured using the direct method, which requires measurements of the weak auroral $\mathrm{[O \textsc{iii}]}\lambda 4363$ line to measure the electron temperature of the gas, and theoretically calibrated strong line methods, which relate ratios of bright emission line fluxes to the gas phase metallicity determined from photoionization models. Theoretical strong line methods systematically measure higher metallicities than the direct method, but each type of technique requires approximations that may bias measurements high or low, respectively (e.g., \citealt{stasinska05, ke08, moustakas10}). Since there is no way to determine which method best approximates the ``true" metallicity of a galaxy, it remains unclear which version of the MZR is the best representation of reality.

Recently, evidence has accumulated for a correlation between the MZR and other galaxy parameters. \citet{ellison08} showed that at a fixed stellar mass, galaxies with higher specific star formation rates (sSFR, the ratio of star formation rate (SFR) to stellar mass) or larger half-light radii have lower metallicity. Following this initial discovery, other authors similarly found that galaxies at fixed mass with higher SFR have lower metallicity (\citealt{lara-lopez10, mannucci10}, hereafter M10). This correlation was dubbed the ``fundamental metallicity relation" by M10, who suggested that this relation is invariant up to $z \sim 2.5$.  These authors found that the scatter about the median MZR was substantially reduced when the correlation with SFR was accounted for. 

Subsequent theoretical work has investigated mechanisms that may drive the observed anti-correlation between metallicity and SFR. Several authors have proposed analytic models in which both metallicity and SFR are governed by the interplay between inflowing gas (either pristine or previously enriched) and feedback and outflows due to star formation \citep{dave12, dayal13, lilly13}; such models naturally account for a relation between $M_\star$, $Z$, and SFR. Given that a source of gas is required to sustain star formation, it is reasonable that higher SFR galaxies might have lower gas phase metallicities if the inflowing gas is metal poor compared to the ambient ISM. Semi-analytic models (e.g., \citealt{yates12}) and hydrodynamical simulations (e.g., \citealt{derossi15}) have also qualitatively reproduced the observed relation.

More recent observational studies have produced discrepant results regarding the shape of the $M_\star-Z-\mathrm{SFR}$ relation. \citet{yates12} found that the sense of the correlation with SFR reverses at high stellar mass in their sample of observed galaxies and also found a similar effect in a cosmological semi-analytical model. However, \citet{salim14} argued that this apparent turnover is an artifact of signal-to-noise cuts imposed on their observational sample by using metallicities from \citet{tremonti04}, which require confident detections of weak forbidden lines. \citet{am13} used stacked spectra of nearby galaxies to detect the auroral $\mathrm{[O \textsc{iii}]}\lambda 4363$ line, enabling ``average" metallicity measurements for galaxies with similar masses and SFRs using the direct method. This technique cannot be used to measure metallicities of individual galaxies in large galaxy samples because the $\mathrm{[O \textsc{iii}]}\lambda 4363$ line becomes too weak to detect at high $Z$. They found an even stronger correlation of metallicity with SFR than any of the studies mentioned previously, all of which used strong line metallicities. 

The studies discussed above made use of fiber spectroscopy from the Sloan Digital Sky Survey (SDSS), which integrates over large areas for typical galaxies in the local universe. This inherent averaging over many $\mathrm{H} \, \textsc{ii}$ regions can be avoided using integral field spectroscopy. Fiber spectroscopy is also limited to the centers of galaxies, whereas both integral field and drift-scan spectroscopy collect larger fractions of the total light from galaxies, avoiding potential biases in metallicity and SFR measurements due to variations in the covering fraction with redshift. These methods have recently been used to measure the characteristic metallicities of galaxies and to study the correlation between the MZR and SFR \citep{hughes13, sanchez13}

The results from these integral field studies are mixed. \citet{hughes13} studied the correlation between the MZR and both SFR and $\mathrm{H \, \textsc{i}}$ mass and concluded that the MZR does not significantly depend on SFR. Further, accounting for the correlation with SFR actually increased the scatter in their best fit relation, though this may be a result of their modest sample size ($\sim 200$ galaxies). Likewise, \citet{sanchez13} found no correlation between the MZR and SFR using data from the CALIFA survey \citep{sanchez12} to measure metallicities of individual $\mathrm{H} \, \textsc{ii}$ regions. However, a new analysis of the same data by \citet{salim14} did find evidence for a correlation with SFR in the CALIFA data, though again only for a small sample (150 galaxies).

The disagreement between different studies regarding the strength of the correlation between the mass-metallicity relation and SFR motivates careful analysis of systematic errors that enter into the measurements of these three quantities. Recently, \citet{salim14} studied the effects of using different SFR indicators and metallicity measurements on the $M_\star-Z-\mathrm{SFR}$ relation. They found that no changes in the method of measuring these quantities caused the SFR correlation to disappear, so this relation is unlikely to be spurious. However, it is still quite possible that correlations with SFR might be induced by the methods of measuring these quantities; e.g., the choice of metallicity calibration and method of accounting for the degeneracy between metallicity and ionization parameter could be biased in a way that correlates with SFR. 

Since systematic errors can certainly alter the observed strength of this relation, the true strength of the correlation between the mass-metallicity relation and SFR remains quite unconstrained. This limitation compromises any attempt to theoretically interpret the observed relation. Since galaxy chemical evolution models have begun to use the $M_\star-Z-\mathrm{SFR}$ relation as an input prescription or to constrain model parameters (e.g., \citealt{ lilly13,mp15}), it is crucial to understand the range of possible strengths of this relation.

The aim of this paper is to study potential sources of systematic error that can affect the observed strength of the $M_\star-Z-\mathrm{SFR}$ relation. We present the first analysis of this relation using metallicity estimators from \citet{dopita13} (hereafter D13), which we compare to the methods of M10 and \citet{dopita16} (hereafter D16). Further, since the three parameters studied here are all derived quantities and could all suffer from significant systematic errors, we search for measurement biases that could alter the apparent strength of the correlation with SFR.

The paper is organized as follows: in Section~\ref{data} we review our sample selection and methods for measuring the quantities of interest; in Section~\ref{results} we present our $M_\star-Z-\mathrm{SFR}$ relation for SDSS galaxies using metallicity calibrations from D13 and investigate sources of bias in this relation; in Section~\ref{discussion} we discuss the implications of the strength of the correlation with SFR for theoretical analyses of galaxy evolution; and finally in Section~\ref{conclusions} we summarize our findings and conclusions. Throughout the paper we use ``metallicity," ``oxygen abundance," and $Z$ interchangeably to mean the gas-phase oxygen abundance relative to hydrogen ($12+ \log \mathrm{(O/H)}$). We assume a $\Lambda \mathrm{CDM}$ cosmology with $H_0 = 70 \, \mathrm{km} \, \mathrm{s}^{-1} \, \mathrm{Mpc}^{-1}$, $\Omega_m=0.3$, and $\Omega_\Lambda=0.7$.

\section{Data and Calculations}
\label{data}

\subsection{Sample Selection and Properties}
\label{sample}

\begin{figure*}
\begin{centering}
\includegraphics[width=0.9\textwidth]{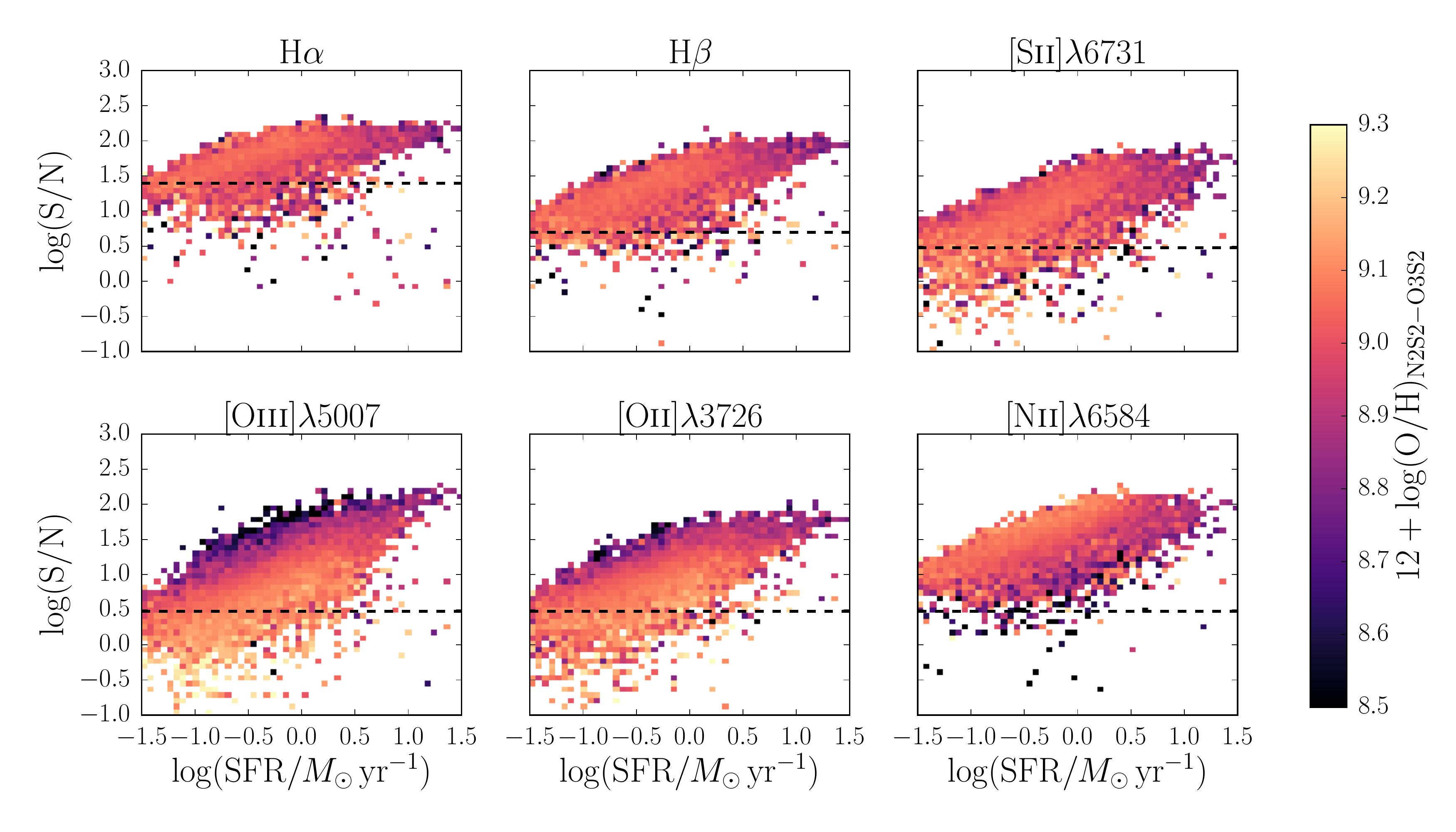}
\caption{Demonstrating the effect of signal-to-noise ratio cuts on sample bias for a subset of our initial galaxy sample with $10.0 < \log(M_\star/M_\odot) < 10.15$. Galaxies are binned by SFR and S/N in six emission lines, labeled in the title of each panel. Color indicates metallicity as measured from the D13 N2S2--O3S2 grid (introduced in Section~\ref{d13_grid_sec}). Black dashed lines show a proposed minimum S/N for each line: 25 for H$\alpha$, 5 for H$\beta$, and 3 for all other lines. For doublets, we show only one of the two lines because both behave similarly. S/N cuts on emission lines shown in the bottom row would induce bias in median metallicity as a function of SFR.
 \label{bias_fig}}
 \end{centering}
\end{figure*}

The data in this analysis are derived from galaxy spectra from the Sloan Digital Sky Survey Data Release 7 (SDSS DR7; \citealt{york00, abazajian09}). We use measurements of stellar masses and emission line fluxes from the publicly available SDSS DR7 MPA/JHU catalog\footnote{http://www.mpa-garching.mpg.de/SDSS/DR7/}\citep{kauffmann03a, brinchmann04, salim07}.  We select our main star-forming galaxy sample following the selection criteria of M10. We require the galaxies to have redshifts between 0.07 and 0.30. The median covering fractions are 22\% and 49\% of the flux at the lower and upper redshift bounds, respectively, as calculated from the $r$-band Petrosian and fiber magnitudes. To avoid large or unphysical reddening corrections, we require the foreground Milky Way $A_V$ to be less than 2.5 and the Balmer decrement (the ratio of $\mathrm{H}\alpha$ to $\mathrm{H}\beta$ emission line flux) to be greater than 2.5. We use the observed Balmer decrement and the \citet{cardelli89} extinction law to correct all line fluxes for dust extinction, assuming $R_V$ of 3.1 and an intrinsic Balmer decrement of 2.86. We remove active galactic nuclei (AGN) from the sample according to the empirical BPT diagram classification of \citet{kauffmann03c}, leaving us with an initial sample of 229,179 galaxies.

We scale down total stellar masses from the MPA/JHU catalog from a \citet{kroupa01} to a \citet{chabrier03} initial mass function. We use the \citet{kennicutt98} relation to calculate the SFR from the extinction-corrected $\mathrm{H}\alpha$ luminosity inside the fiber, again, scaled down to a \citet{chabrier03} IMF. These SFRs only sample the central regions of the galaxies falling within the $3''$ fiber. 

We use fiber-based SFRs to avoid introducing uncertainty from aperture corrections, which require assumptions about the distribution of light and star formation within a galaxy. These SFRs will be biased low with respect to the true total SFRs of the galaxies. However, they will still provide reliable relative ranking of galaxy SFRs, even if they are biased low as an ensemble. They are thus sufficient for identifying trends with SFR, in that they can reliably be used to sort galaxies into bins of SFR. Variations in covering fraction will produce some uncertainty that scatters galaxies into neighboring bins. However, in practice this uncertainty is less than is produced by applying aperture corrections, or by using SFRs derived from broadband colors alone.

We define sSFR to be the ratio of the SFR inside the fiber to the total stellar mass of the galaxy. We use this mix of fiber and total quantities rather than using the stellar mass within the fiber to avoid introducing bias into our sample. The ratio of the fiber to total stellar mass varies systematically with SFR, in the sense that higher SFR galaxies have higher fiber masses at fixed total mass, so mixing the two different types of stellar mass measurements in the same analysis of the $M_\star-Z-\mathrm{SFR}$ relation is problematic. The sSFRs reported for the galaxies in our sample are lower than the true values of sSFR by a factor of $\sim 2-3$, since the total stellar mass includes more of the galaxy than the star formation rate measurement. Again, because we primarily use sSFR as a ranking parameter, our analysis is robust to systematic offsets such as those that affect sSFR.

To ensure that metallicities are well-measured, we exclude galaxies with low signal-to-noise ratios (S/Ns) in the emission lines used to measure metallicity. However, since line strengths correlate with metallicity, we must check that imposing S/N cuts does not induce biases against high or low metallicity galaxies in a way that correlates with $M_\star$ or SFR. 

Figure~\ref{bias_fig} demonstrates the effect of requiring a minimum S/N in metallicity-sensitive lines, for a subset of our initial galaxy sample in a narrow range of stellar mass, $10.0 < \log(M_\star/M_\odot) < 10.15$. Each panel corresponds to one of six different emission lines, and within each panel galaxies are binned by SFR and S/N in that line. Color indicates the median metallicity in each bin, as measured in Section~\ref{d13_grid_sec} below. The dashed line in each panel shows a possible S/N cut (25 for H$\alpha$, 5 for H$\beta$, and 3 for all other lines).

For the lines in the top row of Figure~\ref{bias_fig}, there is no strong variation in $Z$ with S/N, suggesting that S/N cuts on these lines would not introduce bias in the median metallicity as a function of SFR. However, for the lines in the bottom row, there are clear trends between $Z$ and S/N. Higher metallicity galaxies have lower S/N in oxygen lines, whereas lower metallicity galaxies have lower S/N in nitrogen lines. These trends hold regardless of which narrow bin of $M_\star$ we choose, though more galaxies would be removed by such S/N cuts in higher $M_\star$ bins.

We therefore require that galaxies have S/N of at least 25 in the $\mathrm{H}\alpha$ line (following M10), at least 5 in the $\mathrm{H}\beta$ line and at least 3 in the $[\mathrm{S \, \textsc{ii}}] \lambda 6717$ and $[\mathrm{S \, \textsc{ii}}] \lambda 6731$ lines. We do not cut galaxies with low S/N in oxygen or nitrogen lines to ensure that our sample is not biased against galaxies of high or low metallicities at low SFR. Following these S/N cuts, our main sample contains 135,194 galaxies. 90\% of these galaxies have S/N of at least 13.8 in the $[\mathrm{N \, \textsc{ii}}] \lambda 6584$ line and at least $2.5-3.5$ in most of the oxygen lines. The only line of interest with relatively low S/N is $[\mathrm{O \, \textsc{iii}}] \lambda 4959$; 90\% of galaxies have S/N of at least 0.6 in this line.

\begin{figure*}
\minipage{0.5\textwidth}
  \includegraphics[width=\linewidth]{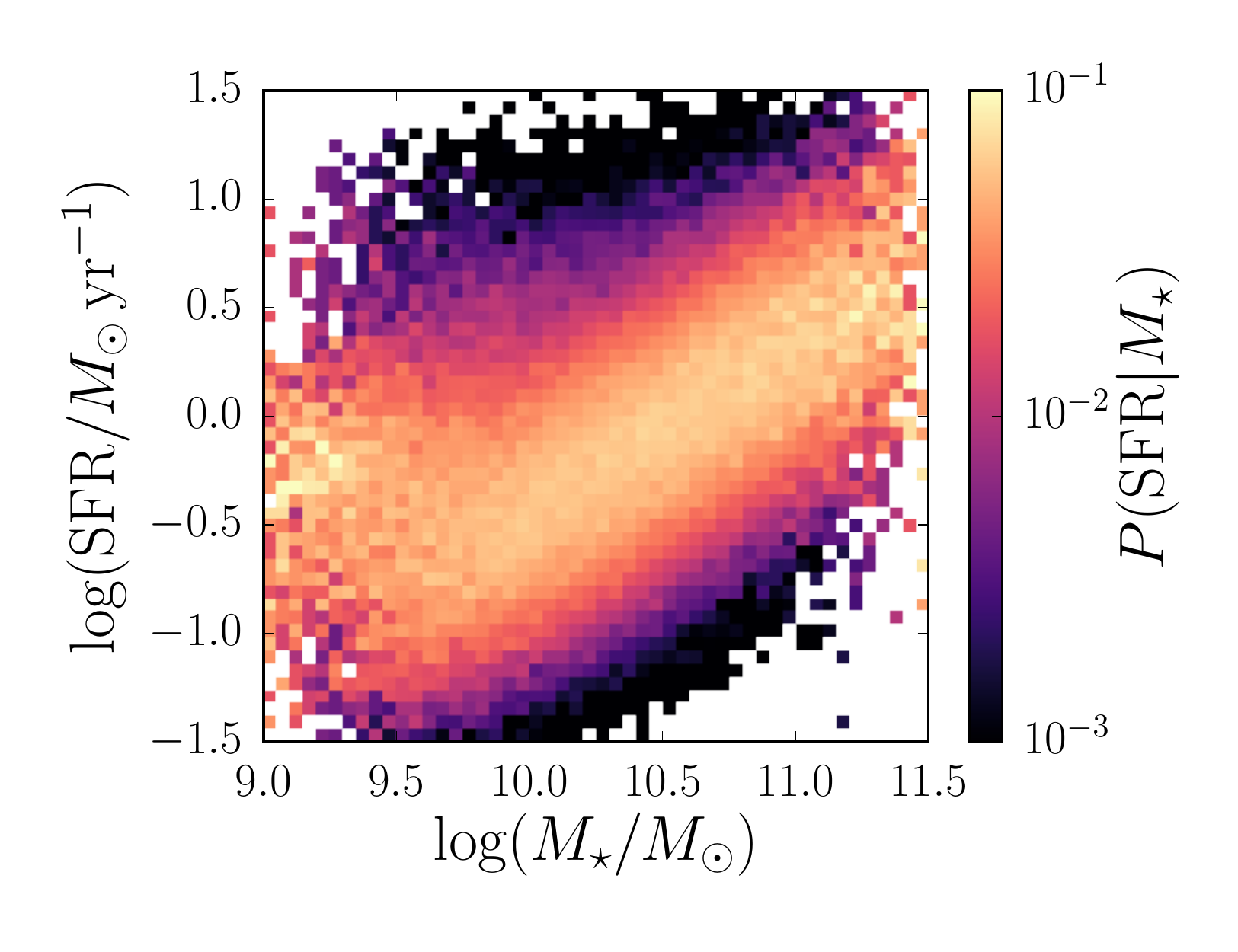}
\endminipage\hfill
\minipage{0.5\textwidth}
  \includegraphics[width=\linewidth]{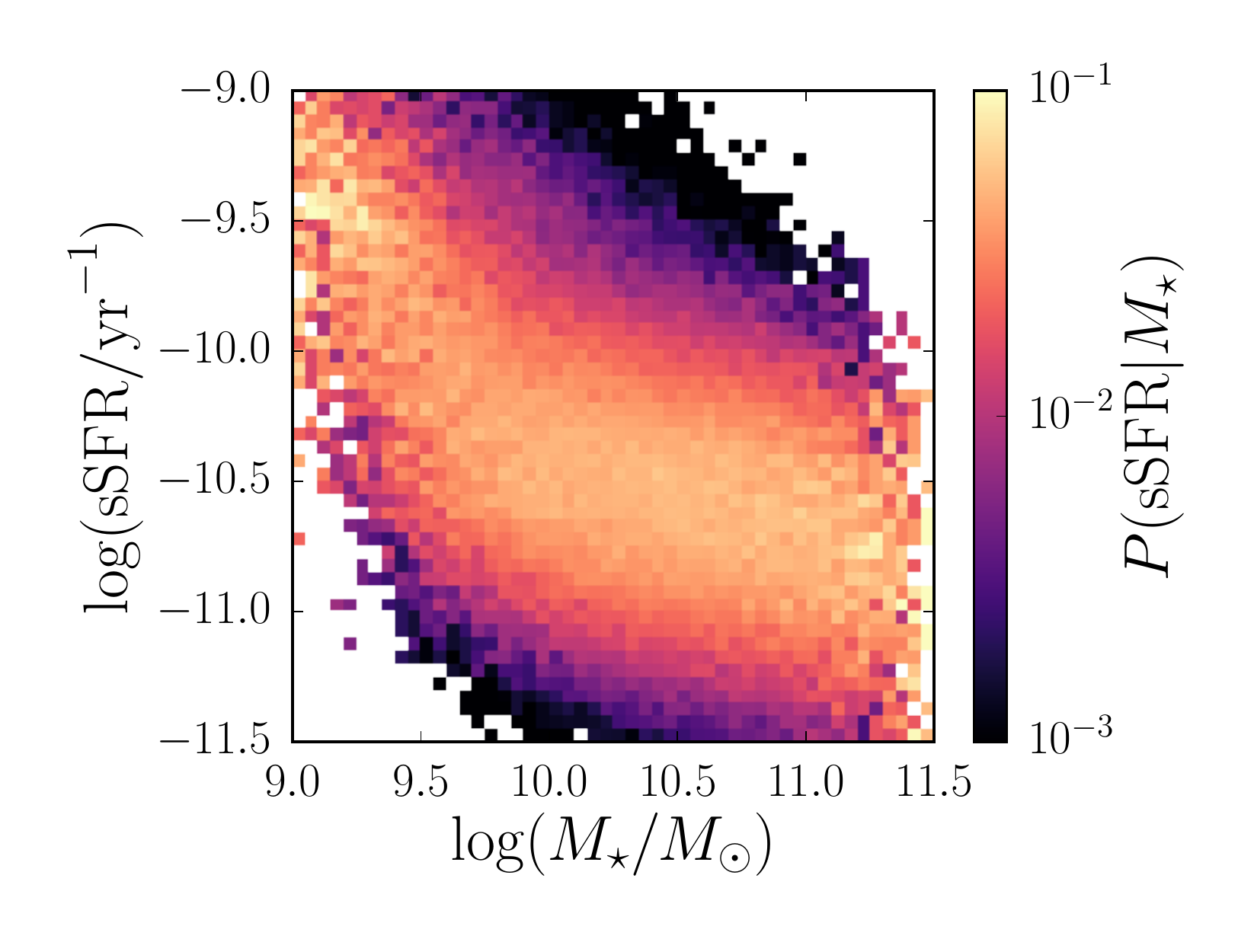}
\endminipage\hfill
\caption{Characterization of the SDSS galaxy sample. Left: distribution of $\log (\mathrm{SFR})$ given $\log (M_\star)$. The columns have been normalized separately so that each column gives the conditional probability distribution function of $\log (\mathrm{SFR})$ for a narrow range of values of $\log (M_\star)$. Right: distribution of $\log (\mathrm{sSFR})$ given $\log (M_\star)$. It is clear that the galaxy sample is biased toward higher (s)SFR at low stellar masses.
\label{pdfs}}
\end{figure*}

Figure~\ref{pdfs} gives an overview of the distribution of galaxy properties in our main galaxy sample. These two-dimensional histograms show how SFR and sSFR vary with $M_\star$. Each column is normalized separately such that a vertical slice gives the conditional probability distribution of the SFR or sSFR, given a value of stellar mass. The left panel shows a clear tendency for lower $M_\star$ galaxies to have lower SFRs, though at the lowest masses, higher SFRs become more likely. This reflects the fact that more vigorously star-forming galaxies are more luminous and are therefore more likely to be included in the sample near the magnitude and surface brightness limits of the spectroscopic survey. If the trend at higher masses is extended down to $\log (M_\star/M_\odot) \sim 9.0$, then the expected $\log (\mathrm{SFR}/ M_\odot \mathrm{\, yr}^{-1})$ would be $\sim -1.5$, much lower than is observed. Similarly, the right panel shows that the conditional probability distribution in sSFR is skewed toward high sSFR at the lowest masses. \textit{The inherent sample bias toward high (s)SFR at $\log (M_\star/M_\odot) < 10.0$ should be kept in mind.}

For some analyses in this paper, we require galaxy structural parameters. We use parameters derived from pure S\'{e}rsic model fits to the $r$-band SDSS images from \citet{simard11}. These structural parameters are only available for $111,982$ of the galaxies in our main galaxy sample. We refer to this subset of galaxies as the \textit{structural sample}.

\subsection{Gas Phase Metallicity Measurements}

We calculate gas phase metallicities (oxygen abundances) using three different strong line methods: (1) the prescription employed by M10, (2) the diagnostic grids of D13, and (3) the calibration of D16. The first metallicity calculation is done to ensure that we can reproduce the results of M10, and we use this relation as a baseline when we compare to the results from other diagnostics. We focus on results from the D13 diagnostic grids, but include a comparison to the updated calibration of D16 for completeness. Our aim here is to assess the dependence of the strength of the correlation between the mass-metallicity relation and SFR on the particular abundance diagnostic used. 

When using strong emission line methods to measure metallicity, one must be wary of potential degeneracies between $Z$ and ionization parameter $q$, defined to be the ratio of ionizing photon number flux to the density of hydrogen atoms, in units of cm s$^{-1}$. A given emission line ratio can correspond to many different pairs of $Z$ and $q$ values, so methods that determine $Z$ from a single line ratio implicitly assume some value of $q$ (e.g., \citealt{kd02, no14}). The dependence on ionization parameter is particularly important when considering variations in metallicity with SFR, since variations in the latter may well affect $q$ (e.g., \citealt{dopita14}), leading to spurious correlations between $Z$ and SFR if not taken into account.

\subsubsection{M10 Metallicity Diagnostic}
\label{mannucci_method}

We first check that we are able to reproduce the relation between metallicity, stellar mass, and star formation rate found by M10. The empirical/theoretical calibrations of \citet{maiolino08} are used to obtain two different measures of oxygen abundance from the R23 index \citep{pagel79}, defined as
\begin{equation}
\mathrm{R23} = \frac{\mathrm{[O \, \textsc{ii}]}\lambda \lambda 3726,3729 + \mathrm{[O \, \textsc{iii}]}\lambda \lambda 4959,5007}{\mathrm{H}\beta},
\end{equation}
and from the N2 index \citep{storchi-bergmann94}, $\mathrm{[N \, \textsc{ii}]}\lambda 6584/\mathrm{H}\alpha$. If both ratios are within the metallicity range within which the conversions were calibrated, only galaxies for which the two metallicity measurements agree within 0.25 dex are kept in the sample, and the metallicities of such galaxies are taken to be the average of the two metallicity measurements. This cut excludes $\sim 3\%$ of galaxies with very discrepant metallicity measurements, leaving us with $130,768$ star-forming galaxies. This is the main star-forming galaxy sample that we use throughout the paper.

The \citet{maiolino08} metallicity calibrations are polynomial fits to the relations between the R23 and N2 indices and metallicities determined from the theoretically calibrated \citet{kd02} technique, for the high-mass SDSS galaxies (direct method metallicities are used for low-mass galaxies). While the \citet{kd02} method does iteratively solve for both ionization parameter and metallicity, the polynomial fit relating the metallicity to a value of a given emission line ratio erases any information about the ionization parameter. This empirical/theoretical calibration effectively fixes the value of the ionization parameter, which introduces some bias into the metallicity measurements in a way that will correlate with SFR. Furthermore, the \citet{kd02} method was calibrated using an older version of the \texttt{MAPPINGS} photoionization code that used now outdated atomic data. This has a significant effect on the derived metallicities, as the old atomic data caused an overestimate of the electron temperature and therefore an underestimate of the metallicity \citep{nicholls13}. Taking these issues into account, it is quite likely that the M10 metallicities suffer from systematic errors that could potentially impact the observed strength of the correlation between metallicity and SFR. 

\subsubsection{D13 Diagnostic Grids}
\label{d13_grid_sec}

\begin{figure*}
\minipage{0.5\textwidth}
  \includegraphics[width=\linewidth]{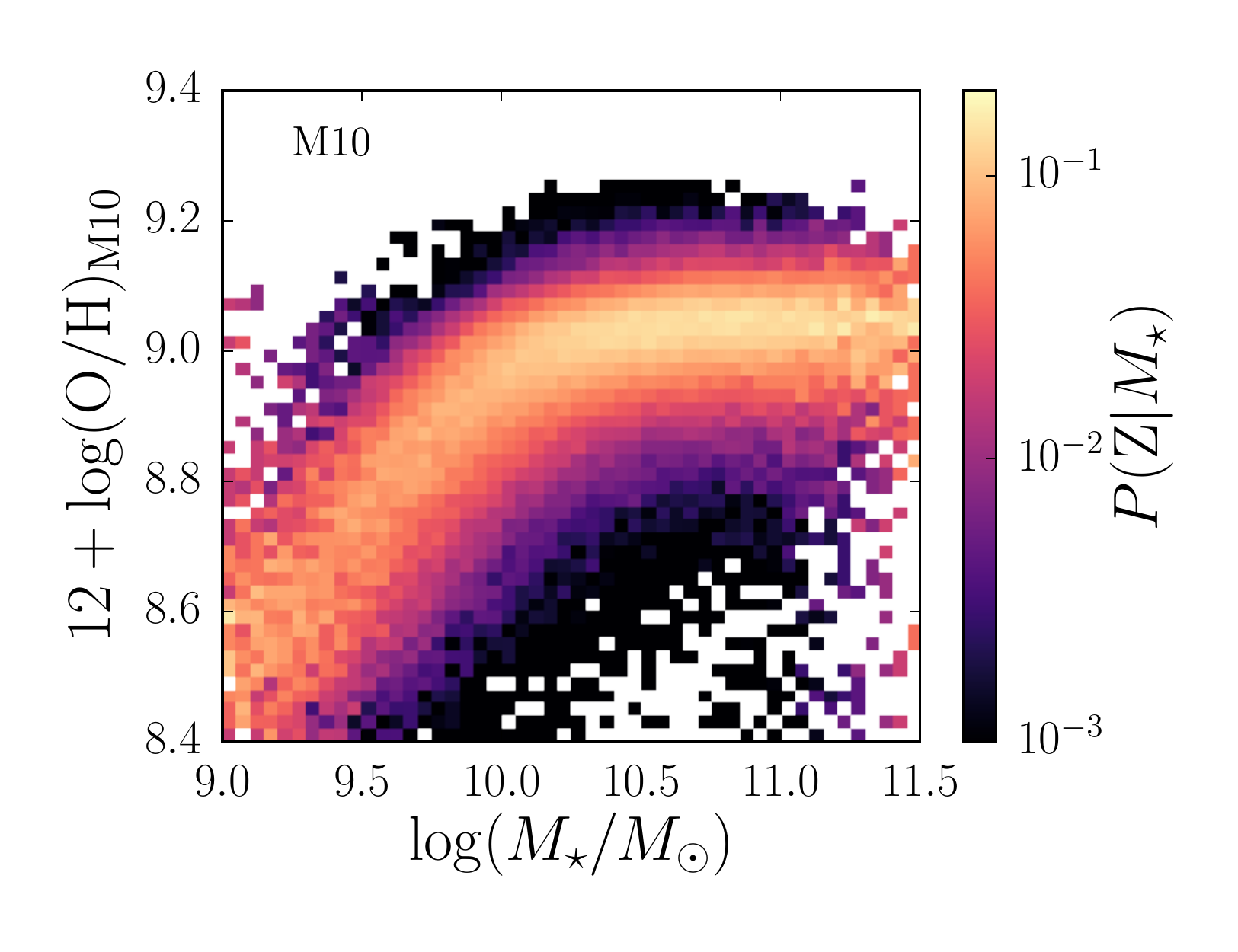}
\endminipage\hfill
\minipage{0.5\textwidth}
  \includegraphics[width=\linewidth]{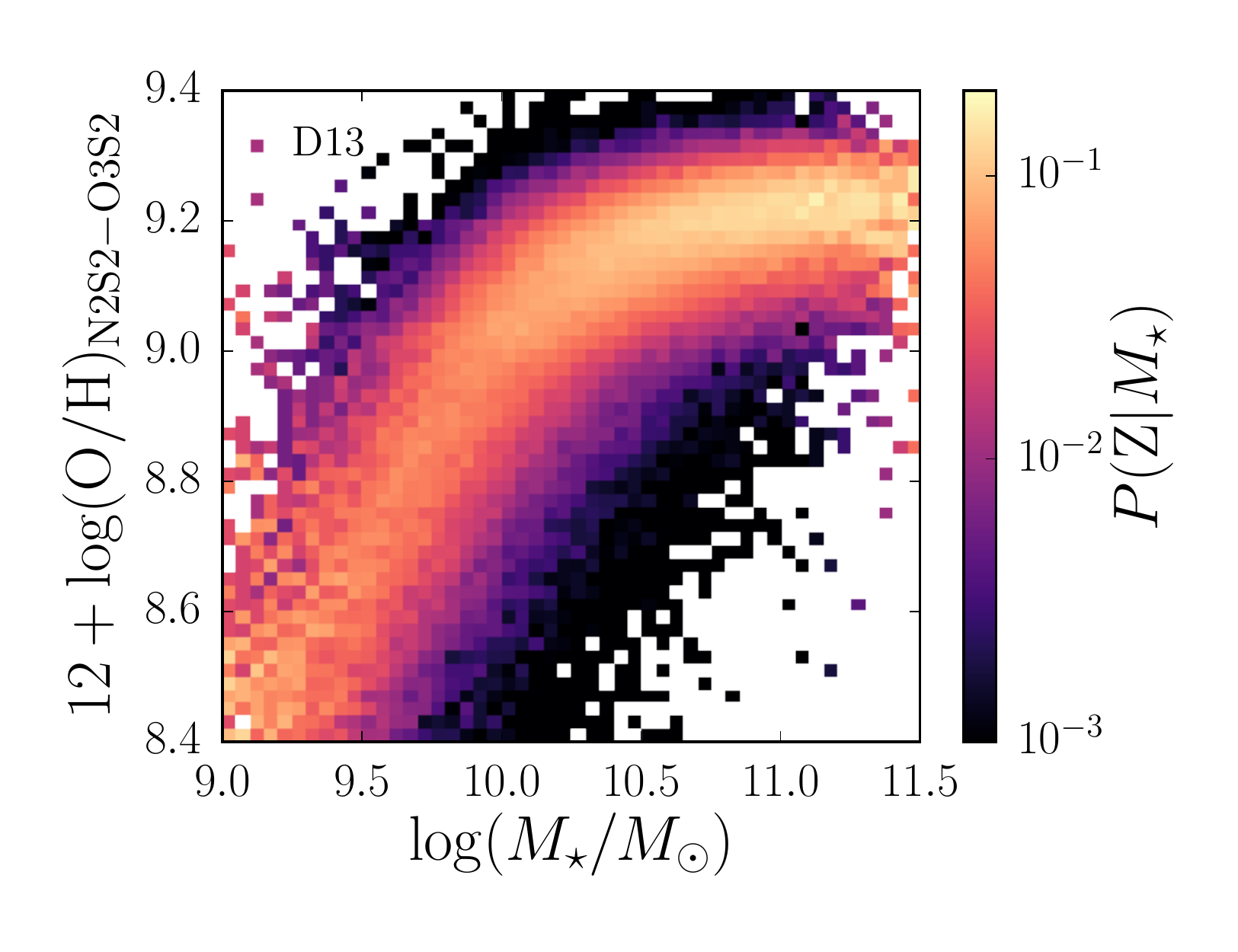}
\endminipage\hfill
\caption{Comparison of M10 and D13 mass-metallicity relations (MZRs). Left: Distribution of the M10 metallicities given $\log (M_\star)$. The columns have been normalized separately so that each column gives the conditional probability distribution function for $12 + \log (\mathrm{O/H})$ for a narrow range of values of $\log (M_\star)$. Right: Distribution of the fiducial D13 metallicities (from the N2S2-O3S2 grid) given $\log (M_\star)$. The D13 MZR reaches higher metallicities and has a steeper slope at low stellar masses than the M10 MZR.
\label{mzrs}}
\end{figure*}

\begin{table}
\caption{Emission line ratios used in the four abundance diagnostic grids from D13 considered in this paper.}
\label{grids_table}
\begin{center}
\begin{tabular}{ccc}
\bf D13 Grid & \bf Abundance-Sensitive & \bf Ionization-Sensitive \\
 \bf Name & \bf Ratio & \bf Ratio
\\ \hline \\ 

$\mathrm{N2S2-O3S2}$ & $\dfrac{\mathrm{[N \, \textsc{ii}]}\lambda 6584}{\mathrm{[S \, \textsc{ii}]}\lambda \lambda 6717,6731}$ &  $\dfrac{\mathrm{[O \, \textsc{iii}]}\lambda 5007}{\mathrm{[S \, \textsc{ii}]}\lambda \lambda 6717,6731}$ \\

$\mathrm{N2S2-O3O2}$ & $\dfrac{\mathrm{[N \, \textsc{ii}]}\lambda 6584}{\mathrm{[S \, \textsc{ii}]}\lambda \lambda 6717,6731}$ &  $\dfrac{\mathrm{[O \, \textsc{iii}]}\lambda 5007}{\mathrm{[O \, \textsc{ii}]}\lambda \lambda 3726,3729}$ \\

$\mathrm{N2O2-O3O2}$ & $\dfrac{\mathrm{[N \, \textsc{ii}]}\lambda 6584}{\mathrm{[O \, \textsc{ii}]}\lambda \lambda 3726,3729}$ &  $\dfrac{\mathrm{[O \, \textsc{iii}]}\lambda 5007}{\mathrm{[O \, \textsc{ii}]}\lambda \lambda 3726,3729}$ \\

$\mathrm{N2O2-O3S2}$ & $\dfrac{\mathrm{[N \, \textsc{ii}]}\lambda 6584}{\mathrm{[O \, \textsc{ii}]}\lambda \lambda 3726,3729}$ &  $\dfrac{\mathrm{[O \, \textsc{iii}]}\lambda 5007}{\mathrm{[S \, \textsc{ii}]}\lambda \lambda 6717,6731}$ \\

\end{tabular} 
\end{center}
\end{table}

Recently, D13 put forth a set of theoretically calibrated strong line abundance determination methods. These grids each map two emission line ratios to values of the metallicity and the ionization parameter. The \texttt{MAPPINGS} code used to calibrate the D13 grids includes up-to-date atomic data, allowing for more accurate determination of electron temperatures. The code also allows for $\kappa \mathrm{-distributed}$ electron energies, following the suggestion of \citet{nicholls12} that this type of energy distribution with a tail toward high energies may be common in astrophysical plasmas. D13 find a likely value of $\kappa = 20$ for H \textsc{ii} regions, which we adopt in our analysis. We verify that our results are unchanged by assuming a Maxwell-Boltzmann distribution ($\kappa = \infty$).

We obtain metallicities and ionization parameters for our sample from the \texttt{pyqz} Python module, made publicly available by D13, using four different grids that provide a clean separation of $Z$ and $q$. Each grid consists of one line ratio that is more sensitive to variations in abundance and one that is more sensitive to the ionization parameter; these are listed in Table~\ref{grids_table}. Because the grids use different combinations of emission line ratios, each has a different sensitivity to reddening and to the relative abundances of nitrogen and sulfur to oxygen. There is also no guarantee that all four grids give the same answer for $Z$ and $q$, given the many uncertainties in applying theoretical ionization models to real galaxy spectra. In an appendix, we present a detailed comparison of the results from all four D13 grids.

Of the four grids we consider in this paper, only the N2S2--O3S2 grid is insensitive to reddening corrections, because it involves line ratios that span a small range in wavelength. The other three grids all depend on the $\mathrm{[O \, \textsc{ii}]}\lambda \lambda 3726,3729$ doublet (Table~\ref{grids_table}), making parameters derived from those grids susceptible to systematics induced by assuming a reddening law and fixed intrinsic Balmer decrement. For this reason, we choose metallicities derived from the N2S2--O3S2 grid to be our ``fiducial" D13 metallicities, and use these in all plots shown below. We also report results from the N2O2--O3O2 grid, which uses two different emission line ratios from the fiducial grid. The metallicities derived from the two grids depending on the $\mathrm{[N \, \textsc{ii}]}/\mathrm{[O \, \textsc{ii}]}$ ratio are nearly identical. We find that the model fits for the N2S2--O3O2 grid may be problematic, as many observed galaxy emission line ratios lie outside of that model grid. Results from the N2S2--O3S2 and N2O2--O3O2 grids therefore span the range of reliable results from the D13 diagnostics; the appendix discusses these issues in detail.

Figure~\ref{mzrs} compares the mass-metallicity relations using the M10 (left panel) and the fiducial D13 (right panel) metallicity diagnostics. The characteristic shape of the MZR -- metallicity increasing with $M_\star$ at low $M_\star$ and then flattening at high $M_\star$ -- is clearly seen for both measures of metallicity, but the slopes and normalizations are different. The D13 diagnostic gives a steeper increase of metallicity with $M_\star$ than the M10 diagnostic, spanning a wider range of metallicities and reaching a maximum metallicity that is $\sim 0.15\, \mathrm{dex}$ higher. The higher maximum metallicity is likely due to a combination of the updated atomic data in the D13 models, the assumption of $\kappa \mathrm{-distributed}$ of electron energies, and the fact that the R23 and N2 indicators used in the M10 diagnostic are known to saturate at high metallicity.

\subsubsection{D16 Metallicity Diagnostic}
\label{d16_sec}

D16 presented a new metallicity diagnostic specifically designed to measure metallicities in high redshift galaxies. This diagnostic uses only $\mathrm{[N \, \textsc{ii}]}\lambda 6584/\mathrm{[S \, \textsc{ii}]}\lambda \lambda 6717,6731$ and the N2 index, so the lines involved span a narrow range in wavelength, making reddening corrections negligible. These line ratios are sensitive to metallicity, but only weakly depend on the ionization parameter and gas pressure. The D16 diagnostic was calibrated using the most recent version of the \texttt{MAPPINGS} code and assuming $\kappa = \infty$, so slight differences from the models used to calibrate the D13 grids are expected. 

\section{Systematics Affecting the Strength of the Correlation with SFR}
\label{results}

We will now investigate potential systematic uncertainties and their effects on the $M_\star-Z-\mathrm{SFR}$ relation. We start with systematic uncertainties in the metallicity determinations, then investigate biases due to the stellar mass estimates, aperture coverage, and dust. For each of these factors studied, we find indications of the potential presence of systematic effects in the $M_\star-Z-\mathrm{SFR}$ relation. 

\subsection{Metallicity Uncertainties}

We begin our investigation of potential systematic errors affecting the $M_\star-Z-\mathrm{SFR}$ relation with a comparison of results obtained using different metallicity measurements, which are known to yield different strengths of correlation with SFR \citep{am13, salim14}. Here, we present the first analysis of this relation using the new D13 abundance diagnostics and compare to the results using the methods of M10 and D16.

\subsubsection{Metallicity Measurement Technique}
\label{z_measurements}

\begin{figure*}
\plottwo{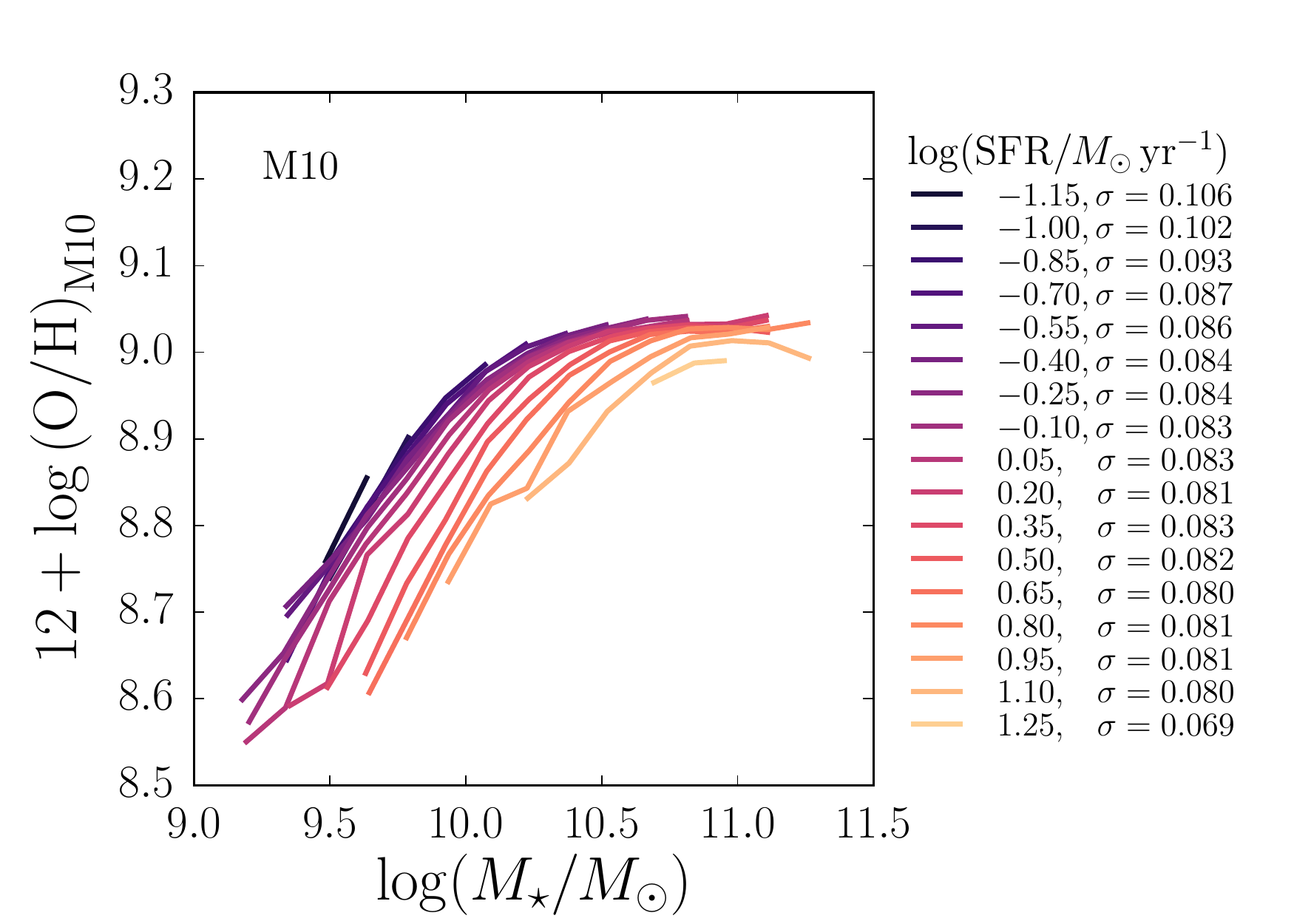}{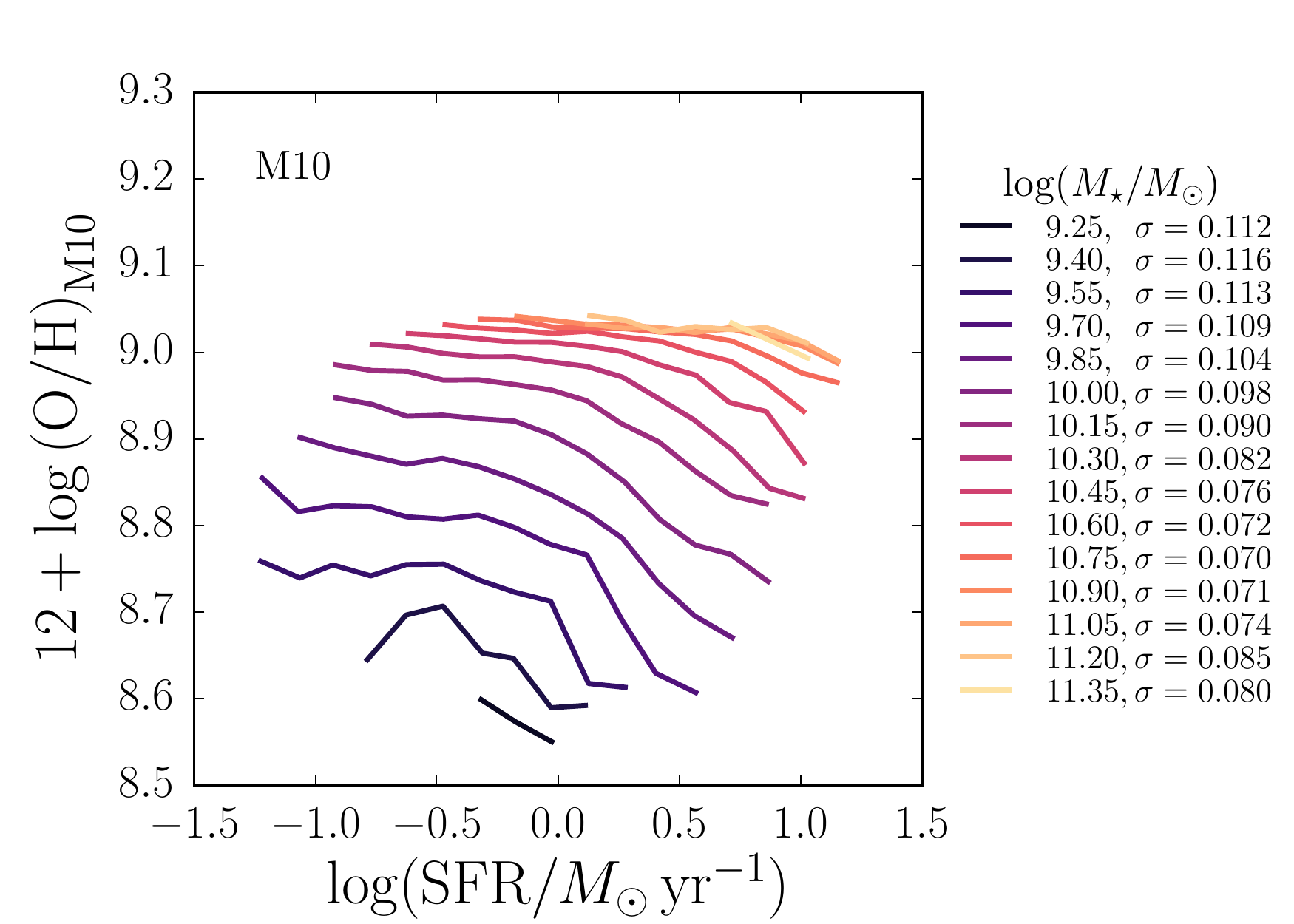}
\caption{Reproduction of the M10 $M_\star-Z-\mathrm{SFR}$ relation. Median gas phase metallicity is plotted against $\log (M_\star)$ in bins of $\log (\mathrm{SFR})$ (left) and against $\log (\mathrm{SFR})$ in bins of $\log (M_\star)$ (right). All bins have 0.15 dex width in each $\log (M_\star)$ and $\log (\mathrm{SFR})$ and each bin contains at least 50 galaxies. Metallicities are calculated following M10 using the R23 and N2 indices with calibrations from \citet{maiolino08}. The dispersion in metallicity about each median line is reported in the legends. We recover the same correlation with SFR reported by M10.}
\label{m10_fmr}
\end{figure*}

\begin{figure*}
\plottwo{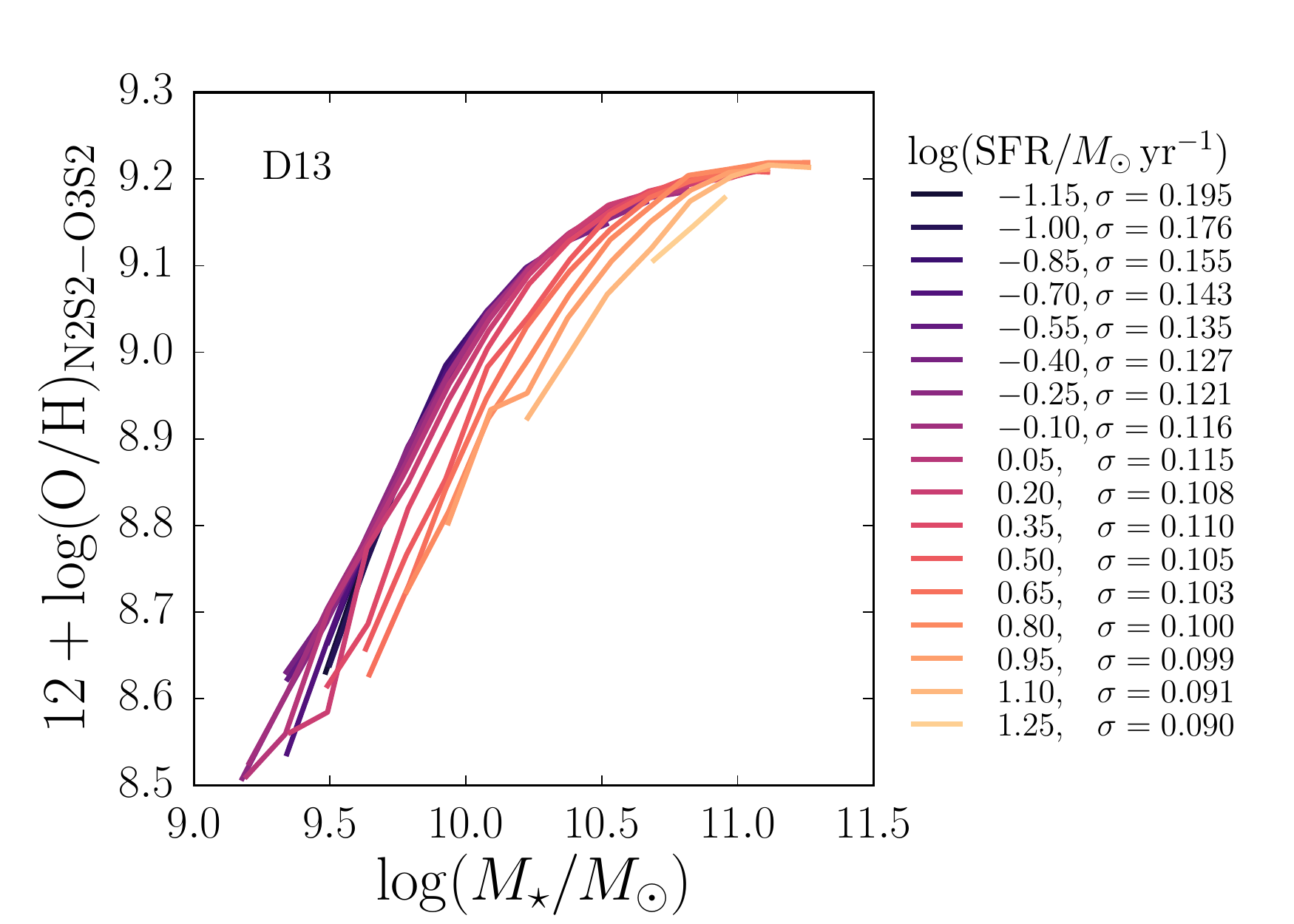}{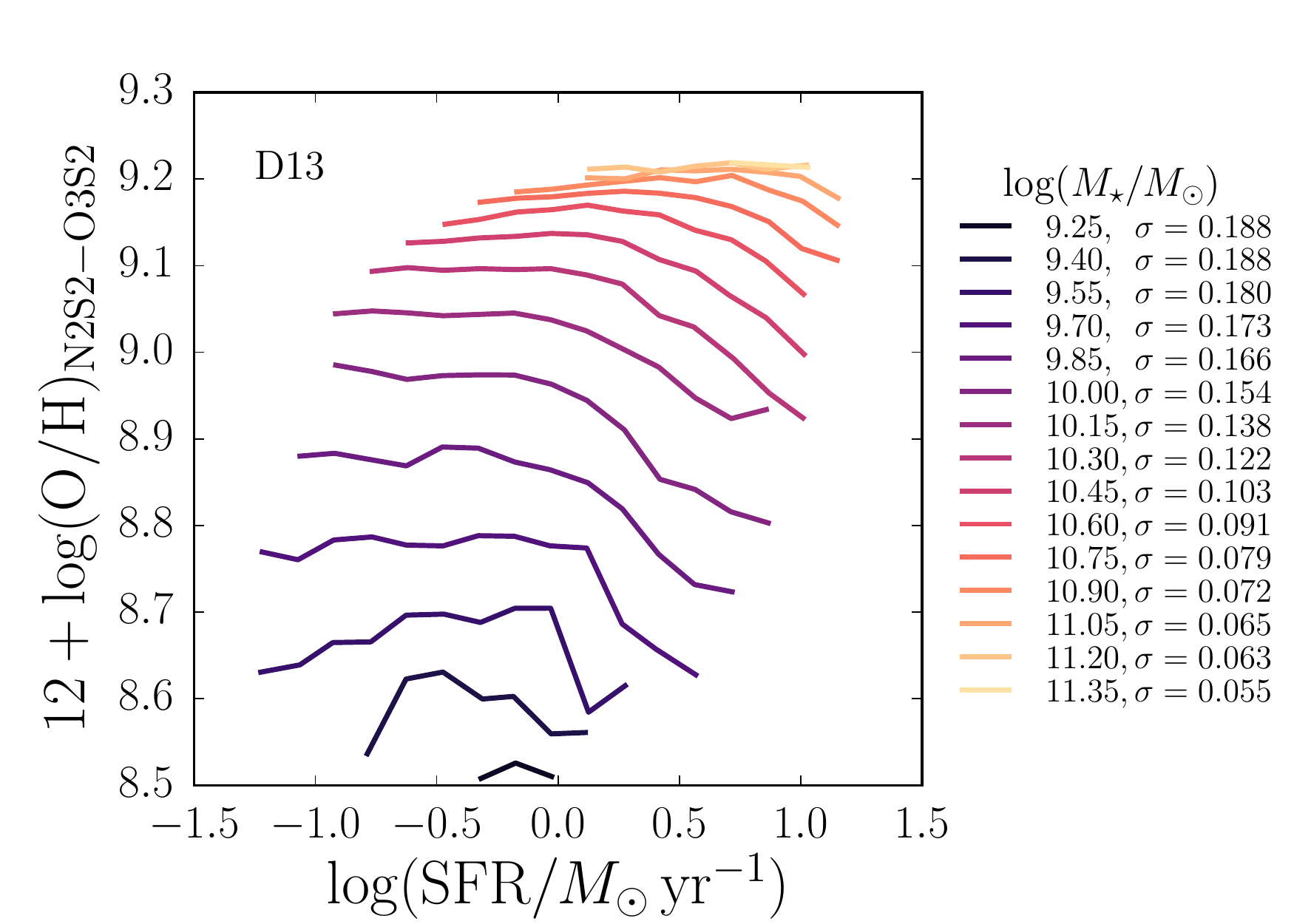}
\caption{$M_\star-Z-\mathrm{SFR}$ relation using metallicities from the fiducial D13 abundance diagnostic (N2S2--O3S2). Binning is performed as in Figure~\ref{m10_fmr}. The correlation with SFR using this abundance diagnostic is weaker than that found by M10.}
\label{d13_fmr}
\end{figure*}

\begin{figure*}
\plottwo{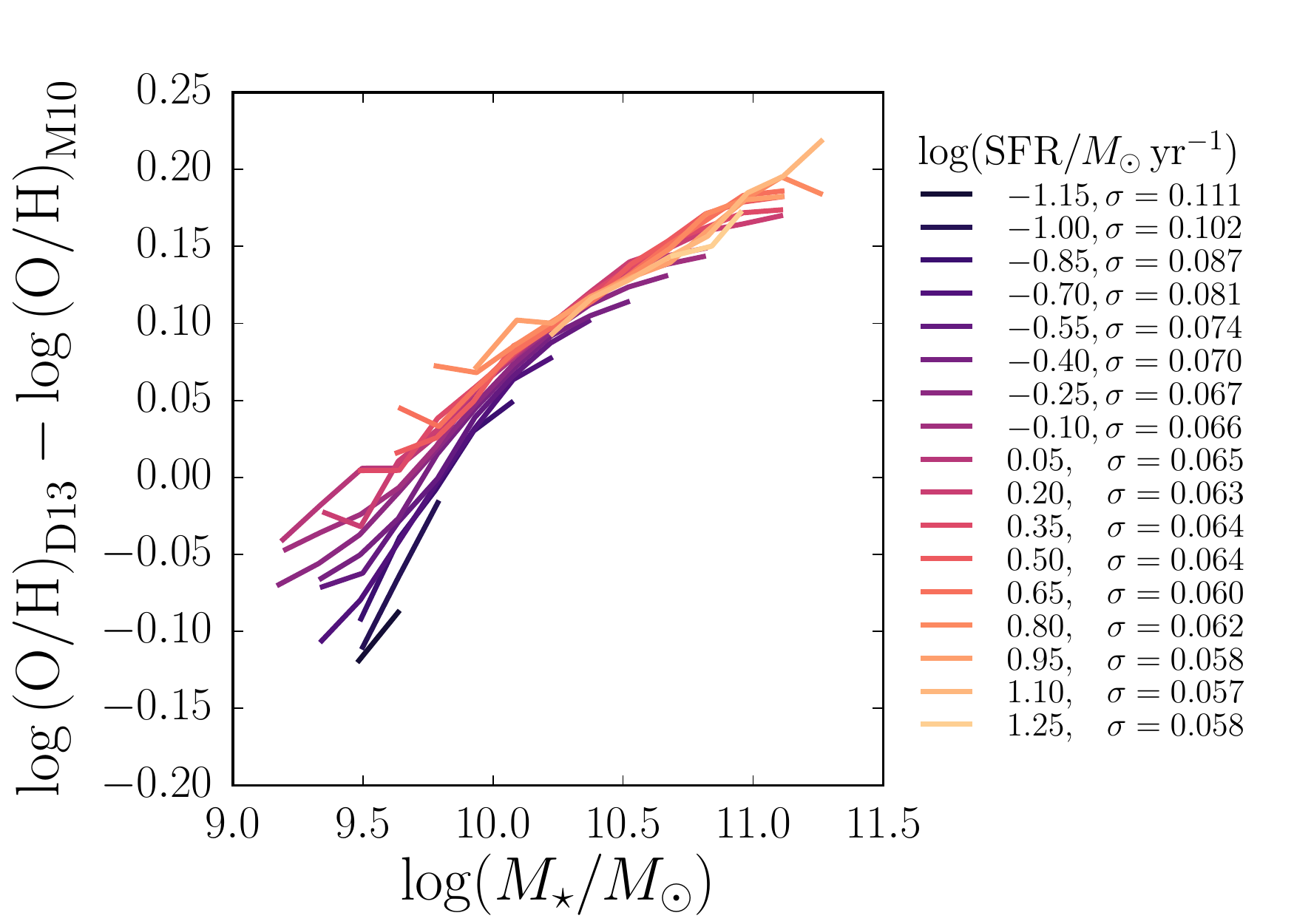}{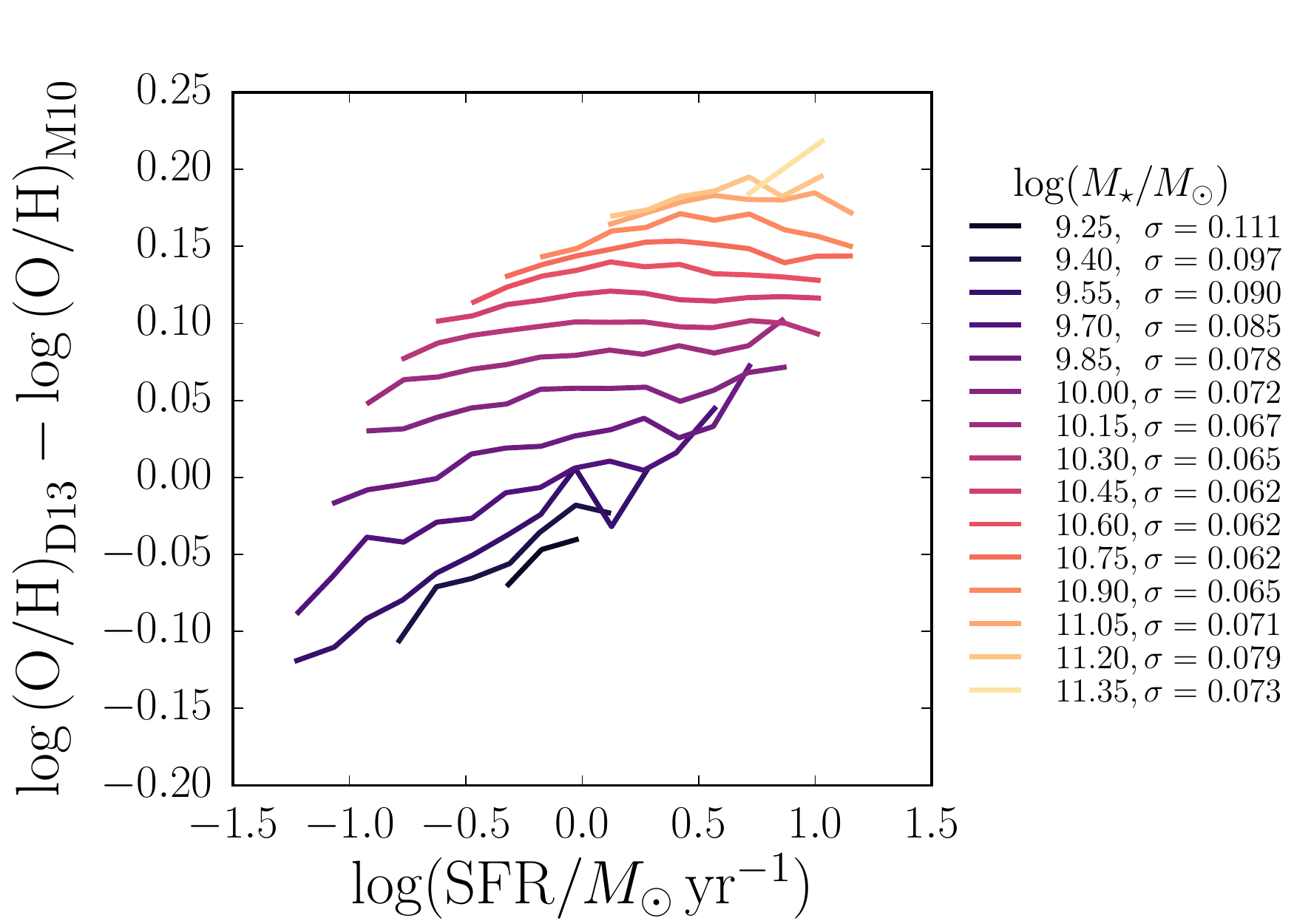}
\caption{Difference between the M10 and D13 $M_\star-Z-\mathrm{SFR}$ relations. We plot the median difference between metallicity calculated from the fiducial D13 N2S2--O3S2 abundance diagnostic grid and metallicity calculated following M10 against $\log (M_\star)$ (left) and $\log (\mathrm{SFR})$ (right). Again, binning is performed as in Figure~\ref{m10_fmr}. The difference is only a function of stellar mass at high masses, but also varies with SFR at the low mass end. }
\label{deltaZ}
\end{figure*}

\begin{figure*}[!ht]
\minipage{0.5\textwidth}
  \includegraphics[width=\linewidth]{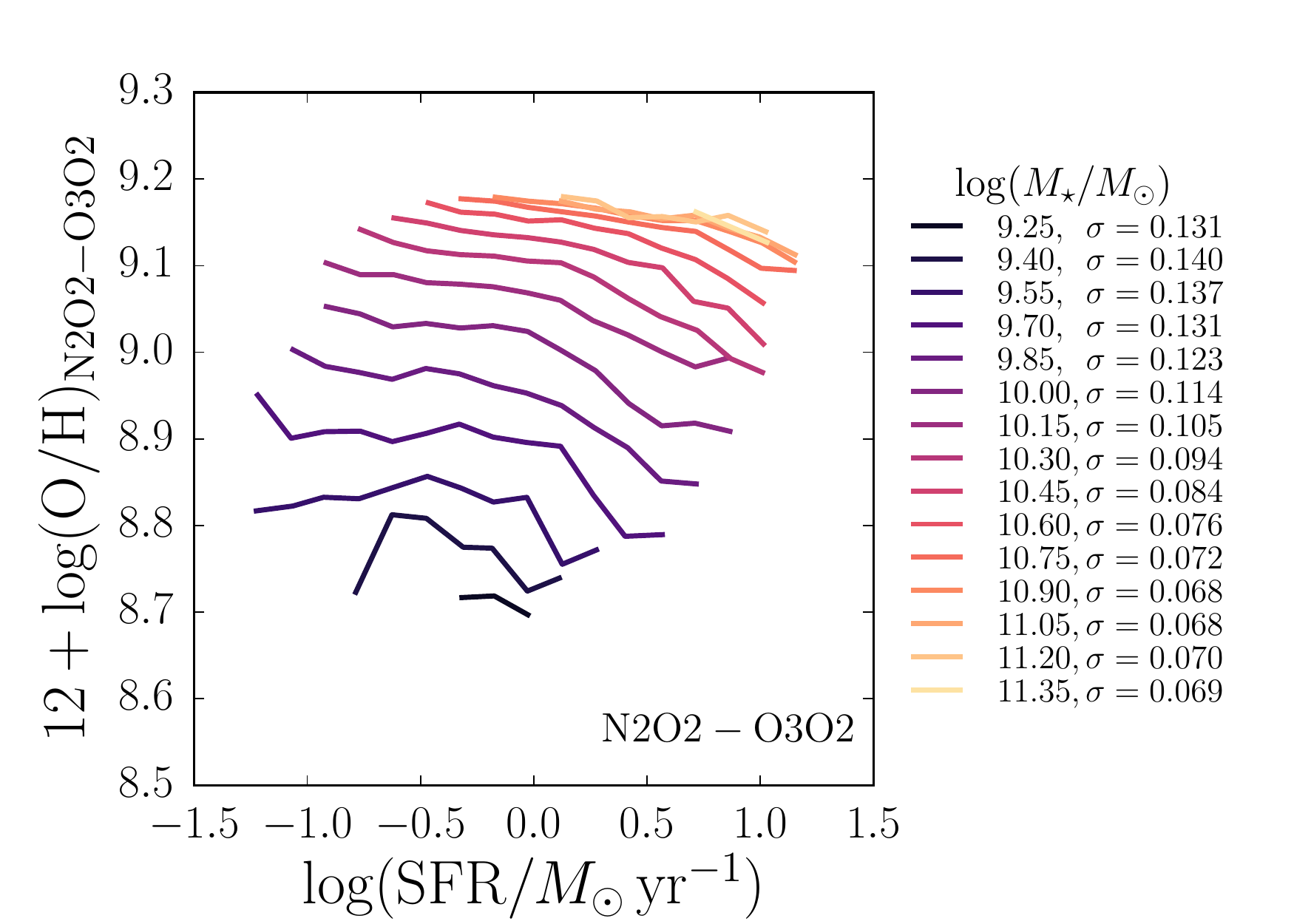}
\endminipage\hfill
\minipage{0.5\textwidth}
  \includegraphics[width=\linewidth]{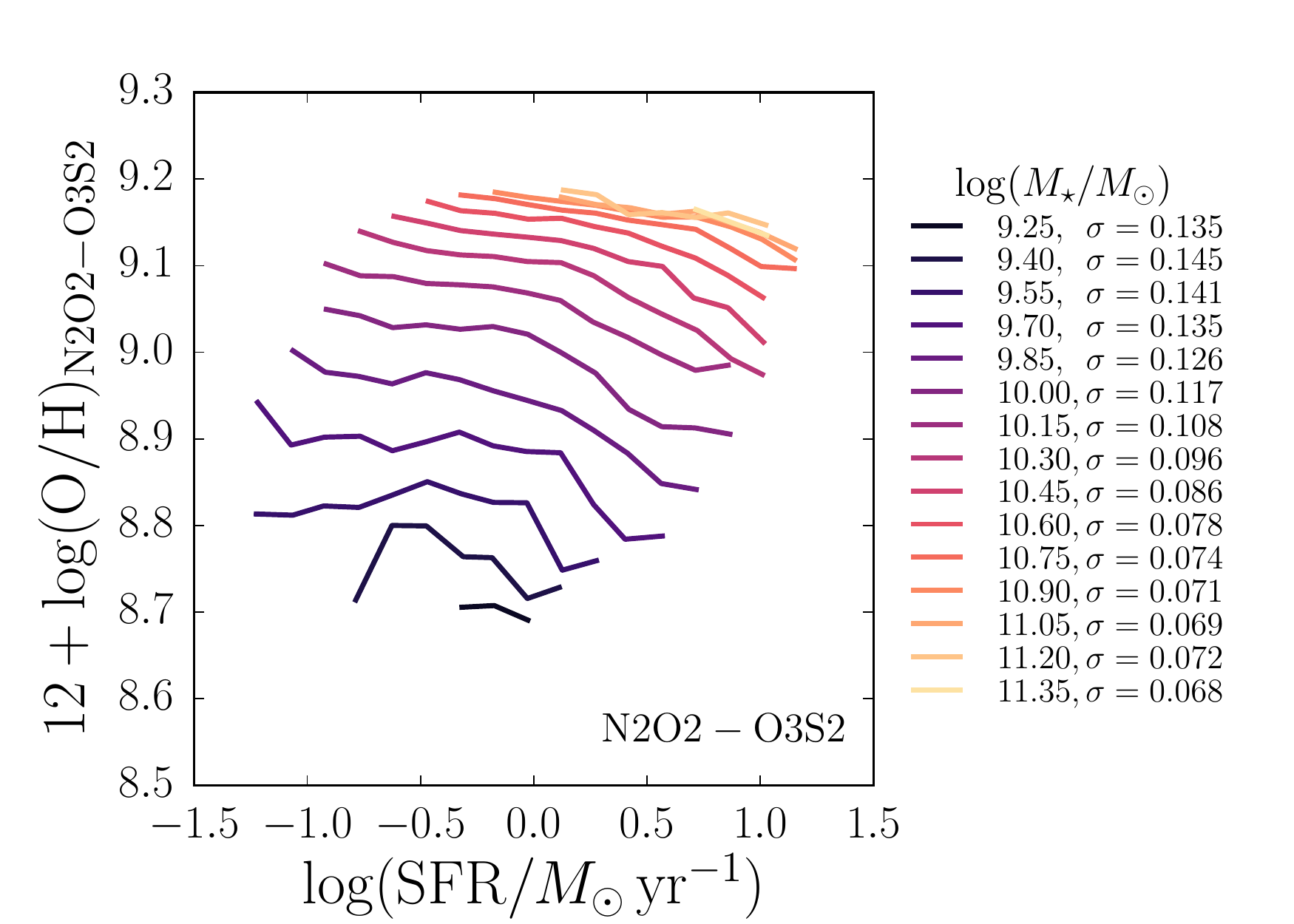}
\endminipage\hfill
\minipage{0.5\textwidth}
  \includegraphics[width=\linewidth]{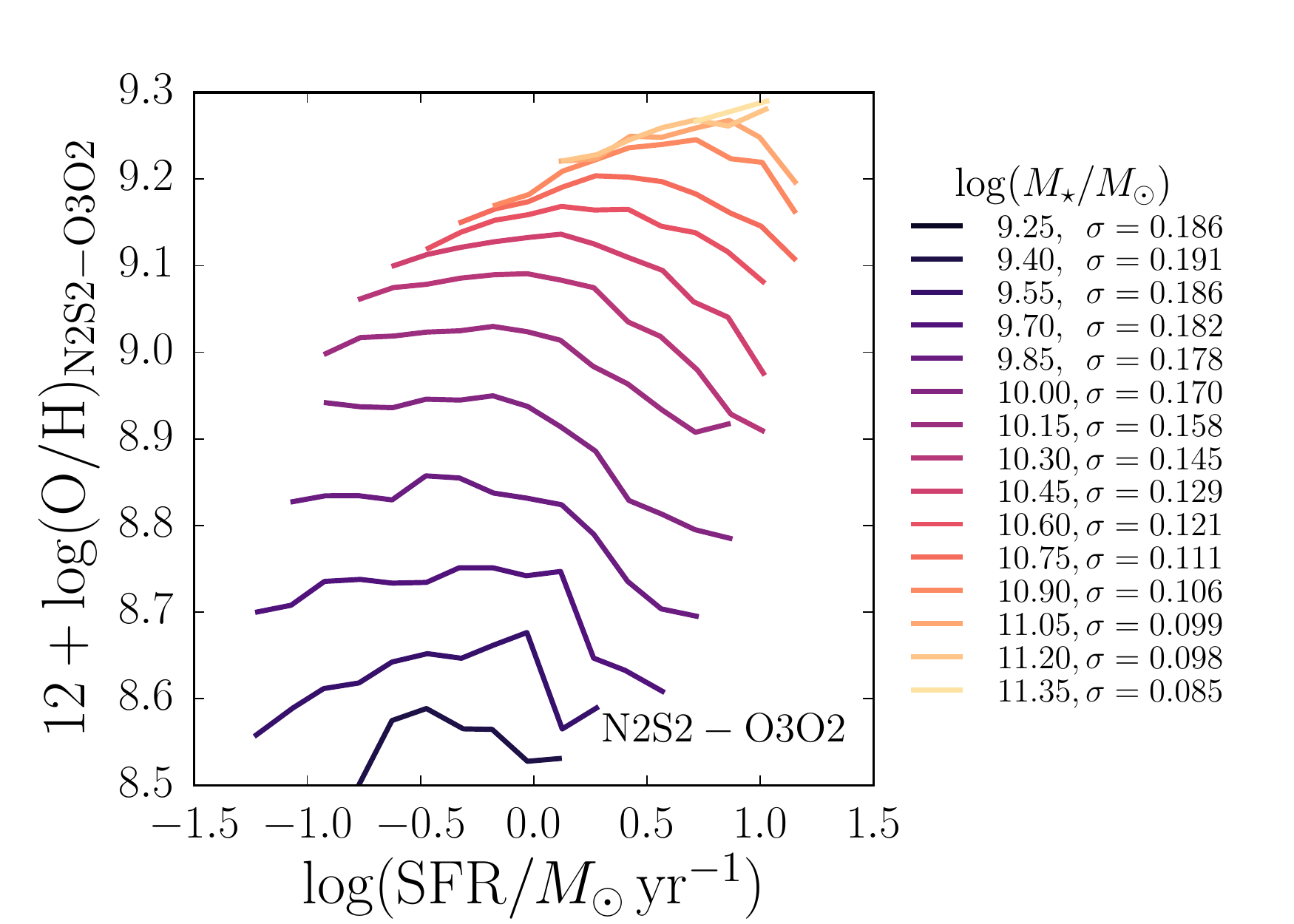}
\endminipage
\minipage{0.5\textwidth}
  \includegraphics[width=\linewidth]{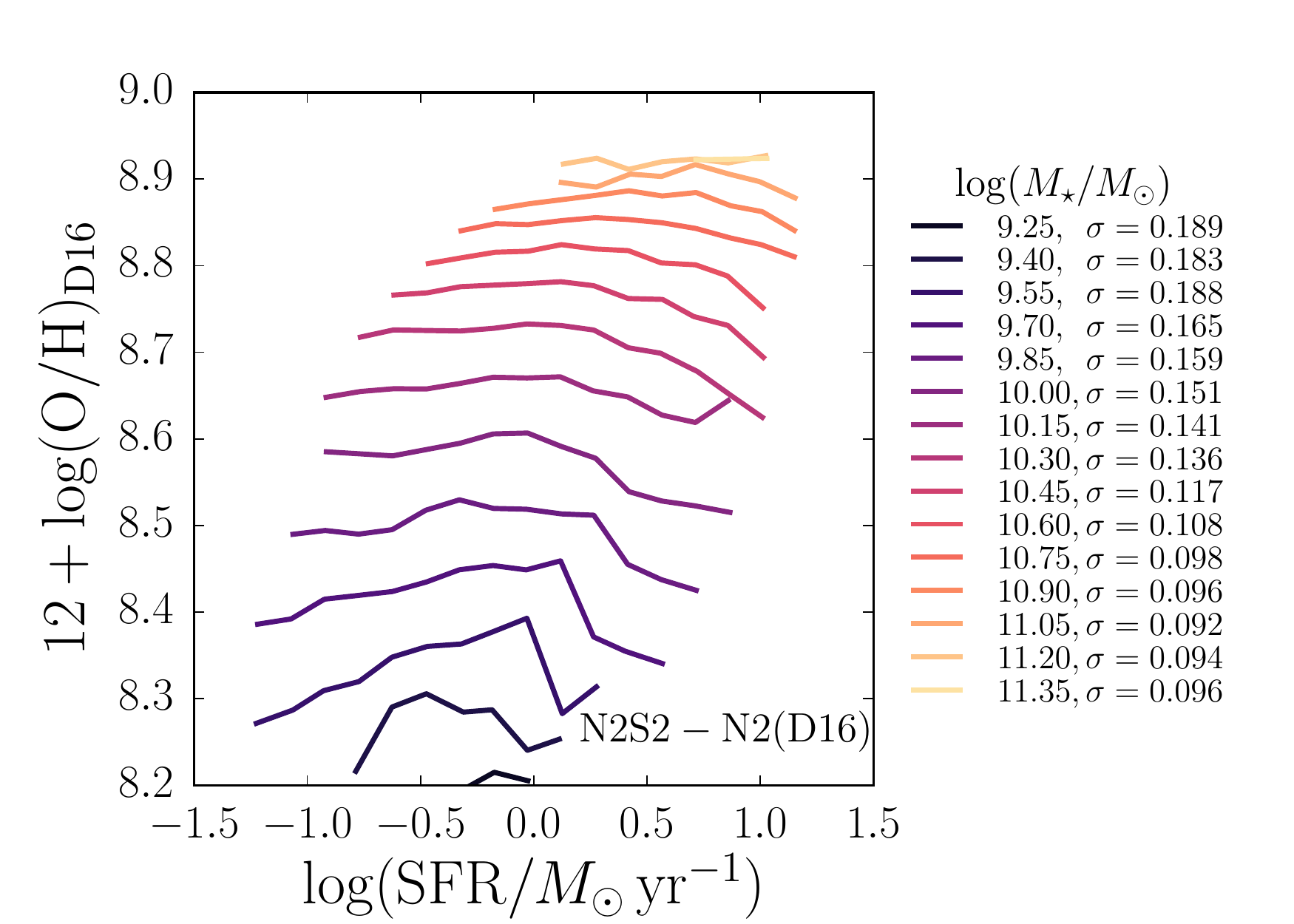}
\endminipage
\caption{Comparison of $M_\star-Z-\mathrm{SFR}$ relations using metallicities from other D13 abundance diagnostic grids and the D16 abundance diagnostic. Each plot is analogous to the right panel of Figure~\ref{d13_fmr} but with metallicities measured using different diagnostics: D13 N2O2--O3O2 (top left), D13 N2O2--O3S2 (top right), D13 N2S2--O3O2 (bottom left), and D16 (bottom right). Recall that the N2S2-O3O2 grid (bottom left) is problematic and is likely not a reliable metallicity estimator for SDSS galaxies (see the appendix). The metallicity axis spans a different range in the bottom right panel because the D16 metallicity calibration produces lower metallicities than either the M10 or D13 diagnostics. For all of these diagnostics, we find a weaker correlation with SFR than was found by M10. \label{d13_comp}}
\end{figure*}

In Figure~\ref{m10_fmr}, we show our reproduction of the $M_\star-Z-\mathrm{SFR}$ relation found by M10. Galaxies are binned in both $M_\star$ and SFR in bins of 0.15 dex width in each quantity. Bins containing fewer than 50 galaxies and bins with large median fractional errors in the Balmer decrement are excluded from the plots. The latter cut is made to ensure that spurious reddening corrections are not driving the observed trends and avoids biasing the sample toward higher S/N galaxies. Metallicities are calculated using the \citet{maiolino08} calibrations of the R23 and N2 emission line ratios as described in Section~\ref{mannucci_method}. We plot the median value of $12 + \log \mathrm{(O/H)}$ as a function of $M_\star$ (SFR) for each bin of SFR ($M_\star$). The dispersions in individual metallicities about the median relations are listed in the legends; these are larger than the spacing between the lines. Our results using this metallicity calibration are consistent with those found by M10.

Figure~\ref{d13_fmr} shows the same $M_\star-Z-\mathrm{SFR}$ relation, but using metallicities calculated from the fiducial D13 abundance diagnostic (N2S2--O3S2). The binning procedure is exactly the same as in Figure~\ref{m10_fmr}. There are two important differences between these sets of plots. When the D13 abundance diagnostic is used, (1) metallicity is more weakly correlated with SFR and (2) the dispersion within a given $M_\star$ or SFR bin is larger. The weaker SFR correlation can be seen in the smaller spread of the lines in the left panel of Figure~\ref{d13_fmr} and in the flatter lines of constant $M_\star$ in the right panel. 

In the right panels of both Figure~\ref{m10_fmr} and Figure~\ref{d13_fmr}, the lines of constant $M_\star$ are much less smooth and even turn over at low SFR. This is due to the paucity of low $M_\star$ galaxies in our sample and the large spread in metallicity within each $M_\star$ bin. Because the lowest $M_\star$ bins are not very well populated, the median metallicity in each bin of $M_\star$ and SFR is less stable against stochastic sampling of the galaxy population. These decreases in $Z$ at low SFR for the lowest $M_\star$ bins are not physically meaningful.

Figure~\ref{deltaZ} gives a quantitative comparison of the two metallicity diagnostics as a function of $M_\star$ and SFR. We plot the median difference between the abundances calculated from the the fiducial D13 grid and the M10 diagnostic. The difference in metallicity between the two grids is correlated with $M_\star$ across the full range of stellar mass and is also correlated with SFR at small $M_\star$. The dispersion about the median metallicity difference becomes large at low SFR and low $M_\star$. 

M10 established the convention of quantifying the strength of correlation with SFR using the parameter $\alpha$ that gives the least scatter in median metallicities in bins of $\mu_\alpha$ and SFR, where $\mu_\alpha$ is defined as 
\begin{equation}
\mu_\alpha = \log \left( M_\star \right) - \alpha \log \left( \mathrm{SFR} \right).
\end{equation}
In this definition, $\alpha = 0$ indicates that metallicity is independent of SFR, while larger values of $\alpha$ indicate stronger correlation with SFR. The $\mu_\alpha$ parameter defines a two-dimensional projection of the three-dimensional $M_\star-Z-\mathrm{SFR}$ relation. The mass-metallicity relation is such a projection, corresponding to $\alpha = 0$, where the $M_\star-Z-\mathrm{SFR}$ relation is collapsed along the SFR axis. If there were no secondary correlation between metallicity and SFR, this would be the two-dimensional projection that gives minimum scatter of median metallicity in bins of $\mu_\alpha$ and SFR about the median $Z$ vs. $\mu_\alpha$ relation. However, if metallicity is anti-correlated with SFR at fixed stellar mass, then a different projection, Z vs. $\mu_\alpha$, where $\alpha > 0$, will yield smaller scatter in median metallicities in bins of $\mu_\alpha$ and SFR. Therefore, the $\alpha$ parameter is a measure of the strength of the anti-correlation between metallicity and SFR in the data. 

We compare the strengths of correlation with SFR given by different metallicity measurement techniques by calculating the value of $\alpha$ for each version of the $M_\star-Z-\mathrm{SFR}$ relation considered here, with results listed in Table~\ref{alphas_table}. We find $\alpha=0.28$ using the M10 definition of metallicity\footnote{M10 find $\alpha = 0.32$; we find that the difference in scatter between $\alpha = 0.28$ and $\alpha = 0.32$ is just 0.002 dex, which we consider to be insignificant.} (Figure~\ref{m10_fmr}). When the fiducial D13 N2S2--O3S2 grid is used to determine metallicities (Figure~\ref{d13_fmr}), we obtain $\alpha = 0.11$, indicating more than a factor of 2 weaker anti-correlation with SFR. 

In Figure~\ref{d13_comp} we show the $M_\star-Z-\mathrm{SFR}$ relation using metallicities calculated from three other D13 abundance diagnostic grids and from the D16 diagnostic. We find that for all D13 grids and for the D16 diagnostic, the correlation between the MZR and SFR is always weaker than that found by M10 (Table~\ref{alphas_table}). However, the shape of the relation depends on the specific diagnostic that is used. Clearly, the different D13 grids give different metallicities, even though the theoretical grids are self-consistent when applied to individual $\mathrm{H \, \textsc{ii}}$ regions. Discrepancies between the grids are therefore likely due to the inability of a single value of $Z$ or $q$ to appropriately describe the conditions across the region of a galaxy spanned by the SDSS fiber.

The two D13 grids depending on [N \textsc{ii}]/[O \textsc{ii}] (top row of Figure~\ref{d13_comp}) yield a stronger correlation between $Z$ and SFR than any of the diagnostics depending on [N \textsc{ii}]/[S \textsc{ii}]. The N2O2--O3O2 grid yields $\alpha = 0.19$, which is still $\sim 30\%$ weaker than the M10 result. Within bins of $M_\star$ and SFR, N2O2--O3O2 metallicities are always greater than M10 metallicities. The difference between the two metallicity measures is largely independent of $M_\star$, but increases slightly at low SFR.

The key point here is that the \textit{choice of metallicity measurement technique affects the overall shape of the relationship between $M_\star$, SFR, and $Z$ and, in particular, changes the strength of the correlation between the MZR and SFR.}  This result is consistent with previous findings that the correlation between $Z$ and SFR changes depending on the metallicity calibration used \citep{am13, salim14}. 

\begin{table}
\caption{Best fitting value of $\alpha$ for each version of the $M_\star-Z-\mathrm{SFR}$ relation considered in this paper. }
\label{alphas_table}
\begin{center}
\begin{tabular}{lc}
\bf Relation & \boldmath$\alpha$
\\ \hline \\
M10 R23 \& N2 & 0.28\\
D13 $\mathrm{N2S2 \, -O3S2}$ (fiducial) & 0.11\\
D13 $\mathrm{N2O2-O3S2}$ & 0.19\\
D13 $\mathrm{N2O2-O3O2}$ & 0.19\\
D13 $\mathrm{N2S2 \, -O3O2}$\tablenotemark{a} & 0.06\\
D16 N2S2 \& N2 & 0.00\\
M10, $\log{(\mathrm{sSFR/yr}^{-1})} < -10.0$ & 0.13\\
M10, $\log{(\mathrm{sSFR/yr}^{-1})} > -10.0$ & 0.42\\
D13 $\mathrm{N2S2 \, -O3S2}$, $\log{(\mathrm{sSFR/yr}^{-1})} < -10.0$ & 0.00\\
D13 $\mathrm{N2S2 \, -O3S2}$, $\log{(\mathrm{sSFR/yr}^{-1})} > -10.0$ & 0.27\\
D13 $\mathrm{N2O2-O3O2}$, $\log{(\mathrm{sSFR/yr}^{-1})} < -10.0$ & 0.14\\
D13 $\mathrm{N2O2-O3O2}$, $\log{(\mathrm{sSFR/yr}^{-1})} > -10.0$ & 0.35\\
D16, $\log{(\mathrm{sSFR/yr}^{-1})} < -10.0$ & 0.00\\
D16, $\log{(\mathrm{sSFR/yr}^{-1})} > -10.0$ & 0.13\\
Simulation: $M_\star$ noisy at high sSFR & 0.02\\
Simulation: $M_\star$ overestimated up to 0.2 dex at high sSFR & 0.05\\
Simulation: $M_\star$ overestimated up to 0.4 dex at high sSFR & 0.20\\
M10 R23 \& N2, structural sample & 0.26\\
D13 $\mathrm{N2S2 \, -O3S2}$, structural sample & 0.09\\
D13 $\mathrm{N2O2-O3O2}$, structural sample & 0.20\\
M10 R23 \& N2, slightly reddened & 0.28\\
M10 R23 \& N2, highly reddened & 0.18\\
D13 $\mathrm{N2S2 \, -O3S2}$, slightly reddened & 0.15\\
D13 $\mathrm{N2S2 \, -O3S2}$, highly reddened & 0.04\\
D13 $\mathrm{N2O2-O3O2}$, slightly reddened & 0.21\\
D13 $\mathrm{N2O2-O3O2}$, highly reddened & 0.17\\
\end{tabular}

\tablenotetext{1}{We find that the D13 N2S2--O3O2 grid produces problematic fits to the galaxy emission line ratios and is therefore not a reliable metallicity estimator for the SDSS sample (see the appendix).}
\tablecomments{Larger values of $\alpha$ indicate stronger anti-correlation between $Z$ and SFR; $\alpha=0$ indicates no correlation.}

\end{center}
\end{table}

\subsubsection{Determination of $\alpha$ and the Reduction in Scatter}

The values of $\alpha$ reported in Table~\ref{alphas_table} are those that give the minimum scatter in the \textit{median} values of $Z$ in bins of $\mu_\alpha$ and SFR (on the order of 100 measurements) about the median $Z-\mu_\alpha$ relation. For these $\alpha$ measurements, the scatter is reduced relative to the scatter about the mass-metallicity (i.e., $Z$ vs. $\mu_0$) relation by $\sim 20 - 40\%$. One can also calculate the value of $\alpha$ that minimizes the scatter in the values of $Z$ for individual galaxies (on the order of $100,000$ measurements). As the values of scatter reported in our figure legends show, there is always some scatter in individual metallicities about the median values in a given $M_\star$ or SFR bin at the level of $\sim 0.05-0.2 \mathrm{\, dex}$. When the best fitting $\alpha$ is chosen to minimize the scatter in individual metallicities, the reduction in scatter is on the order of just a few percent \citep{salim14}, making the measured $\alpha$ both smaller and less reliable. For comparisons between observations and theoretical models of the $M_\star-Z-\mathrm{SFR}$ relation, it is essential that the strength of the secondary correlation between $Z$ and SFR be quantified in a consistent way.

\subsubsection{The $M_\star-Z-\mathrm{sSFR}$ Relation}

\begin{figure}
\includegraphics[width=0.48\textwidth]{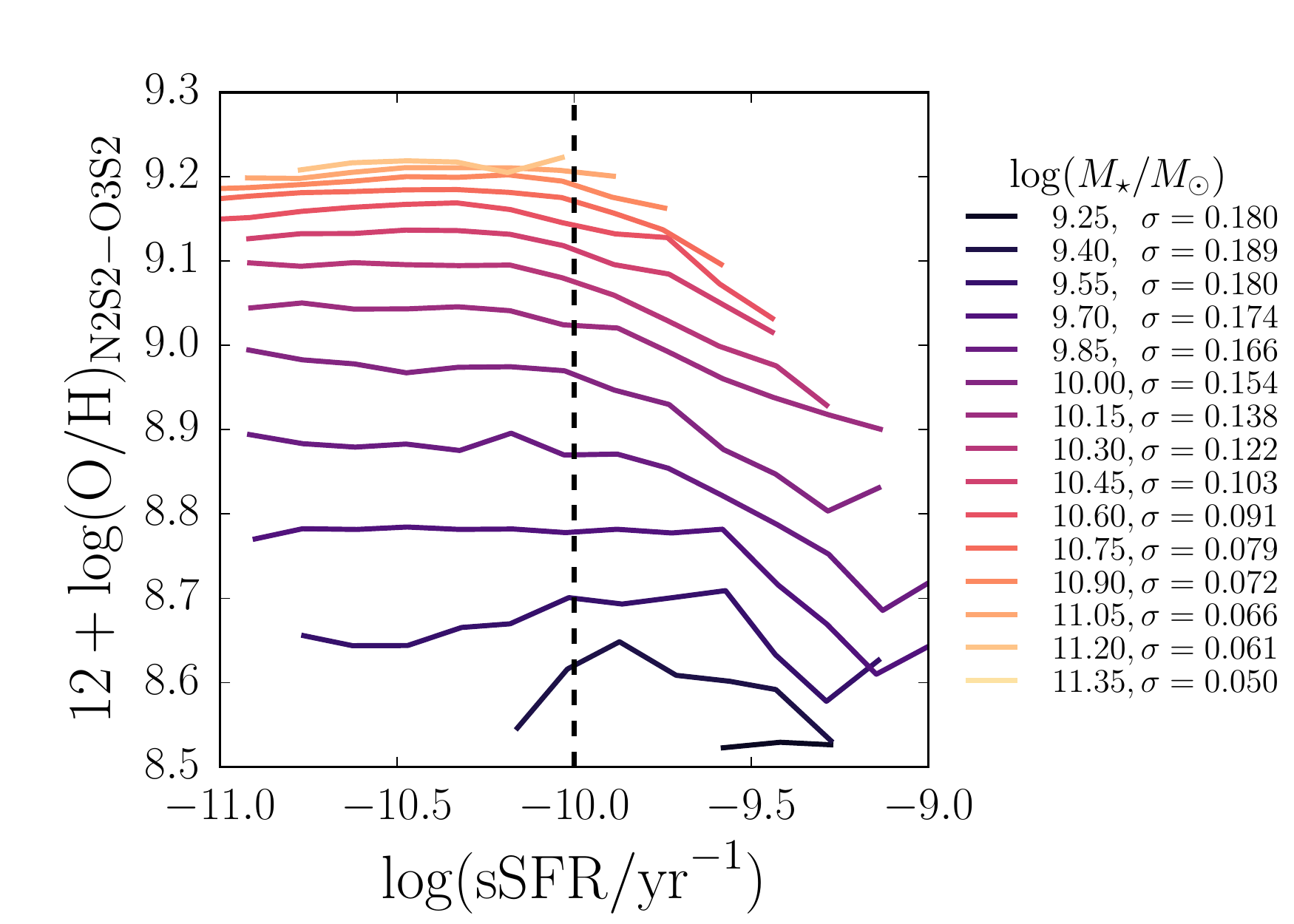}
\caption{Recasting the $M_\star-Z-\mathrm{SFR}$ relation in terms of specific SFR. Analogous to the right panel of Figure~\ref{d13_fmr} but using $\log (\mathrm{sSFR})$ instead of $\log (\mathrm{SFR})$. Metallicities are calculated using the fiducial D13 abundance diagnostic grid (N2S2--O3S2). All bins have 0.15 dex width in each $\log (M_\star)$ and $\log (\mathrm{sSFR})$ and each bin contains at least 50 galaxies. The dashed line at $\log (\mathrm{sSFR/yr}^{-1}) = -10$ is shown for reference; $18,960$ galaxies (14.5 \% of the sample) lie to the right of this line. This figure demonstrates that metallicity is only strongly anti-correlated with sSFR in the high sSFR regime.
\label{ssfr_fig}}
\end{figure}

We transition to casting this relation in terms of specific star formation rate (sSFR), defined as $ \mathrm{SFR}/M_\star$. This quantity describes the relative importance of recent and past star formation, and is potentially more sensitive to signatures of recent gas infall than SFR. Figure~\ref{ssfr_fig} shows the median metallicity in bins of stellar mass and sSFR as a function of sSFR, with metallicity calculated using the fiducial D13 grid. At small sSFR, the metallicity is less dependent on sSFR; i.e., the lines of constant mass are flatter at low sSFR. There is a more notable decrease in $Z$ with increasing sSFR at high sSFR, though again, the decrease we find is smaller than that found by M10. Throughout the remainder of the paper, a boundary between these low and high sSFR regimes is indicated for reference by a dashed line at $\log{(\mathrm{sSFR/yr}^{-1})} = -10.0$ wherever a quantity is plotted against sSFR. This is a rough, qualitative split, and should not be taken to mean that a sharp transition in star formation behavior occurs at exactly this sSFR boundary.

A starburst galaxy is typically defined as a galaxy with $b = \mathrm{SFR}/\langle \mathrm{SFR} \rangle \geq 2$. Defining $\langle \mathrm{SFR} \rangle = M_\star / \tau_{age}$ and assuming a typical galaxy age $\tau_{age} = 10 \, \mathrm{Gyr}$, we see that $b=1$ for $\log{(\mathrm{sSFR/yr}^{-1})} = -10.0$. Effectively then, most galaxies to the right of the dashed line in Figure~\ref{ssfr_fig} are starburst galaxies. The measure of sSFR used here, fiber SFR divided by total $M_\star$, underestimates the sSFR by a factor of $\sim 2-3$, so the galaxies in the high sSFR regime are actually even more starburst-like than they seem. These are rarer than the galaxies in the low sSFR regime, both in the local universe and in our sample. Only $18,960$ galaxies, or 14.5\% of the sample, lie in the high sSFR regime.

We quantify the different strengths of correlation between metallicity and SFR in these two regimes of low and high specific SFR by computing the best fitting values of $\alpha$ for two subsamples of galaxies: (1) those with $\log{(\mathrm{sSFR/yr}^{-1})} < -10.0$, and (2) those with $\log{(\mathrm{sSFR/yr}^{-1})} > -10.0$ for both the M10 and D13 abundance diagnostics. Selecting high (low) sSFR galaxies does, by definition, remove low (high) SFR galaxies at a fixed mass, which reduces the scatter in median metallicities in bins of $M_\star$ (or $\mu_0$) and SFR about the median mass-metallicity relation for each of these subsamples. However, if there is any secondary correlation between metallicity and SFR within one of these subsamples, a nonzero value of $\alpha$ will further reduce the scatter about the median $Z-\mu_\alpha$ relation.

We find, for all methods of calculating metallicity, that there is a weaker anti-correlation between $Z$ and SFR in the low sSFR regime than in the sparsely populated high sSFR regime (Table~\ref{alphas_table}). Similarly, \citet{salim14} found variation in the strength of anti-correlation between metallicity and offset in sSFR from the star-forming main sequence. The interesting result here is that the $\sim 15\%$ of galaxies in this sample that exhibit strong anti-correlation between $Z$ and SFR lie in the tails of the sSFR distribution at high $M_\star$ or in a biased regime of parameter space at low $M_\star$ (Figure~\ref{pdfs}). \textit{The observed correlation of the MZR with SFR and sSFR across the local galaxy population is driven by the relatively rare galaxies in the high sSFR regime,} with SFRs that are elevated compared to their past average. 

\subsubsection{D13 vs. D16 Diagnostics, and Comparison to \citet{kashino16b}}

Recently, \citet{kashino16b} (hereafter K16) argued that there is no anti-correlation between $Z$ and SFR at fixed $M_\star$ when the D16 metallicity diagnostic is used. They interpret this to mean that abundance diagnostics using the [N \textsc{ii}]/[S \textsc{ii}] ratio cannot be used to detect the $M_\star-Z-\mathrm{SFR}$ relation, assuming it is driven by infall of pristine gas causing increased star formation. In this scenario, the overall metallicity decreases due to gas infall, but the relative abundances of N and S remain unchanged, making the [N \textsc{ii}]/[S \textsc{ii}] ratio insensitive to the change in metallicity. 

However, we show here that the D13 grids using [N \textsc{ii}]/[S \textsc{ii}] and [N \textsc{ii}]/[O \textsc{ii}] as abundance-sensitive lines do produce weak, but nonzero, anti-correlations between $Z$ and SFR. We compare our results from the D13 grids to those from the D16 diagnostic in Figure~\ref{d13_comp} and in Table~\ref{alphas_table}. We do find a best fitting $\alpha = 0$ for the D16 diagnostic, indicating no anti-correlation with SFR across the entire sample. This result is comparable to the low $\alpha = 0.11$ derived for the D13 N2S2--O3S2 grid (which also uses the [N \textsc{ii}]/[S \textsc{ii}] ratio), as can be seen by comparing the bottom right panel of Figure~\ref{d13_comp} to the right panel of Figure~\ref{d13_fmr}.  However, while the overall values of $\alpha$ are low for both the D13 N2S2--O3S2 grid and D16 diagnostic, both show stronger correlations between $Z$ and SFR at high sSFR, in agreement with other D13 grids. The downturns in $Z$ at high SFR are weaker, but still present, in the $M_\star-Z-\mathrm{SFR}$ relation produced by the D16 diagnostic.

We compute the best fitting values of $\alpha$ for the D16 diagnostic in the low and high sSFR regimes, and find that there is a weak anti-correlation between $Z$ and SFR in the high sSFR regime for the D16 grid (Table~\ref{alphas_table}). This suggests that diagnostics relying on [N \textsc{ii}]/[S \textsc{ii}] actually are capable of detecting the $M_\star-Z-\mathrm{SFR}$ relation. This ratio may, in fact, be sensitive to changes in the overall metallicity even if the relative abundances of N and S are unchanged, or the mechanism driving the $M_\star-Z-\mathrm{SFR}$ relation may cause variation in the N/S ratio. 

Our results using the D16 grid are qualitatively different from those of K16, in that we see weak anti-correlation between $Z$ and SFR in the high sSFR regime, and they do not. Further, they report a reversal in the sense of the correlation between $Z$ and SFR at high stellar masses. Both of these differences can be attributed to different choices made in sample selection and the measurement of the various quantities. We demonstrate in Section~\ref{sample} above that applying S/N cuts on the forbidden lines induces bias against high metallicity galaxies at low SFR. K16 use S/N cuts on oxygen lines, which cause the apparent reversal at high stellar masses. They also use their own stellar mass determinations, which, as we discuss in Section~\ref{mass_sec} below, likely affects the apparent strength of the $M_\star-Z-\mathrm{SFR}$ relation.

\subsection{Stellar Mass Uncertainties}
\label{mass_sec}

We now discuss the systematic errors that may enter into the measurement of stellar masses and what role such biases might play in the observed strength of the $M_\star-Z-\mathrm{SFR}$ relation.

\subsubsection{Stellar Mass Measurement Technique}

Stellar masses in the MPA/JHU catalog are calculated by fitting the galaxy's photometry with a large library of model spectral energy distributions (SEDs) constructed using a range of values for parameters such as dust attenuation, stellar mass, star formation history (SFH), metallicity, and age. Probability density functions (PDFs) are constructed for each galaxy parameter such that the values of parameters corresponding to better fits are assigned higher likelihood. The median of the resulting stellar mass PDF for each galaxy is used as the best estimate of its stellar mass in the present study. 

Many ingredients that are required for any SED fitting procedure are highly uncertain, including the initial mass function, models of stellar evolution (in particular, the treatment of luminous, poorly understood phases of post main sequence evolution; see \citealt{conroy09}), the relative geometry of the dust and stars, and the form of the SFH. Stellar masses derived using different fitting methods and/or stellar evolutionary models vary by up to a factor of 2 \citep{kg07, conroy09, moustakas13}. The chosen parametrization of the SFH can change the derived logarithmic stellar mass even more, up to $0.6 \, \mathrm{dex}$ at low redshift \citep{pforr12}. For thorough discussions of the uncertainties that enter into the derivation of stellar masses from SED fitting, see reviews by \citet{conroy13} and \citet{courteau14}.

Given these substantial uncertainties, it is possible that stellar masses are biased high or low in cases where fits of the template spectra to the photometry yield uncertain stellar mass measurements. When the data cannot be used to place strong constraints on model parameters, the derived stellar mass becomes increasingly dependent on input assumptions, e.g., the form of the SFH and priors on the various parameters. 

\begin{figure}
\includegraphics[width=0.48\textwidth]{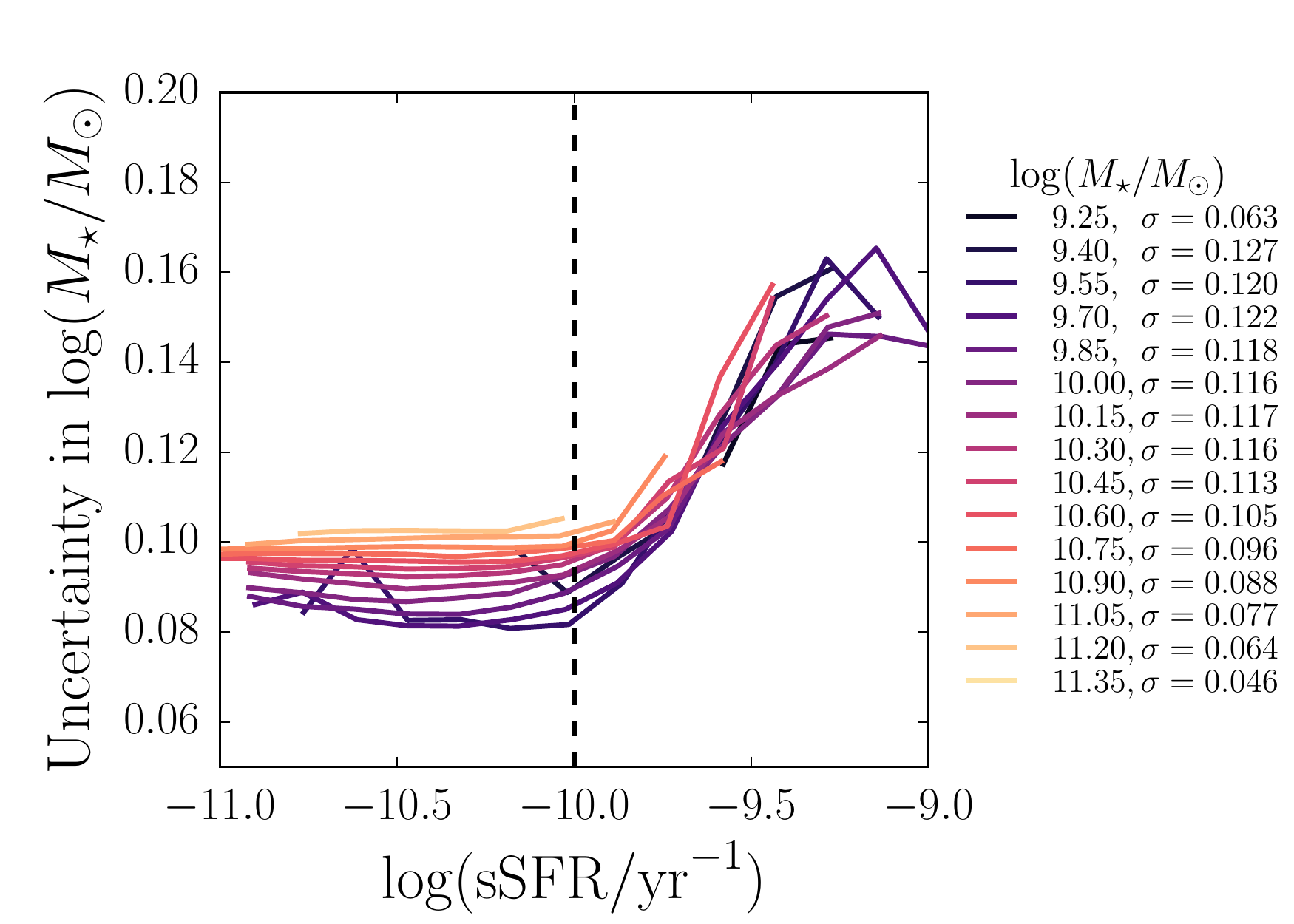}
\caption{Median uncertainty in the measurement of $\log (M_\star)$ as a function of sSFR. Uncertainty is defined as half of the difference between the values of $\log (M_\star)$ at the 84th percentile of the probability distribution and the 16th percentile reported in the MPA/JHU catalog. Binning is performed as in Figure~\ref{ssfr_fig}. Again, a dashed line at $\log (\mathrm{sSFR/yr}^{-1}) = -10$ is shown for reference. Stellar masses become more uncertain in the high sSFR regime across the full range of stellar mass. The true uncertainties are likely to be even larger, since the reported uncertainties do not include systematic errors due to assumptions in the modeling.\label{m_err_fig}}
\end{figure}

These issues motivate us to investigate whether uncertainties in the stellar mass determinations may play a role in the observed correlation between the MZR and SFR. Figure~\ref{m_err_fig} shows the median uncertainty in the stellar mass in bins of $M_\star $ and sSFR as a function of sSFR. Here, uncertainty is defined as half of the difference between the 84th percentile and 16th percentile values of the $\log{(M_\star)}$ PDF given in the MPA/JHU catalog. This value is comparable to a $1 \, \sigma$ uncertainty. These values reflect the formal errors in the model fits, but do not include any estimates of systematic uncertainties. In the high sSFR regime, the uncertainty in the stellar mass increases across all mass bins, meaning that the PDF widens. This behavior is reasonable, as the mass-to-light ratio becomes more uncertain when galaxy light is dominated by young, blue stars which obscure the older, fainter stellar population.

It is particularly interesting that this upturn in stellar mass uncertainty occurs at roughly the same value of sSFR where the anti-correlation between $Z$ and SFR begins to strengthen, indicated by the vertical dashed line in Figure~\ref{m_err_fig}. This raises the question of whether bias in the stellar mass measurements at high sSFR could drive the downturn in metallicity. In the case that stellar mass is preferentially overestimated relative to the true value, galaxies assigned to a given mass bin would in reality have smaller masses and therefore smaller metallicities (according to the MZR). If enough lower mass galaxies contaminate a higher mass bin at high sSFR, this could decrease the median metallicity, causing the appearance of a stronger correlation between metallicity and SFR.

\subsubsection{Simulation of Overestimated $M_\star$ at High sSFR}
\label{mass_sim}

\begin{figure}
\includegraphics[width=0.48\textwidth]{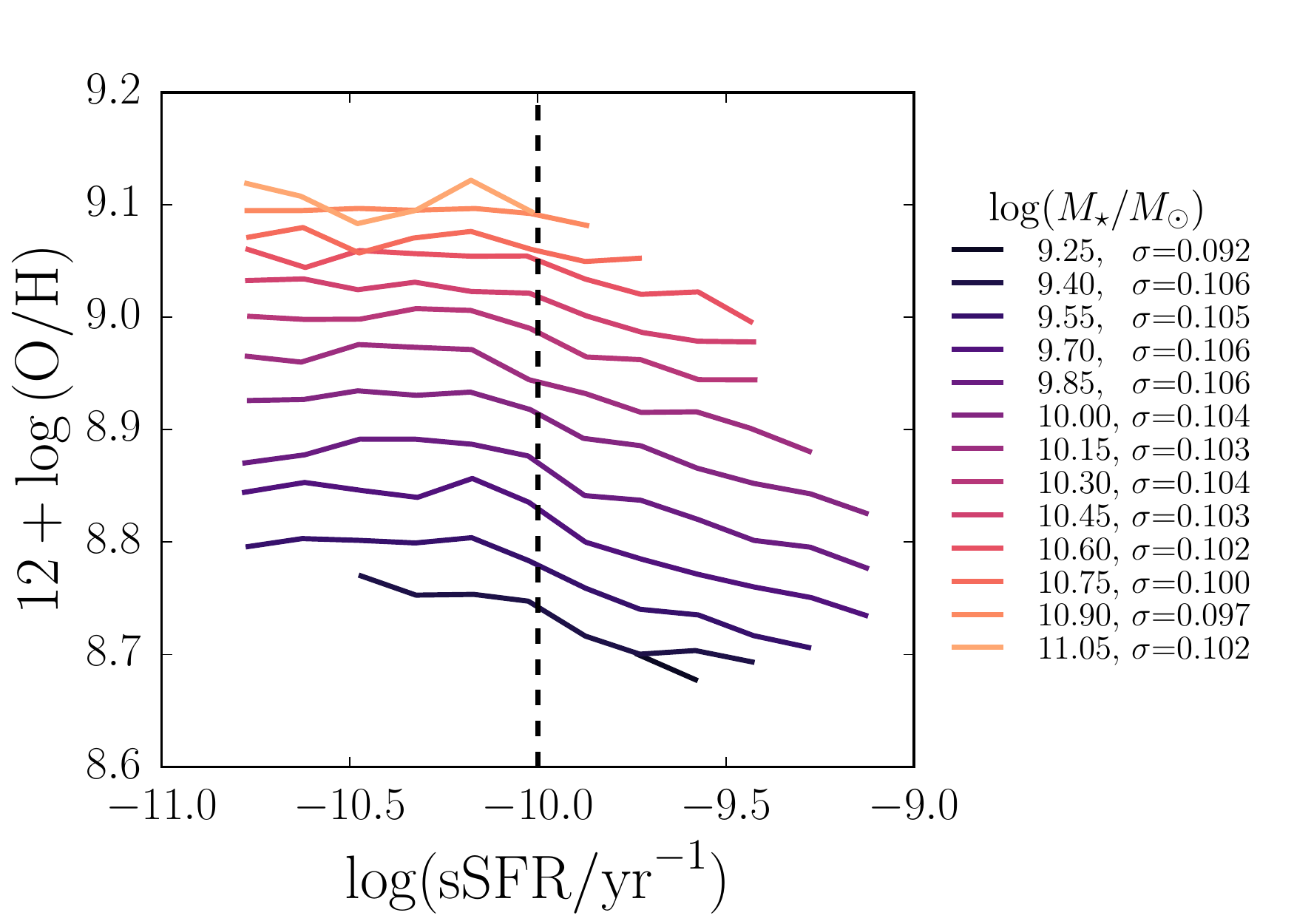}
\caption{$M-Z-\mathrm{sSFR}$ relation for a simulated galaxy population where stellar mass is biased high in the high sSFR regime. Stellar masses, SFRs, and metallicities are generated assuming $Z$ is independent of SFR. Above $\log (\mathrm{sSFR} /\mathrm{yr}^{-1}) = -10$, a noisy offset is added to the stellar mass assuming that the systematic upward bias in $\log (M_\star)$ at a given $\log (\mathrm{sSFR})$ increases linearly from 0.1 dex to 0.4 dex at $\log (\mathrm{sSFR} /\mathrm{yr}^{-1}) = -9$. The decreased median metallicities at higher sSFR in lines of constant $M_\star$ show that biasing $M_\star$ high at high sSFR can induce a correlation with sSFR comparable to that observed in the SDSS galaxy sample. \label{m_sim_fig}}
\end{figure}

We quantify the potential effects of overestimating $M_\star$ at high sSFR by simulating a population of galaxies where the MZR is independent of SFR, but where there is an upward bias in stellar mass that increases with sSFR only for galaxies in the high sSFR regime. In this model, galaxy masses are increasingly overestimated when the current SFR becomes greater than the past average SFR.

To generate the population of galaxies with no correlation between $Z$ and SFR, we draw 100,000 values of $M_\star$ from the \citet{bell03} stellar mass function to account for the fact that very massive galaxies are uncommon. We calculate metallicity for each galaxy using the best fit relation for the MZR given by \citet{tremonti04} and generate SFRs to match the trend and spread in the $M_\star-\mathrm{SFR}$ relation from \citet{brinchmann04}. Note that these simulated metallicities are on a different scale from the metallicity measurements discussed previously, so the absolute values of $12 + \log{(\mathrm{O/H})}$ in this section are not directly comparable to the values in other figures. No correlation with SFR is built into the metallicities. We confirm that the best fit value of $\alpha$ is 0 for this synthetic population.

We first examine the effect of increased noise but no bias at high sSFR. Since massive galaxies are rare,  in the case of a large amount of scatter, a high mass bin will become preferentially contaminated by lower mass galaxies. This effect could potentially drive the median metallicity in that bin to lower values. We test this possibility by adding Gaussian noise with a mean of 0 and spread of 0.15 dex to the values of $\log (M_\star/M_\odot)$ for galaxies in the high sSFR regime. This level of noise is chosen to be comparable to the uncertainties at high sSFR in Figure~\ref{m_err_fig}. This procedure produces a $M_\star-Z-\mathrm{SFR}$ relation with a best fit $\alpha$ of 0.02 at most and a shape that does not qualitatively match the observed relation (not shown). We conclude that noisier stellar mass measurements with no bias at high sSFR cannot cause the appearance of a stronger observed correlation with SFR.

To simulate masses being overestimated at high sSFR, we add a noisy, positive offset to the value of $\log (M_\star/M_\odot)$ for galaxies with $\log (\mathrm{sSFR} /\mathrm{yr}^{-1}) > -10$. We choose a mean upward bias in stellar mass that increases linearly with sSFR in the high sSFR regime. We first assume that the upward offset is comparable to the uncertainties in Figure~\ref{m_err_fig}, so that the mean bias increases from 0.1 dex at $\log (\mathrm{sSFR} /\mathrm{yr}^{-1}) = -10$ to 0.2 dex at $\log (\mathrm{sSFR} /\mathrm{yr}^{-1}) = -9$. For each galaxy, the offset is drawn from a Gaussian distribution centered at the appropriate mean value for its sSFR and with a standard deviation of 0.1 dex. In this simulated population, galaxies are preferentially scattered into $M_\star$ bins that are too large for their metallicities, causing the median metallicities at high sSFR to decrease. The best fit $\alpha$ is 0.05 for this scenario, indicating a weak, but measurable anti-correlation between $Z$ and SFR. 

Finally, we ask how large the upward bias in stellar mass must be to produce values of $\alpha$ comparable to what is found in observational studies. We find that choosing mean upward offsets that increase from 0.1 dex to 0.4 dex at the highest sSFRs produces $\alpha$ of 0.2. The resulting $M_\star-Z-\mathrm{sSFR}$ relation is shown in Figure~\ref{m_sim_fig}. This correlation with sSFR is stronger than that found with the fiducial D13 grid ($\alpha=0.11$) and nearly as strong as that found with the M10 metallicity diagnostic ($\alpha=0.28$). This level of bias is within the realm of possibility, given that much work remains to be done in understanding systematic issues in stellar population synthesis modeling at high sSFR.

This exercise has shown that it is possible that preferentially overestimating $M_\star$ at high sSFR could induce a spurious anti-correlation between the MZR and SFR. However, there is no reason to expect that stellar masses should be biased high in particular, and there is some evidence that stellar masses are actually underestimated in the high sSFR regime (e.g., \citealt{pforr12}). If stellar masses are instead biased low, that would result in an observed $M_\star-Z-\mathrm{SFR}$ that appears weaker than it is in reality. \textit{The unconstrained systematic errors in stellar mass determination in the high sSFR regime translate to large uncertainties in the strength of the $M_\star-Z-\mathrm{SFR}$ relation.}

\subsection{Aperture Effects}
\label{aperture_effects}

\begin{figure}
\includegraphics[width=0.48\textwidth]{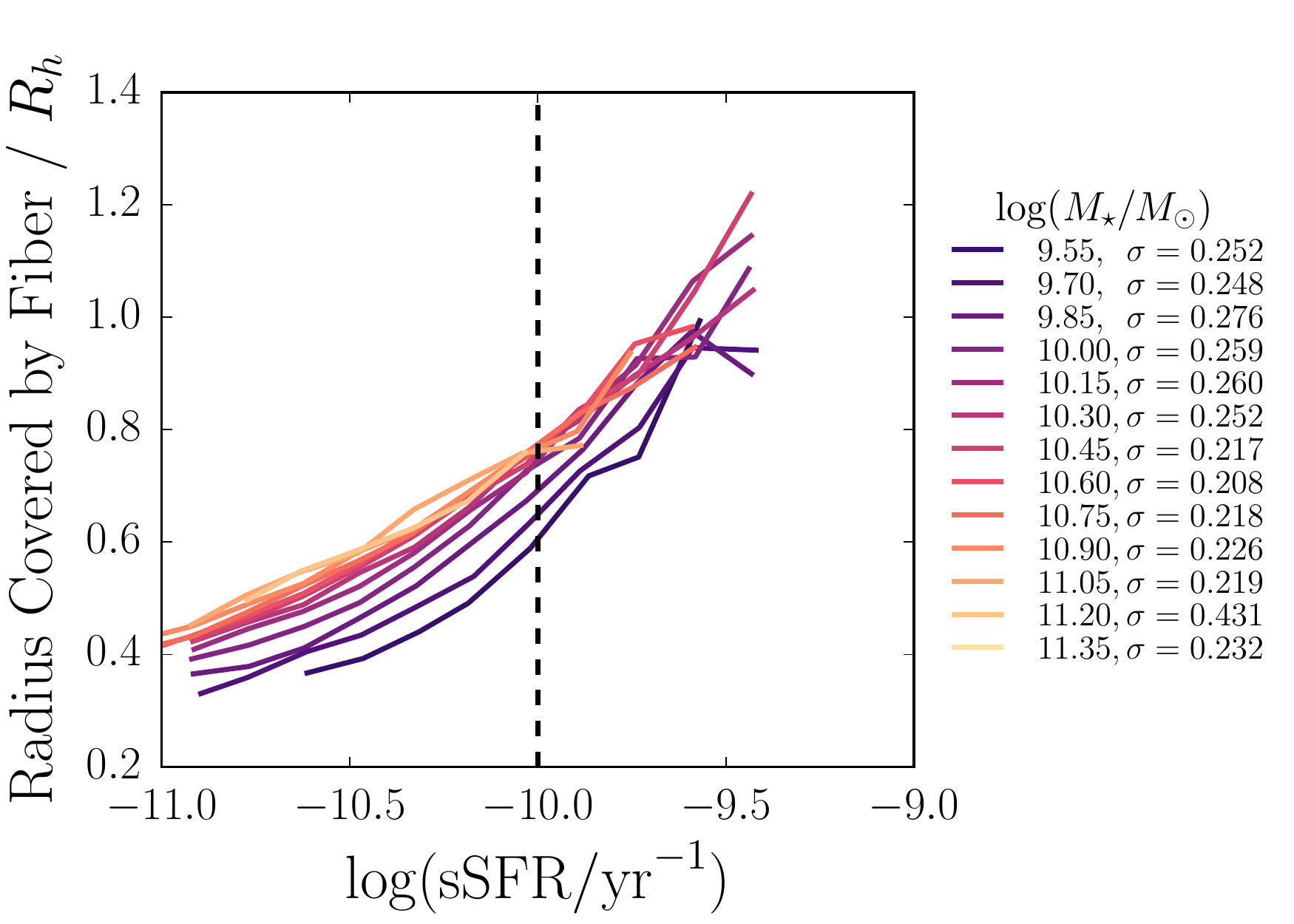}
\caption{Galaxy coverage by the SDSS spectroscopic fiber as a function of sSFR for our structural sample (see text). We plot the median of the radius covered by the $3''$ fiber at the redshift of each galaxy normalized by the half-light radius, $R_h$, of that galaxy. Half-light radii are the values from pure S\'{e}rsic model fits in the $r$-band from \citet{simard11}. Binning is performed as in Figure~\ref{ssfr_fig} and the dashed line at $\log (\mathrm{sSFR/yr}^{-1}) = -10$ is shown for reference. Across the full range of stellar mass, galaxies with higher sSFR also have a larger fraction of their light covered by the SDSS fiber. Therefore light from the outer regions of galaxies, which tend to be metal poor, is included in the measurement of SFR and $Z$ for galaxies in the high sSFR regime. \label{covering_frac_fig}}
\end{figure}

This study, as well as the vast majority of previous studies of the $M_\star-Z-\mathrm{SFR}$ relation, makes use of fiber spectroscopy from SDSS. The $3''$ spectroscopic fiber covers a larger physical area of target galaxies at higher redshifts. For our sample, the fiber covers physical sizes ranging from about $4 \, \mathrm{kpc}$ for the nearest galaxies to $13.4 \, \mathrm{kpc}$ for the farthest. 

The SFRs used in this paper and in M10 are calculated from the $\mathrm{H}\alpha$ flux within the spectroscopic fiber. Therefore, given that star formation is known to also occur in the outskirts of galaxies (e.g., \citealt{moffett12, richards16}), a larger fraction of the total star formation activity in a galaxy will be measured for more distant galaxies. Using photometric SFR indicators instead of fiber spectroscopy does not cause the $M_\star-Z-\mathrm{SFR}$ relation to disappear, and both types of SFR measurements can yield relations with comparable strengths \citep{salim14}. However, the choice of particular SFR indicator does cause the strength of the correlation with SFR to change.

Similarly, gas phase metallicity is known to decrease toward the outer portions of most star-forming galaxies (e.g., \citealt{zaritzky94, sanchez14}), though some galaxies have been found to have large oxygen abundances in their outskirts (e.g., \citealt{bresolin12}). \citet{tremonti04} found that edge-on galaxies have lower metallicities than the MZR would predict for their stellar mass. They argue that the fiber covers more of the disk in such galaxies, so the low metallicity gas in the outskirts contributes to the measured metallicity. 

However, the measured metallicity of a galaxy is a flux-weighted mean. Even if there is low metallicity gas outside of the central region covered by the fiber, it would only contribute to the measured global metallicity in a larger aperture if a significant amount of emission line flux were coming from the outskirts. Therefore, the amount by which aperture effects change the measured metallicity depends on the spatial distribution of star-forming regions within a given galaxy.

Other authors have investigated the covering fraction required for the measured gas phase metallicity to approximate the true global metallicity of a galaxy. Notably, \citet{kewley05} found that a minimum covering fraction of 20\% is needed, corresponding to a minimum redshift $z=0.04$ for SDSS (the minimum in our sample is $z=0.07$). However, they note that a higher minimum redshift is required if the sample contains many late type or high luminosity galaxies, and that the threshold must be higher still for the metallicities of most galaxies in a sample to approximate the global value. 

To test empirically whether aperture effects may be affecting measured metallicities and SFRs, in Figure~\ref{covering_frac_fig} we plot the median radius, normalized by the half-light radius, covered by the SDSS fiber in bins of $M_\star$ and sSFR. The half-light radius $R_h$ is the value computed from a pure S\'{e}rsic model fit to the $r$-band SDSS image from \citet{simard11}. Here we use our structural sample, a subset of our main sample (see Section~\ref{sample}). We verify that the strength of correlation between $Z$ and SFR is similar to that in the main sample (Table~\ref{alphas_table}). The two lowest $M_\star$ bins in previous plots do not appear in Figure~\ref{covering_frac_fig} because structural parameters are available for too few of the low mass, low surface brightness galaxies. We have removed 235 galaxies with anomalously small $R_h$ measurements ($< 0.5 \, \mathrm{kpc}$) from the sample for this plot to avoid artificially large dispersions within $M_\star$ bins; this does not affect our median coverage results.

The clear correlation between the size covered by the fiber and the sSFR seen in Figure~\ref{covering_frac_fig} is due to SDSS sample selection. At a fixed stellar mass, higher SFR galaxies tend to have higher surface brightness, and thus those galaxies sample a larger volume in the magnitude-limited SDSS. Therefore, higher sSFR galaxies are, on average, more distant and a larger fraction of their area is covered by the SDSS fiber.

Figure~\ref{covering_frac_fig} shows that galaxies in the high sSFR regime have at least $0.75 \, R_h$ covered by the spectroscopic fiber in the median, so the derived sSFR and $Z$ sample the conditions in the outskirts of those galaxies. Low sSFR galaxies have a median coverage of less than $0.75 \, R_h$, raising the possibility that low sSFR galaxies may have systematically larger metallicities measured at fixed mass because their metallicities and SFRs are measured for the central regions only.

Given the trend of increasing covering fraction with sSFR observed in our data, it is possible that aperture effects could bias metallicity measurements such that low sSFR galaxies appear to have higher metallicities at fixed stellar mass. This is not likely to be the major driver of the observed relationship between metallicity and SFR \citep{salim14}, but \textit{aperture effects could cause the decrease of metallicity with sSFR at fixed $M_\star$ to appear more dramatic at high sSFR.}

\subsection{Dust Effects}

\begin{figure*}
\plottwo{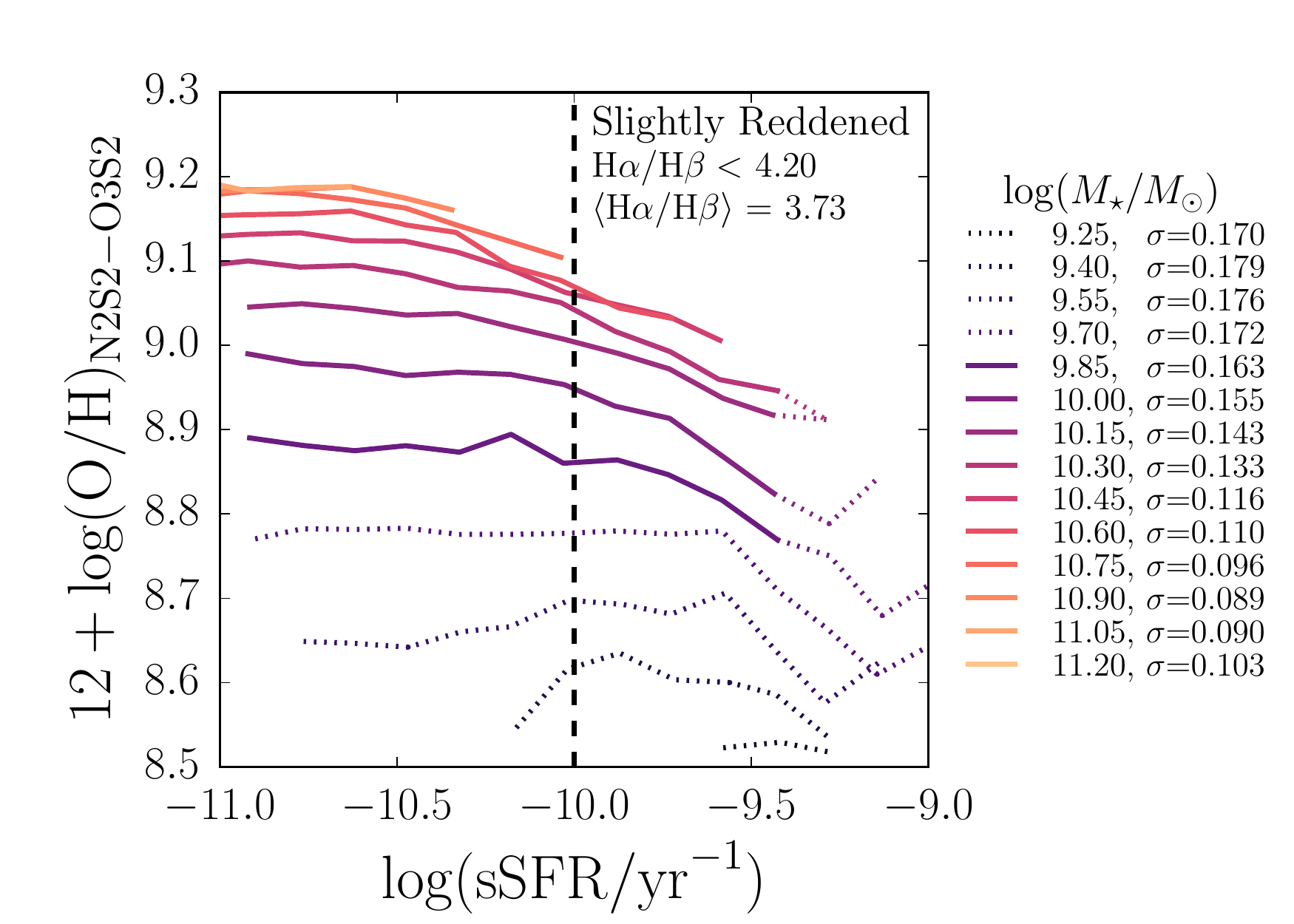}{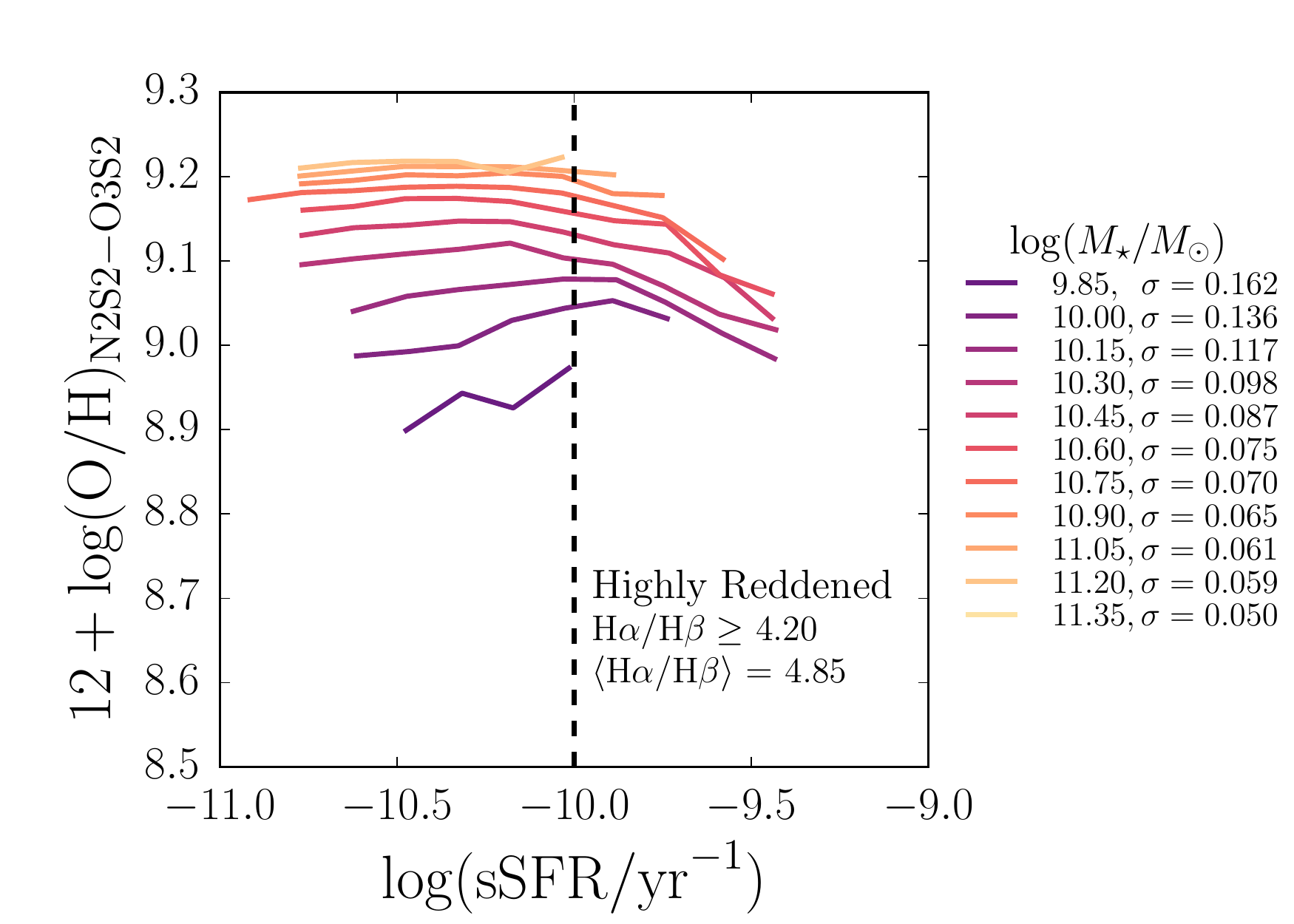}
\caption{Comparison of the $M_\star-Z-\mathrm{sSFR}$ relation for highly reddened and slightly reddened galaxies. We split our galaxy sample in half at the median value of the Balmer decrement, $\mathrm{H} \alpha / \mathrm{H} \beta = 4.20$. These plots are analogous to Figure~\ref{ssfr_fig}, with metallicities calculated using the fiducial D13 grid, but for two smaller samples. The dashed line at $\log (\mathrm{sSFR/yr}^{-1}) = -10$ is shown for reference. In the left panel, low $M_\star$ and high sSFR bins that do not appear in the right panel are shown as dotted lines; the solid lines provide a fair comparison between similar ranges of $M_\star$ and sSFR. These plots show that the shape of the $M_\star-Z-\mathrm{sSFR}$ relation is different for slightly and highly reddened galaxies across the same range of $M_\star$ and sSFR, even though the fiducial D13 grid is not sensitive to reddening.}
\label{dust_fig}
\end{figure*}

We now examine the impact of dust on the $M_\star-Z-\mathrm{SFR}$ relation. Figure~\ref{dust_fig} shows the effect of splitting our main sample of galaxies in half based on the value of the Balmer decrement, $\mathrm{H} \alpha / \mathrm{H} \beta$. The left panel shows the $M_\star-Z-\mathrm{sSFR}$ relation for a ``slightly reddened" sample of galaxies, where we have used metallicities from the fiducial D13 grid, restricted to galaxies with Balmer decrements smaller than the median for the full sample ($\mathrm{H} \alpha / \mathrm{H} \beta = 4.2$). The right panel shows the same for ``highly reddened" galaxies with Balmer decrements larger than the median; these galaxies have spectra that are strongly affected by dust absorption. In the left panel, we have used dotted lines to show ranges of low $M_\star$ and high sSFR that do not appear in the right panel, such that the solid lines provide a fair comparison between similar ranges of $M_\star$ and sSFR. 

Only high mass galaxies appear in the highly reddened sample because lower mass galaxies have not undergone enough metal enrichment to produce much dust. The slightly reddened sample spans almost the same range of the $Z$--sSFR space as the full galaxy sample, except for the high $Z$, high sSFR regime; such galaxies are typically dusty. Note that the bins in the right panel are more populated than the bins in the left panel, since the two samples have equal numbers of galaxies, but the highly reddened sample spans fewer bins in $M_\star$ and sSFR.

\textit{Strikingly, the overall shape of the $M_\star-Z-\mathrm{sSFR}$ relation is different for the slightly and the highly reddened galaxy samples.} The relation for the slightly reddened galaxies is similar to that found for the full sample, though metallicity does decrease somewhat with sSFR even in the low sSFR regime. This trend results in a stronger SFR dependence for the slightly reddened sample ($\alpha = 0.15$ vs. $\alpha = 0.11$ for the main sample). For the highly reddened sample, the metallicity curves are flatter in the low sSFR regime, and the sense of the correlation with sSFR even reverses for the lowest stellar mass bins such that higher sSFR galaxies have higher metallicities.

We have checked that these results do not depend on the particular metallicity calibration used (Table~\ref{alphas_table}). Given that the fiducial D13 grid is the least reddening-sensitive of all the metallicity calibrations analyzed in this paper, this grid should yield the smallest differences in metallicity between slightly and highly reddened galaxies, and should be unaffected by the choice of reddening law. We have also checked that splitting the sample in half based on inclination does not change the strength of correlation with SFR (not shown). We conclude that the changing shape of the $M_\star-Z-\mathrm{sSFR}$ relation for the slightly and highly reddened galaxies is related to the dust content and not to orientation, but the physical interpretation remains unclear.

\section{Discussion}
\label{discussion}

We have measured oxygen abundances and ionization parameters for $130,768$ star-forming galaxies in SDSS DR7 using theoretically calibrated abundance diagnostic grids from D13. For every D13 grid, we found a weaker correlation between metallicity and SFR than was found by previous studies using different metallicity measurement techniques \citep{mannucci10, am13}. Further, we showed that the $M_\star-Z-\mathrm{SFR}$ relation is driven by galaxies whose SFRs are elevated compared to their past average SFRs. We investigated possible biases that might affect the strength of the observed relation due to stellar mass measurement uncertainties, aperture effects, and dust. In this section, we consider the uncertainties that may affect the D13 abundance diagnostics and discuss the implications of the observed strength of the $M_\star-Z-\mathrm{SFR}$ relation for theoretical work.

\subsection{Concerns in Calculating Metallicity}
\label{metallicity_issues}

We have performed the first analysis of the  $M_\star - Z - \mathrm{SFR}$ relation using metallicities calculated using new abundance diagnostics from D13. These grids were calibrated using a new version of the \texttt{MAPPINGS} photoionization code that includes up-to-date atomic data, realistic models of dust and stellar spectra, and $\kappa \mathrm{-distributed}$ electron energies (which may be more realistic than Maxwell-Boltzmann distributed energies; \citealt{nicholls12}). The grids simultaneously determine both the ionization parameter and metallicity for each galaxy, minimizing possible biases in the metallicity measurements due to degeneracies between these parameters.

These improvements should bring the derived metallicities closer to the true metallicities of the SDSS galaxies, but as with any theoretical model, these photoionization models can only provide an approximate, imperfect representation of reality. Simplifying assumptions must be made (e.g., spherical geometry, fixed age stellar population) to make the calculations tractable, and the models can only sample a finite amount of parameter space. Some issues with these abundance--ionization parameter grids are already known; for example, the $\mathrm{[S \, \textsc{ii}]}$ flux is underpredicted by the \texttt{MAPPINGS} code (D13), meaning that three of the four grids we consider, including the fiducial (and least reddening-sensitive) grid, may suffer from systematic error because they use the $\mathrm{[S \, \textsc{ii}]}$ flux. 

As mentioned in Section~\ref{z_measurements}, the four D13 grids we use to measure metallicities do not give identical $M_\star - Z - \mathrm{SFR}$ relations. These theoretical grids are self-consistent when applied to individual $\mathrm{H \, \textsc{ii}}$ regions, so the fact that each grid gives a different strength of correlation between metallicity and SFR means that the emission line fluxes of SDSS galaxies are not perfectly described by the photoionization models of $\mathrm{H \, \textsc{ii}}$ regions. The discrepancies may be due to averaging metallicities, ionization parameters, and/or reddening corrections over many $\mathrm{H \, \textsc{ii}}$ regions.

These problems with characterizing the metallicity and ionization state of entire galaxies using the D13 grids could introduce biases into the measurement of metallicity and ionization parameter and could alter the observed relationship between these quantities and $M_\star$ and SFR; we revisit systematic effects on the ionization parameter in the appendix. We do not argue that the D13 grids provide the best possible measurements of abundance and ionization parameter for galaxies in SDSS. Rather, \textit{this study demonstrates that using a new abundance diagnostic results in a weaker anti-correlation between Z and SFR than found by previous studies} \citep{mannucci10, am13}. Since it is presently not possible to know which abundance determination method gives values closest to reality, one cannot determine the true strength of the correlation between metallicity and SFR. 

\subsection{Comparing Observations to Theoretical Models}

The observed correlation between the mass-metallicity relation and (s)SFR presents a challenge to theories of the chemical evolution of galaxies. While much theoretical work has focused on the physics governing the mass-metallicity relation (e.g., \citealt{larson74, ds86, brooks07, fd08}), relatively few models have specifically attempted to account for the more recently observed $M_\star-Z-\mathrm{SFR}$ relation.

Several authors have used ``equilibrium" or ``bathtub" models, in which galaxies self-regulate in response to changes in their gas mass, to explain the observed SFR dependence \citep{dave11, lilly13, pipino14}. If galaxies exist in a steady state where gas inflows are balanced by outflows and star formation, then variations in the accretion rate cause scatter about the equilibrium MZR. Infall of metal-poor gas initially causes metallicities to decrease, but increased star formation drives galaxies back toward the equilibrium relation \citep{dalcanton07}. \citet{dayal13} argued that the observed $M_\star-Z-\mathrm{SFR}$ relation can be explained by variations in outflow efficiency with stellar mass, assuming that outflows scale with the SFR. The ability of these different models to account for an anti-correlation between metallicity and SFR indicates that such a relationship could be a natural outcome of how galaxies regulate inflows, outflows, and star formation, though no consensus has been reached regarding the mechanisms governing such regulation.

Interestingly, all of these models suggest that the gas content of galaxies is the fundamental driver of the observed $M_\star-Z-\mathrm{SFR}$ relation, and that the (s)SFR of galaxies is changing in response to changes in the gas content. Recent observational studies corroborate the notion that the gas content may be the more fundamental ``third parameter" in the MZR \citep{hughes13, bothwell13, bothwell16}. However, these studies of atomic and molecular gas content are restricted to relatively small sample sizes compared to studies using SFR as the third parameter.

A further challenge to chemical evolution models is our observation that the anti-correlation between $Z$ and SFR is only important in the high sSFR regime, where galaxies' current SFRs are higher than their past average SFRs ($14.5 \, \%$ of our sample). It appears that these galaxies are undergoing a fundamentally different mode of star formation than the majority of the galaxy population (e.g., following major accretion events or mergers; a similar suggestion was made by \citealt{salim14}). A viable chemical evolution model should account for the different behaviors in the low vs. high sSFR regimes.

Some theoretical studies have used the observed $M_\star-Z-\mathrm{SFR}$ relation to predict other properties of the galaxy population, such as the relation between stellar mass and stellar metallicity at $z=0$ \citep{mp15} and variations in accretion and outflow rates \citep{forbes14}. However, since the large systematic uncertainties inherent in metallicity and stellar mass measurements will be propagated into any parameters constrained using the observed $M_\star-Z-\mathrm{SFR}$ relation, models should not be tuned to match any single observational result. Rather, at this stage, models attempting to explain the physical origin of the anti-correlation between $Z$ and SFR or using the observed relation in analytical models to make predictions for other quantities should be flexible enough to account for a wide range of possible strengths of the relation.

We have shown that the strength of the observed $M_\star-Z-\mathrm{SFR}$ relation is highly dependent on the abundance determination method used and may be affected by other sources of systematic error (e.g., bias in stellar mass measurement, dust, and/or aperture effects). Much observational and theoretical work remains to be done in understanding systematic errors, particularly those related to dust and stellar mass determinations. Upcoming IFU surveys with large, statistically powerful sample sizes such as MaNGA \citep{bundy15} and SAMI \citep{bryant15} will shed light on biases that arise when metallicity and SFR are computed using integrated light from galaxies spanning a range of redshifts.

\section{Conclusions}
\label{conclusions}

We have analyzed systematic effects that could affect the observed strength of the anti-correlation between the mass-metallicty relation and star formation rate, using metallicity estimators from D13. Using measurements of stellar masses and emission line fluxes for a sample of $\sim$130,000 galaxies in the SDSS spectroscopic survey, we explored various possible biases in the observed relationship. Our conclusions are summarized below.

\begin{enumerate}

\item We demonstrate that the sample of SDSS star forming galaxies is inherently biased toward high SFR and sSFR at low stellar mass (Figure~\ref{pdfs}).

\item We present the first analysis of the $M_\star-Z-\mathrm{SFR}$ relation computed using metallicity calibration grids from D13 and find that the anti-correlation between $Z$ and SFR derived using these new abundance diagnostics is weaker than that found by M10 (Figure~\ref{d13_fmr}, Table~\ref{alphas_table}). 

\item We show that the anti-correlation between $Z$ and SFR is largely driven by galaxies in the high sSFR regime, where the current SFRs are elevated compared to the past average SFRs (Figure~\ref{ssfr_fig}, Table~\ref{alphas_table}).

\item We compare our results using the D13 grids to the results from K16 using the new D16 abundance diagnostic. We show that the D16 diagnostic recovers a weak anti-correlation between $Z$ and SFR in the high sSFR regime, and that the turnover in the sense of this correlation at high stellar mass found by K16 is caused by their S/N cuts on oxygen lines. (Figure~\ref{bias_fig}, Figure~\ref{d13_comp}, Table~\ref{alphas_table}).

\item By simulating a population of galaxies with stellar mass determinations biased high, we find that systematic uncertainties in stellar mass estimates at high specific star formation rate may change the apparent strength of correlation with SFR. We estimate that the observed correlation with SFR could be driven by systematic errors in stellar mass determinations if stellar masses were systematically overestimated by up to $\sim 0.4$ dex at high sSFR (Figure~\ref{m_err_fig}, Figure~\ref{m_sim_fig}, Table~\ref{alphas_table}).

\item We find that poorly understood systematics with covering fraction and dust attenuation may affect the observed strength of the correlation between the MZR and SFR (Figure~\ref{covering_frac_fig}, Figure~\ref{dust_fig}).

\item Given that the observed $M_\star-Z-\mathrm{SFR}$ relation is compared to and used to inform models of galaxy chemical evolution, a wide range of possible strengths of the correlation with SFR must be accounted for in such theoretical analyses.

\end{enumerate}

\acknowledgments

We thank the referee for thorough and constructive comments that improved the quality of this manuscript. We are grateful to the MPA/JHU group and the authors of D13 and \citet{simard11} for making their data and code publicly available. OGT would like to thank Ben Williams, Phil Rosenfield, and Jessica Werk for helpful conversations. 

OGT is supported by a National Science Foundation IGERT fellowship under grant DGE-1258485, by a National Science Foundation Graduate Research Fellowship under Grant No. DGE-1256082, and by an ARCS Foundation fellowship. OGT gratefully acknowledges support from the Washington Research Foundation Fund for Innovation in Data-Intensive Discovery and the Moore/Sloan Data Science Environments Project at the University of Washington. CC acknowledges support from NASA grant NNX13AI46G, NSF grant AST-1313280, and the Packard Foundation. 

The analysis and plots presented here made use of iPython and packages from Astropy, matplotlib, NumPy, and SciPy \citep{ipython, astropy, matplotlib, scipy}. This research benefitted from NASA?s Astrophysics Data System Bibliographic Services.

Funding for the SDSS and SDSS-II has been provided by the Alfred P. Sloan Foundation, the Participating Institutions, the National Science Foundation, the U.S. Department of Energy, the National Aeronautics and Space Administration, the Japanese Monbukagakusho, the Max Planck Society, and the Higher Education Funding Council for England. The SDSS Web Site is http://www.sdss.org/.

The SDSS is managed by the Astrophysical Research Consortium for the Participating Institutions. The Participating Institutions are the American Museum of Natural History, Astrophysical Institute Potsdam, University of Basel, University of Cambridge, Case Western Reserve University, University of Chicago, Drexel University, Fermilab, the Institute for Advanced Study, the Japan Participation Group, Johns Hopkins University, the Joint Institute for Nuclear Astrophysics, the Kavli Institute for Particle Astrophysics and Cosmology, the Korean Scientist Group, the Chinese Academy of Sciences (LAMOST), Los Alamos National Laboratory, the Max-Planck-Institute for Astronomy (MPIA), the Max-Planck-Institute for Astrophysics (MPA), New Mexico State University, Ohio State University, University of Pittsburgh, University of Portsmouth, Princeton University, the United States Naval Observatory, and the University of Washington.

\begin{appendix}
\section{Detailed Comparison of the D13 Grids}

The D13 abundance diagnostic grids simultaneously determine metallicity and ionization parameter. Here we show the SDSS galaxy data on the four D13 grids that provide good separation between metallicity and ionization parameter that we consider in this paper. We compare the values of $Z$ and $q$ that are derived from the four different grids and show the resulting trends between $q$, $M_\star$, and SFR.

\subsection{Metallicity and Ionization Parameter Measurements}

\begin{figure*}
\minipage{0.5\textwidth}
  \includegraphics[width=\linewidth]{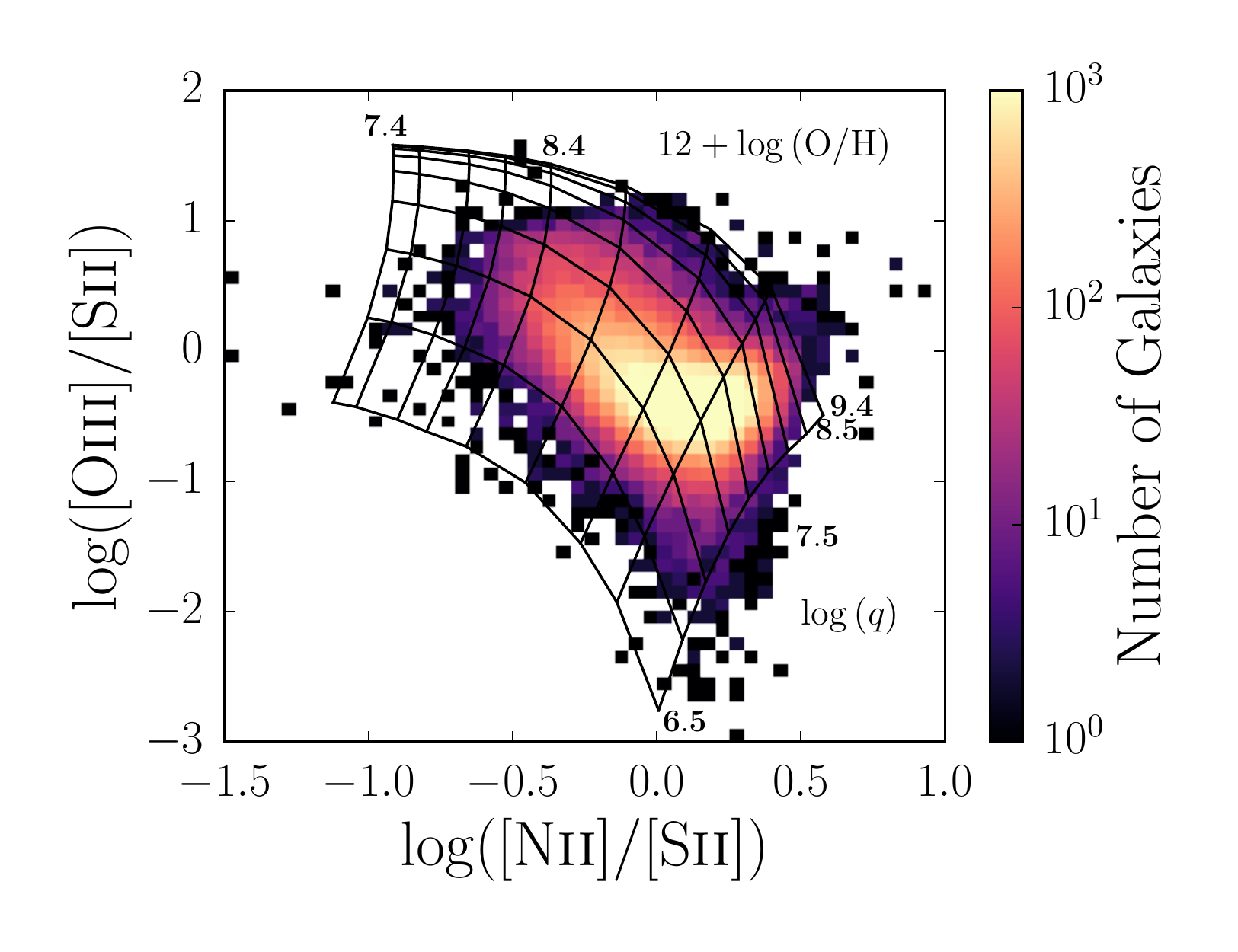}
\endminipage\hfill
\minipage{0.5\textwidth}
  \includegraphics[width=\linewidth]{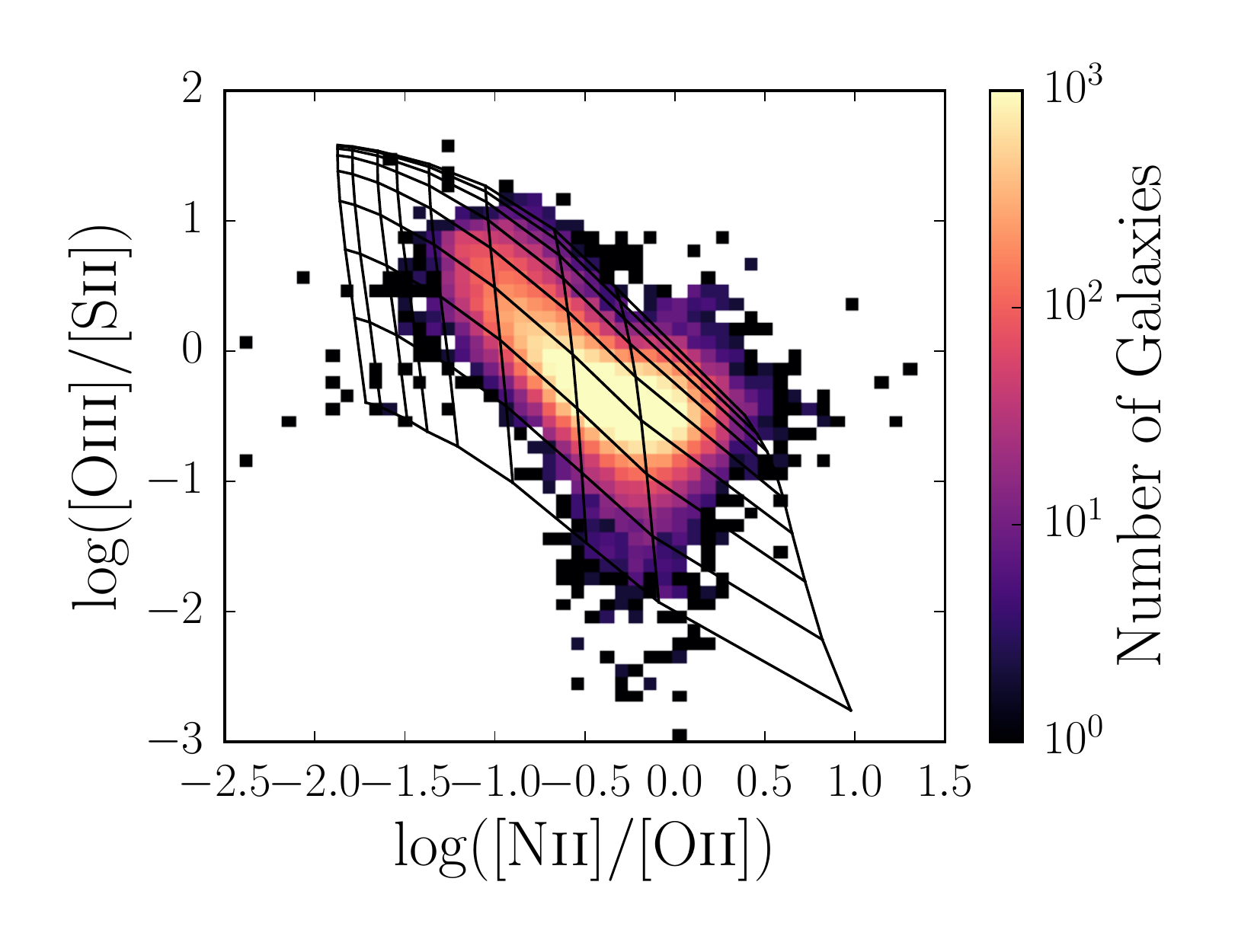}
\endminipage\hfill
\minipage{0.5\textwidth}
  \includegraphics[width=\linewidth]{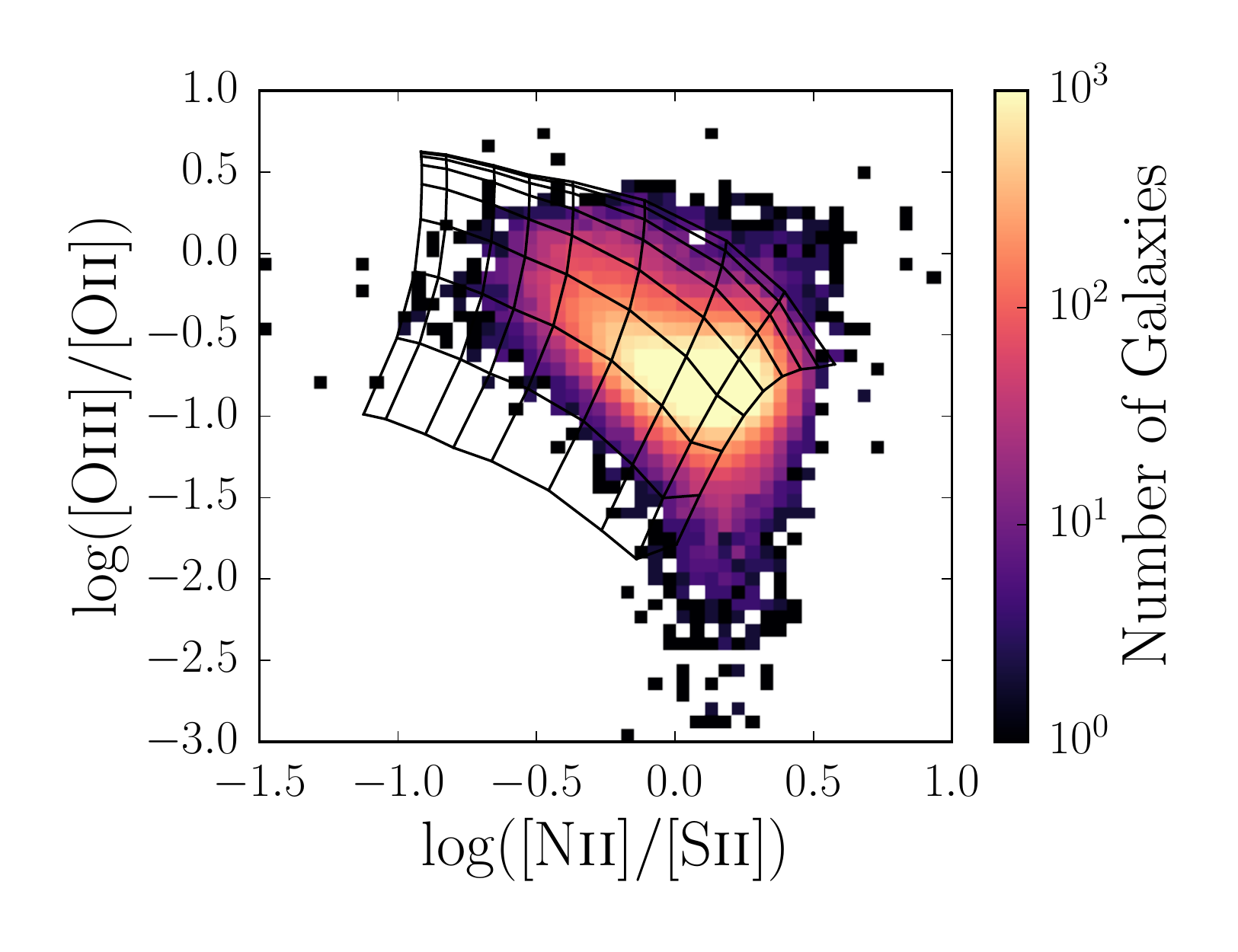}
\endminipage\hfill
\minipage{0.5\textwidth}
  \includegraphics[width=\linewidth]{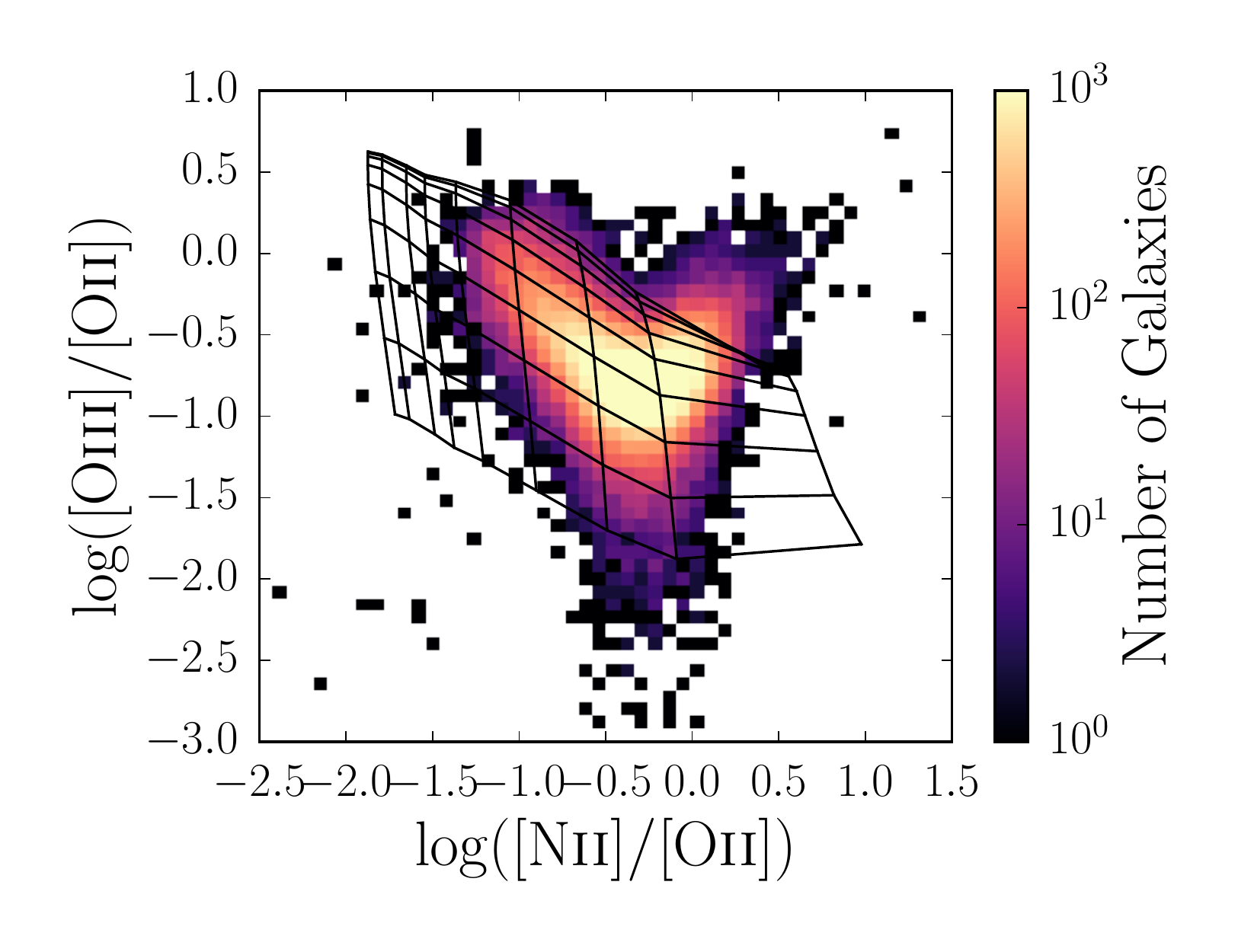}
\endminipage\hfill
\caption{Emission line ratios of SDSS galaxies in our main sample plotted on the four D13 grids considered in this paper. In each grid, $Z$ increases from left to right, and $q$ increases from bottom to top; values of $Z$ and $q$ are shown in the top left panel and are identical for all four grids. The color coding indicates the number of galaxies in each narrow bin in emission line space on a logarithmic scale. For most grids, the well-populated regions of the galaxy distribution lie within the grid lines. 
\label{data_on_grids}}
\end{figure*}

\begin{figure*}
\minipage{0.5\textwidth}
  \includegraphics[width=\linewidth]{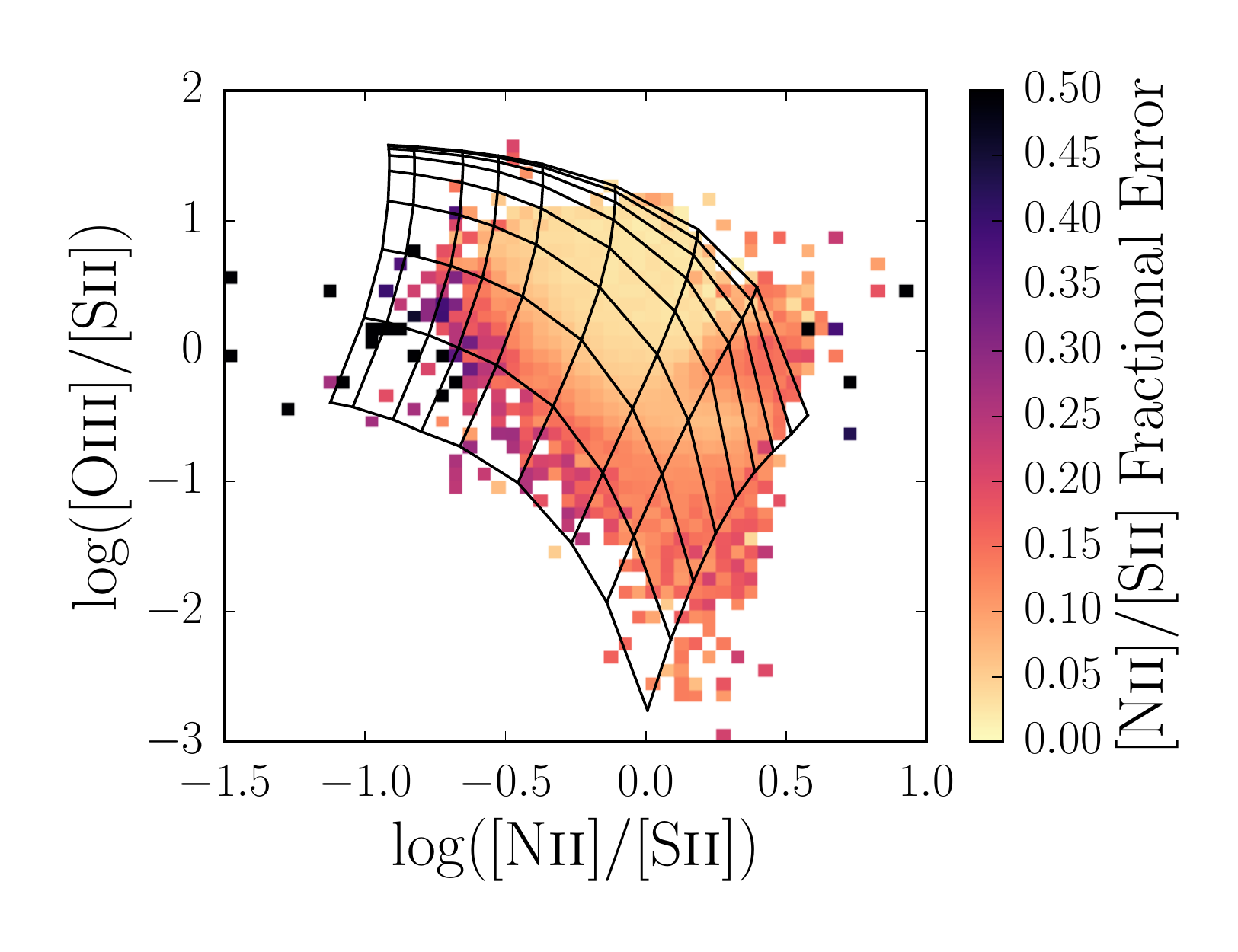}
\endminipage\hfill
\minipage{0.5\textwidth}
  \includegraphics[width=\linewidth]{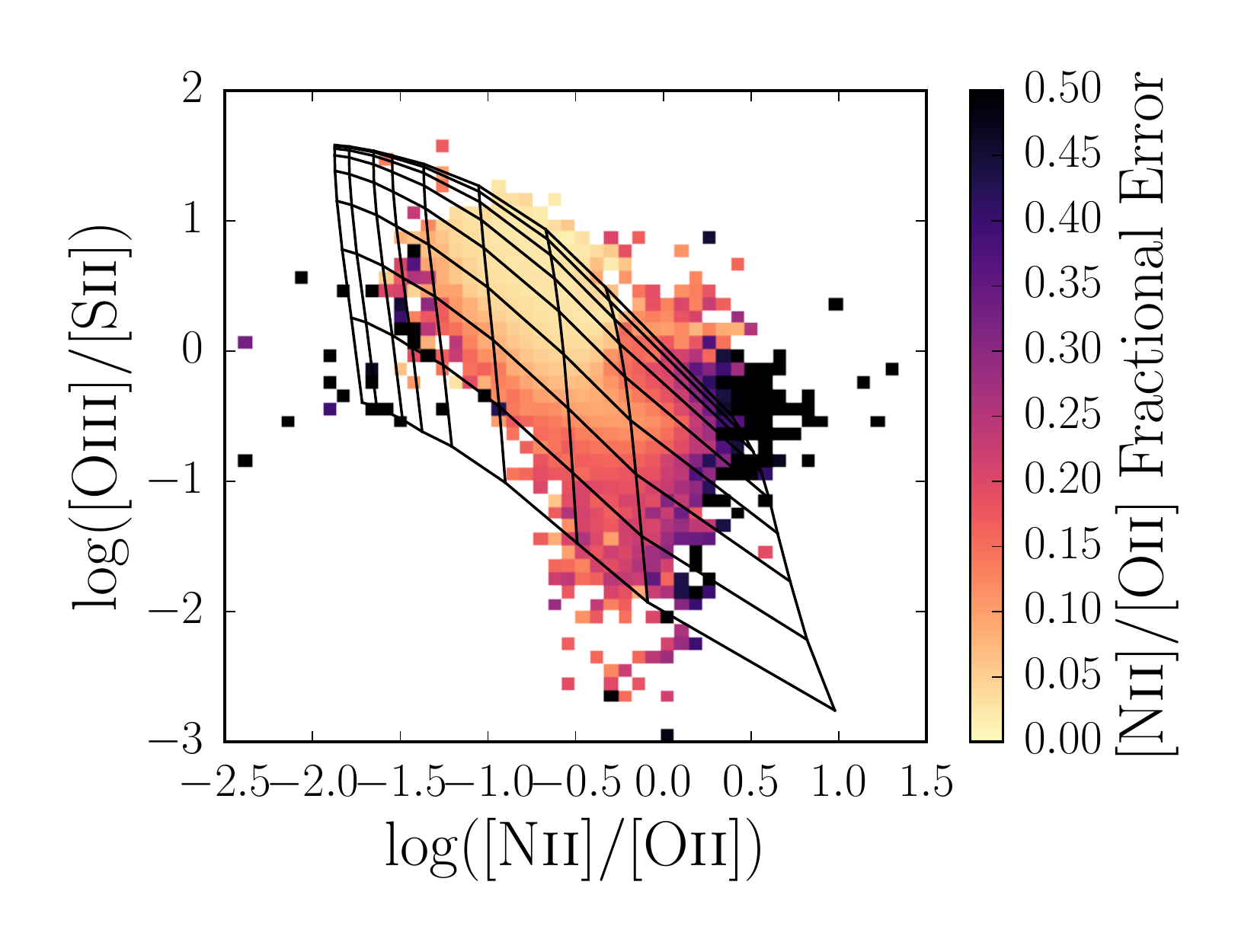}
\endminipage\hfill
\minipage{0.5\textwidth}
  \includegraphics[width=\linewidth]{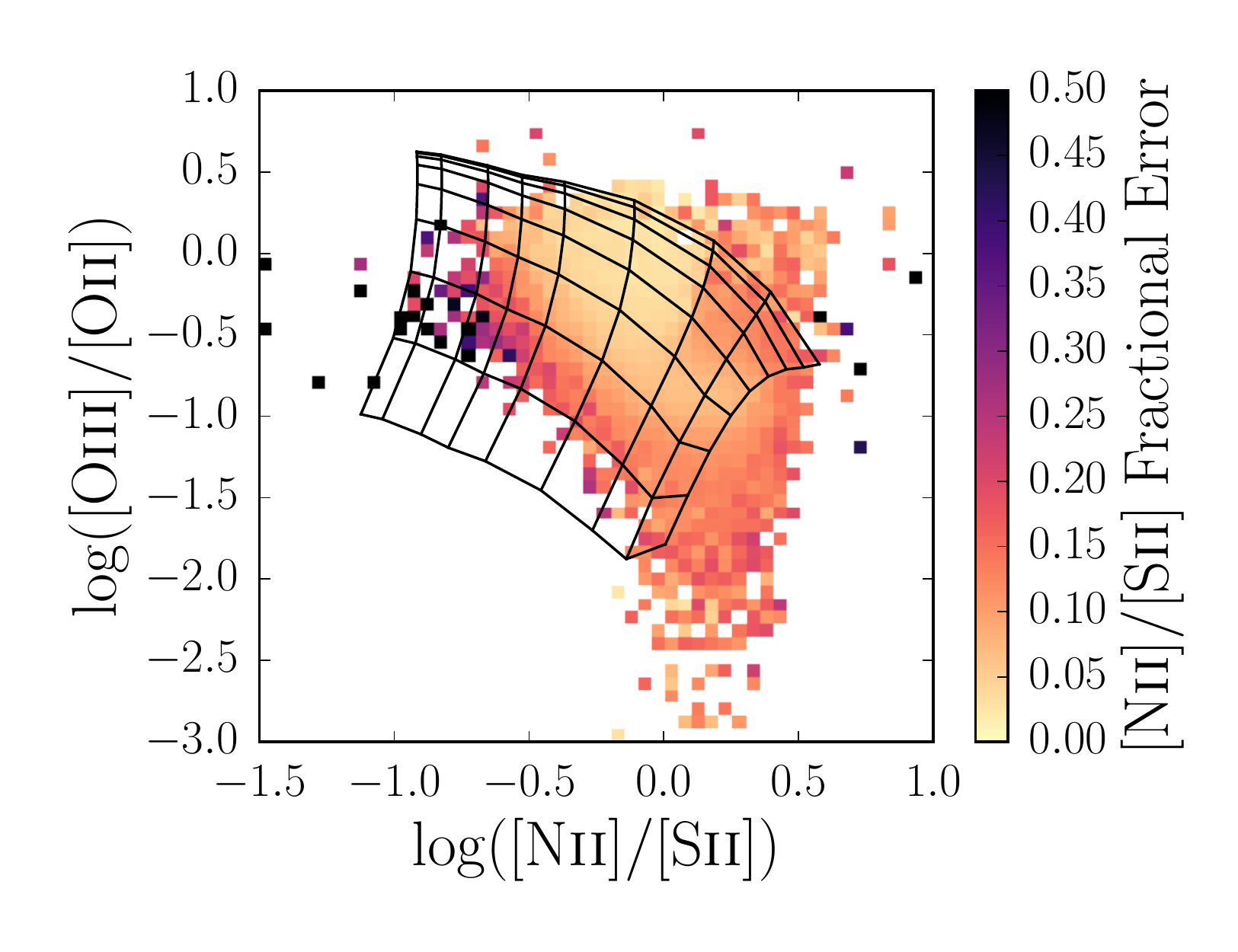}
\endminipage\hfill
\minipage{0.5\textwidth}
  \includegraphics[width=\linewidth]{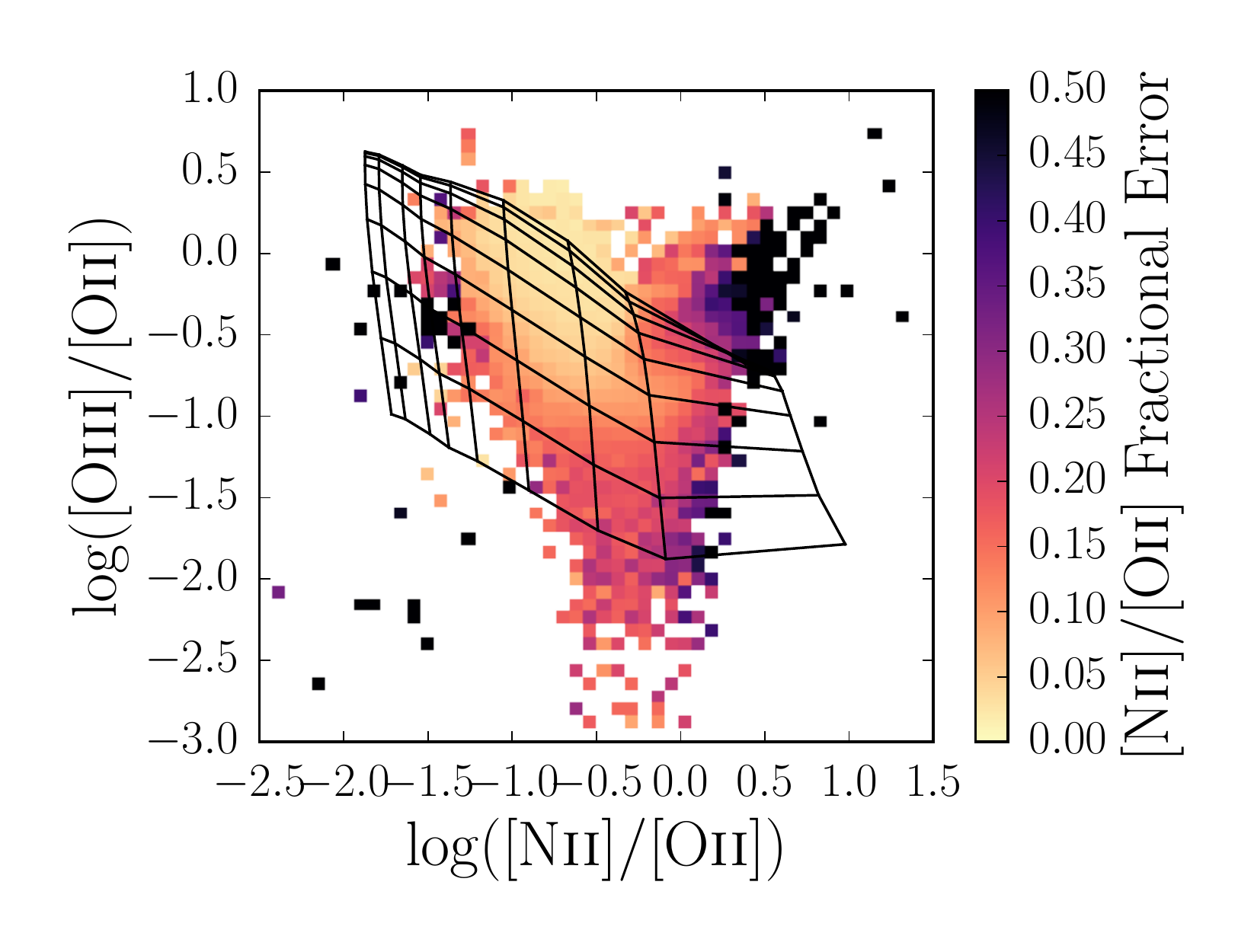}
\endminipage\hfill
\caption{Emission line ratios of SDSS galaxies in our main sample plotted on the four D13 grids considered in this paper. In each grid, $Z$ increases from left to right, and $q$ increases from bottom to top; values of $Z$ and $q$ are shown in the top left panel of Figure~\ref{data_on_grids} and are identical for all four grids. The color coding indicates the median fractional error in the abundance-sensitive ratio within each bin in emission line space.
\label{data_on_grids_xerrs}}
\end{figure*}

\begin{figure*}
\minipage{0.5\textwidth}
  \includegraphics[width=\linewidth]{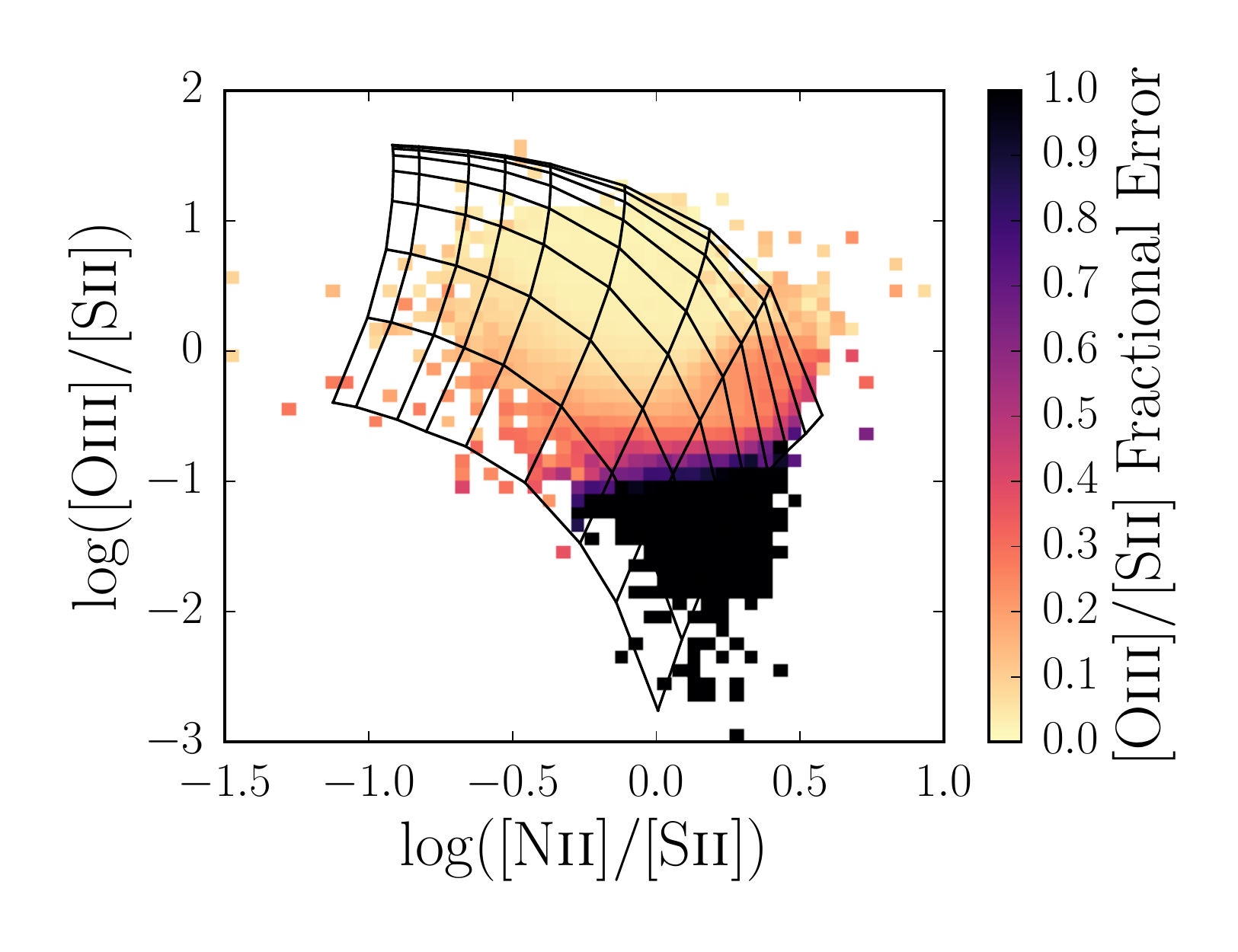}
\endminipage\hfill
\minipage{0.5\textwidth}
  \includegraphics[width=\linewidth]{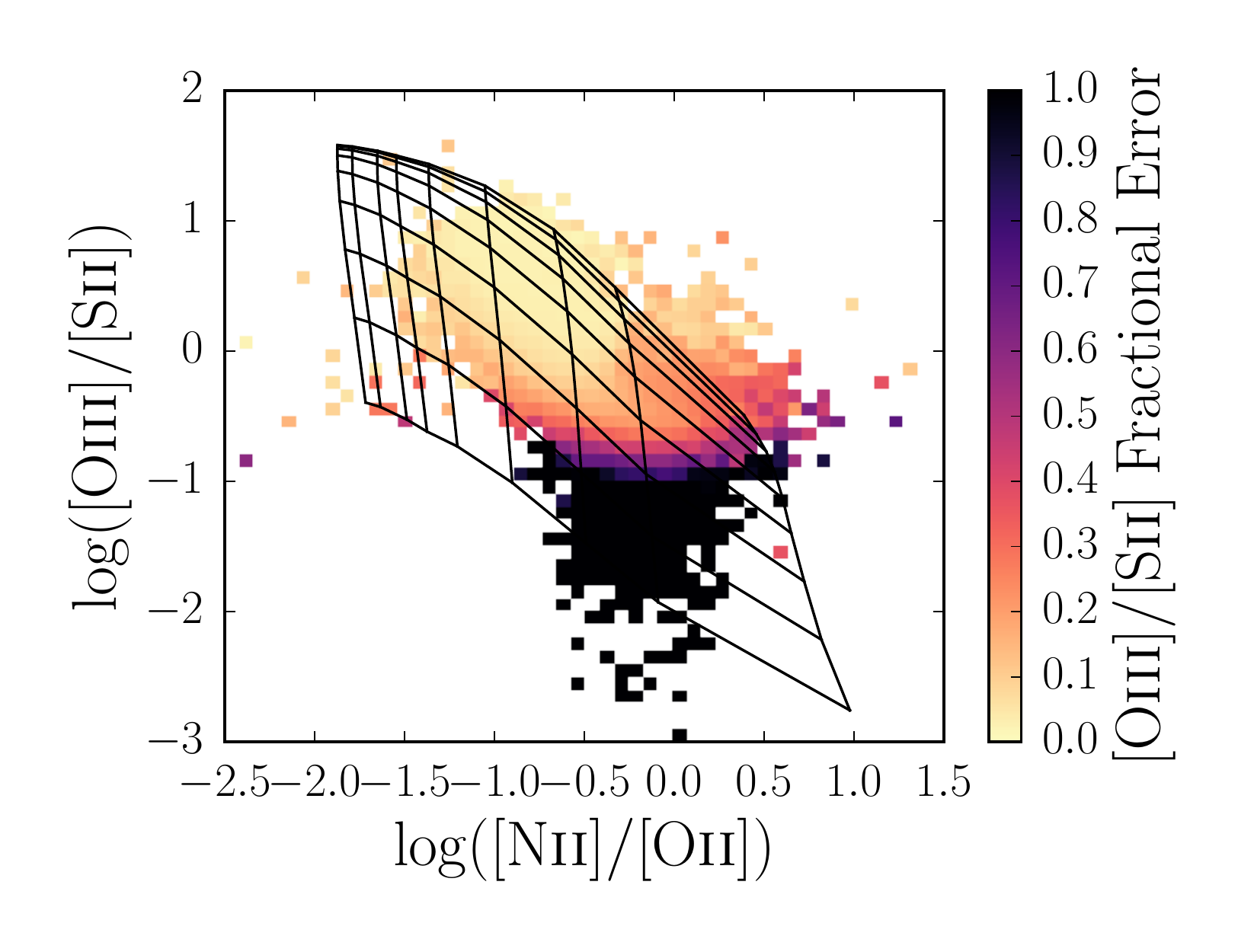}
\endminipage\hfill
\minipage{0.5\textwidth}
  \includegraphics[width=\linewidth]{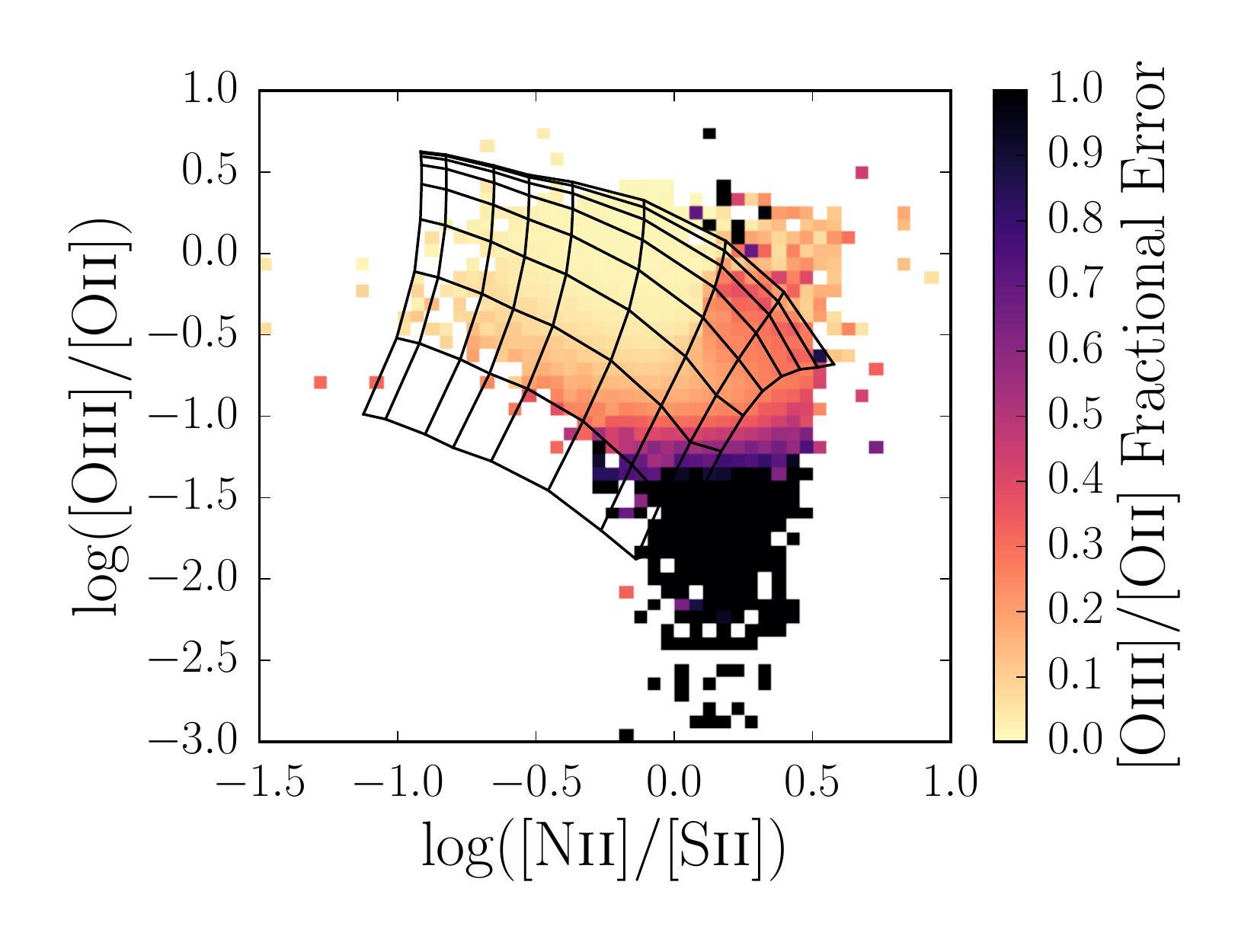}
\endminipage\hfill
\minipage{0.5\textwidth}
  \includegraphics[width=\linewidth]{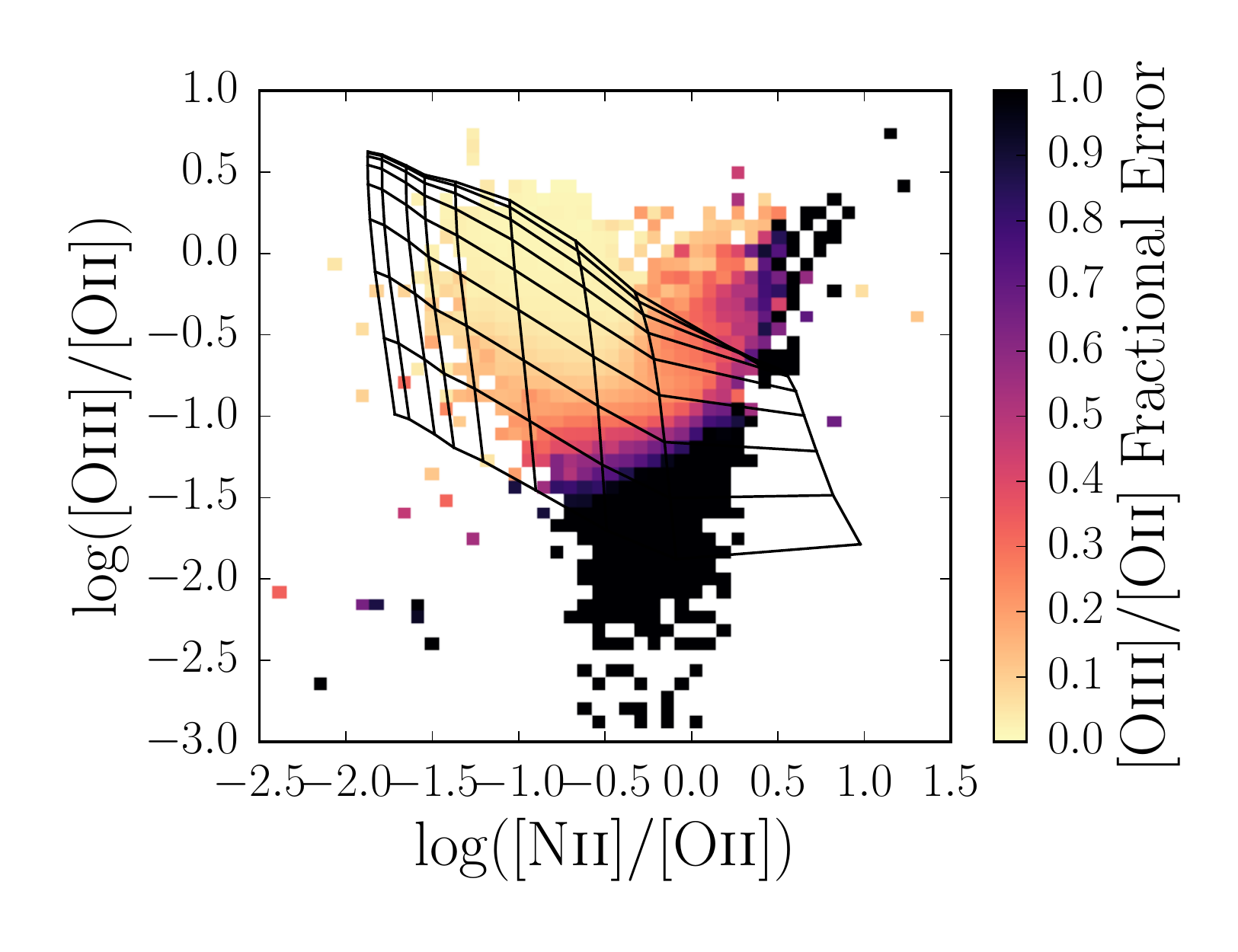}
\endminipage\hfill
\caption{Emission line ratios of SDSS galaxies in our main sample plotted on the four D13 grids considered in this paper. In each grid, $Z$ increases from left to right, and $q$ increases from bottom to top; values of $Z$ and $q$ are shown in the top left panel of Figure~\ref{data_on_grids} and are identical for all four grids. The color coding indicates the median fractional error in the ionization-sensitive ratio within each bin in emission line space. Note that the scale of these color bars is different from Figure~\ref{data_on_grids_xerrs}; this is because the ionization-sensitive ratios tend to have larger fractional errors.
\label{data_on_grids_yerrs}}
\end{figure*}

\begin{figure*}
\begin{centering}
\includegraphics[width=0.8\textwidth]{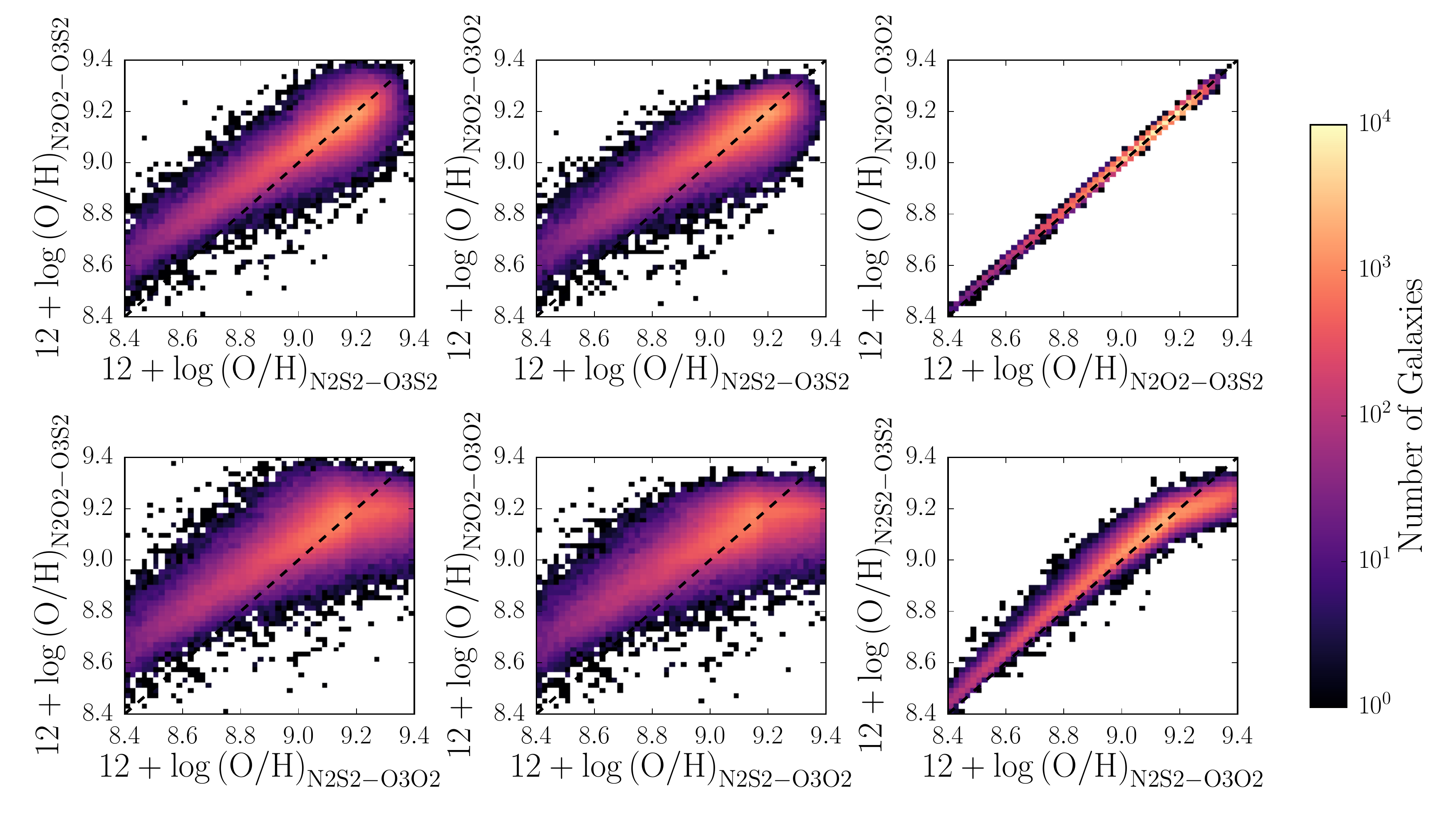}
\caption{Comparing measurements of metallicity from the four D13 diagnostic grids for our main sample of galaxies from SDSS. Galaxies that fall farther away from the black dashed line in each panel have more discrepant metallicity measurements. These plots show that metallicity measurements from different D13 grids generally follow linear relationships with each other. We conclude that all D13 grids are reliable abundance diagnostics, with the exception of the N2S2-O3O2 grid, which may be problematic; this grid is the cause of the turnovers in the bottom row of plots. 
\label{comp_Z_fig}}
\end{centering}
\end{figure*}

\begin{figure*}
\begin{centering}
\includegraphics[width=0.8\textwidth]{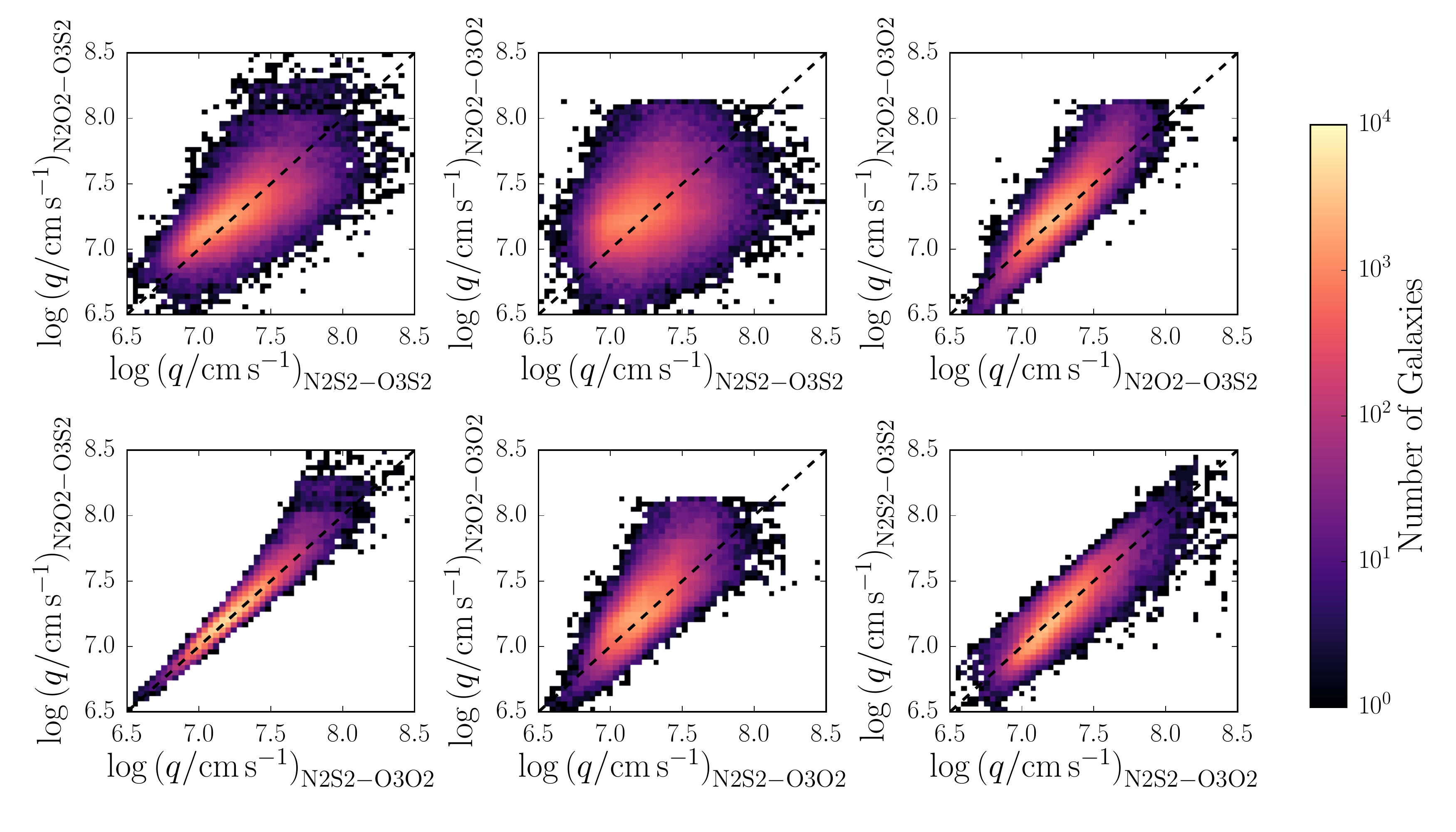}
\caption{Comparing measurements of ionization parameter from the four D13 diagnostic grids for our main sample of galaxies from SDSS. Galaxies that fall farther away from the black dashed line in each panel have more discrepant ionization parameter measurements. These plots demonstrate that ionization parameter measurements for a given galaxy change substantially depending on the choice of D13 diagnostic grid.
\label{comp_q_fig}}
\end{centering}
\end{figure*}

Figure~\ref{data_on_grids} shows all four D13 grids plotted over the corresponding emission line ratios for our main star-forming galaxy sample from SDSS. The data are binned by values of the emission line ratios in each plot, and color indicates the number of galaxies in each bin on a logarithmic scale. Some galaxies (249 out of $130,768$ in the full sample) do not fall within the fiducial grid (top left panel), and are therefore assigned NaN values and excluded from our calculations. 

Figure~\ref{data_on_grids_xerrs} shows the emission line ratios for our main galaxy sample plotted on the four D13 grids, but this time each bin is color-coded by the median linear fractional errors in the abundance-sensitive ratio (so that a fractional error of 0.5 corresponds to a 50\% uncertainty in the value of the emission line ratio). Figure~\ref{data_on_grids_yerrs} is similar, except the bins are color-coded by the median linear fractional errors on the ionization-sensitive ratio. Regions of the grids where lines are closely spaced are more susceptible to measurement errors. In such regions (at high $q$ and small $Z$), a small change in the value of a line ratio would cause a large change in the measured parameters. The measured values of $Z$ and $q$ are therefore more stable in the regions of the grids where the grid lines are farther apart.

From these plots, it is clear that galaxies that lie outside of the D13 diagnostic grids tend to have large fractional errors in both of the emission line ratios used to construct the grids. The galaxies that lie in the well-populated centers of the grids typically have small measurement errors, indicating that the D13 grids are aligned with the observed emission line ratios. The exception to this statement is the N2S2-O3O2 grid, where many of the galaxies in the core of the distribution that do not have large measurement errors fall outside of the grid (bottom left panel). This suggests that real galaxies do not lie in the same region of parameter space as this particular theoretical grid, meaning that the $Z$ and $q$ values derived for SDSS galaxies from the N2S2-O3O2 are likely not reliable. For this reason, we do not report results for the $M_\star-Z-\mathrm{SFR}$ relation using the N2S2-O3O2 grid in the body of this paper.

Interestingly, a hint of the AGN sequence can be seen in the upper right region of these plots, most notably in the N2O2-O3O2 grid (bottom right panel). The median fractional errors on the emission line ratios are small for many (though not all) bins in this region, so the presence of these galaxies on the AGN sequence cannot be explained by scatter. We have applied the \citet{kauffmann03c} AGN cut to the galaxy sample plotted here, so the presence of at least some of the galaxies in this region of parameter space indicates that some galaxies hosting AGN may be improperly classified as star-forming by the standard cuts. We have checked that removing the $\sim3000$ galaxies that lie along the AGN sequence in the N2O2-O3O2 grid does not affect the results of our analysis.

Figure~\ref{comp_Z_fig} compares the metallicities derived from the four D13 grids that we consider in this paper. The dashed line in each panel denotes perfect agreement. Several points are worth noting here. First, only the two grids depending on the [N \textsc{ii}]/[O \textsc{ii}] ratio agree perfectly; this is because the [N \textsc{ii}]/[O \textsc{ii}] ratio is primarily sensitive to metallicity in the D13 modeling, as indicated by the vertical lines of constant metallicity in those two grids. The D13 grids were calibrated to the relationship between the N/O abundance ratio and metallicity using measurements of H \textsc{ii} regions from \citet{vanzee98}. Second, the grids depending on [N \textsc{ii}]/[O \textsc{ii}] give systematically higher  metallicities below $Z \sim 9.0$ than those depending on [N \textsc{ii}]/[S \textsc{ii}]. Third, the metallicities derived from different grids largely follow linear relationships with each other. The exception to this last point is the N2S2--O3O2 grid, which we have already demonstrated is problematic. This grid gives higher metallicities than all other grids at high $Z$, even assigning many galaxies the maximum allowed metallicity. This is expected from the fact that many galaxies ($6662$, 5\% of the sample) fall off of this grid at high $Z$ and indicates that the N2S2-O3O2 grid should not be considered a reliable abundance diagnostic. Since the metallicities from the remaining three grids are linearly related, with differences of $\sim 0.2 \, \mathrm{dex}$ at most, we conclude that these three D13 grids are useful metallicity diagnostics for SDSS galaxies.

Similarly, Figure~\ref{comp_q_fig} shows a comparison of ionization parameters calculated from all four D13 grids. Again, the dashed line in each panel denotes perfect agreement. There is substantial scatter about the one-to-one relations. The derived ionization parameter is very dependent on the choice of D13 grid, with differences of up to $\sim 1 \, \mathrm{dex}$ in the most extreme case. It appears that the determination of an ``average" ionization parameter across an entire galaxy is problematic, and more so than for metallicity. We caution against a strict interpretation of the mean ionization parameter in a galaxy as pointing directly to a physical effect. It is possible that the fitted value of $q$ is very sensitive to inherent differences between modeled and actual spectra.

\subsection{Trends Between Ionization Parameter, $M_\star$, and SFR}

\begin{figure*}
\minipage{0.5\textwidth}
  \includegraphics[width=\linewidth]{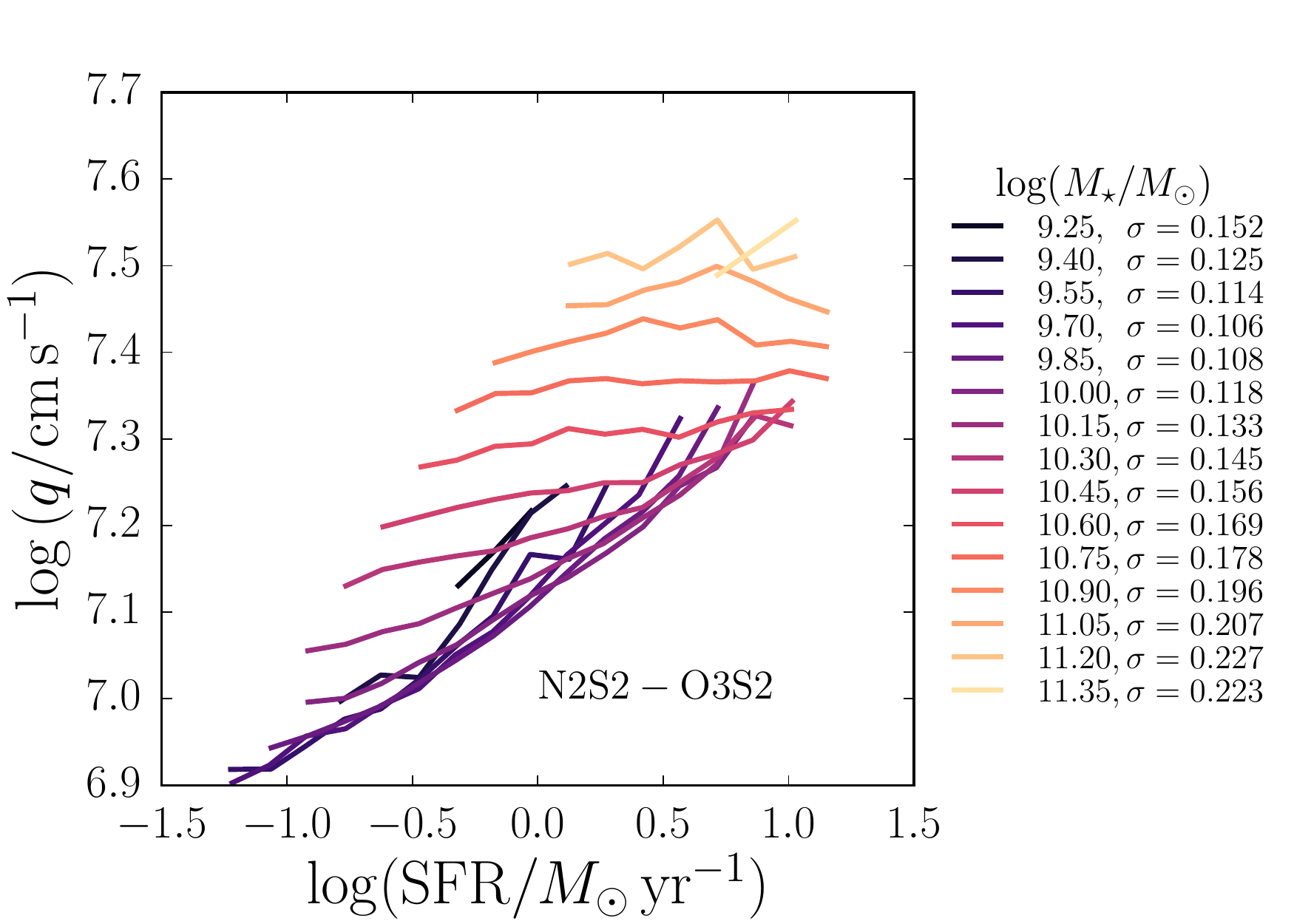}
\endminipage\hfill
\minipage{0.5\textwidth}
  \includegraphics[width=\linewidth]{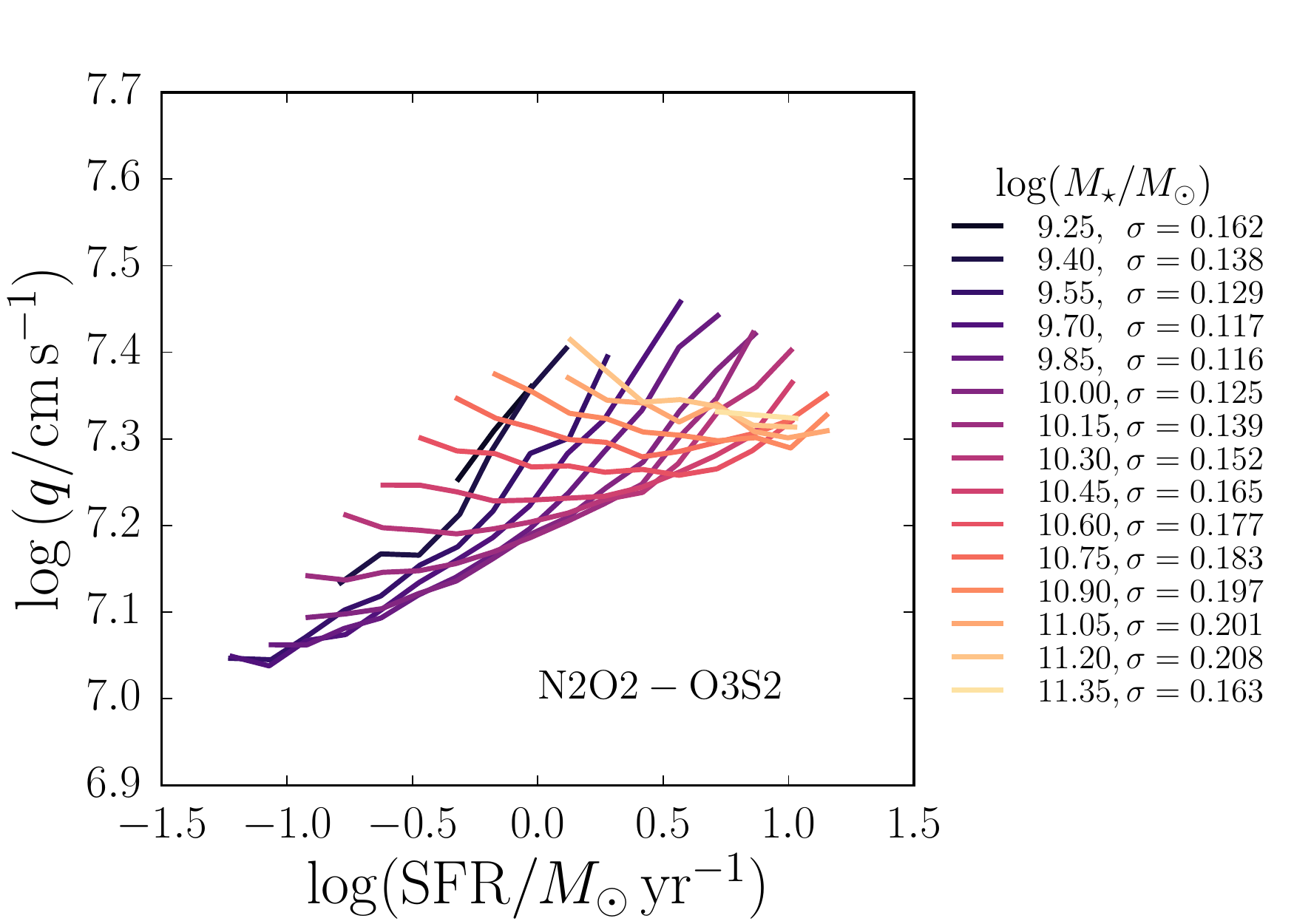}
\endminipage\hfill
\minipage{0.5\textwidth}
  \includegraphics[width=\linewidth]{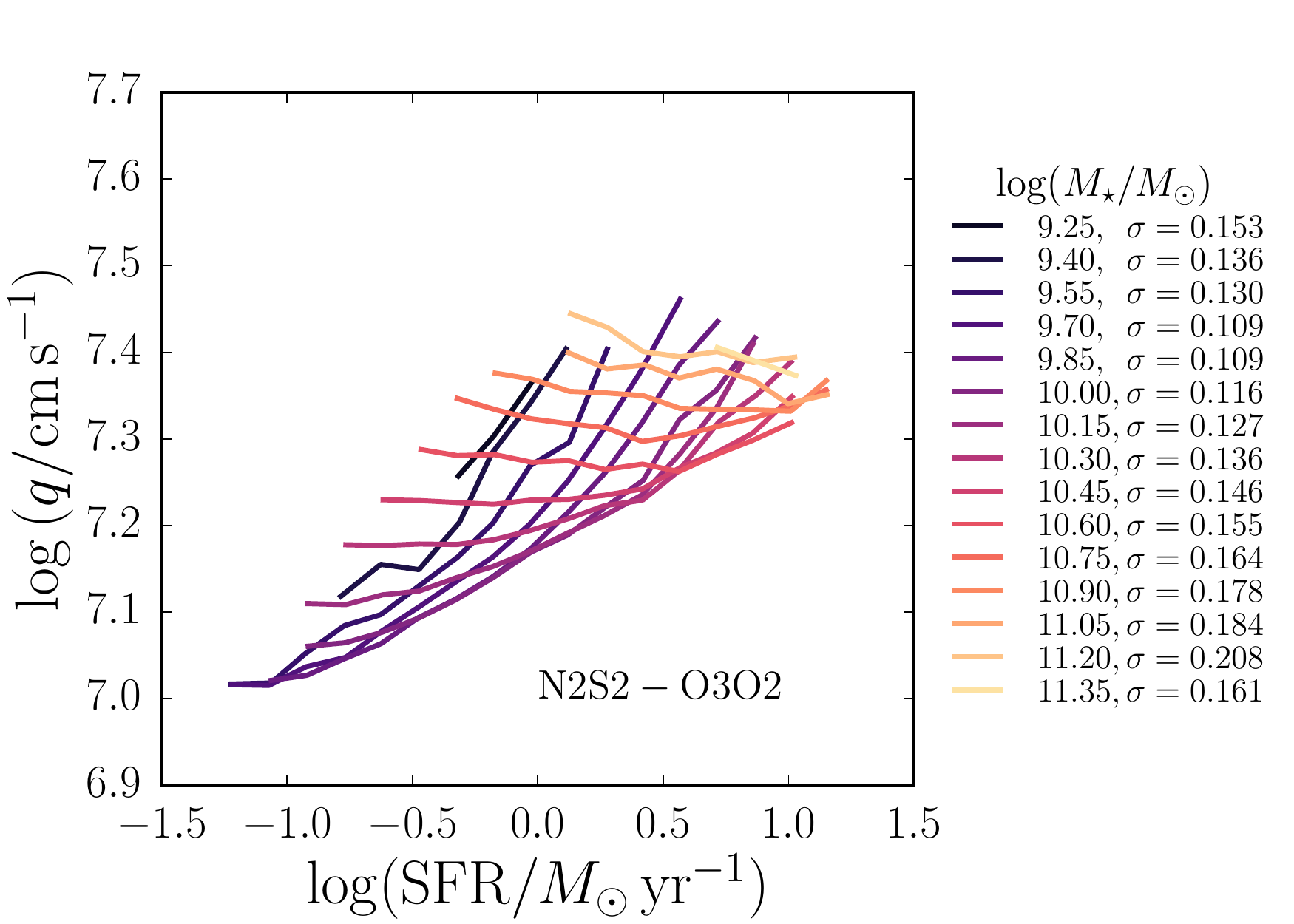}
\endminipage\hfill
\minipage{0.5\textwidth}
  \includegraphics[width=\linewidth]{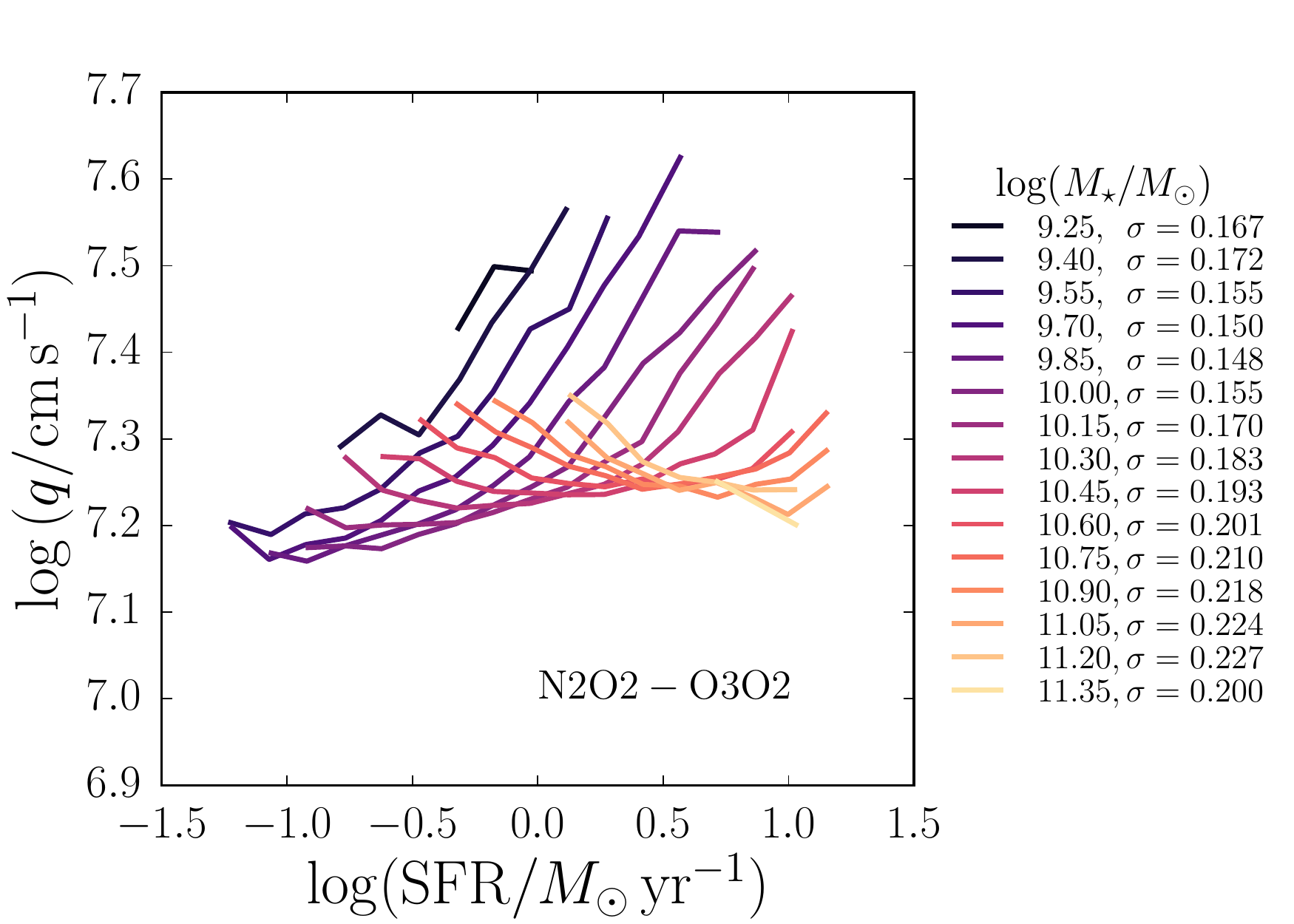}
\endminipage\hfill
\caption{Comparison of correlations between ionization parameter $q$, SFR, and $M_\star$ for the four different D13 diagnostic grids. These plots are analogous to the right panel of Figure~\ref{m10_fmr} but with median values of $q$ plotted instead of median metallicity. The choice of D13 diagnostic grid dramatically changes the sense and strength of observed correlations between $q$, SFR, and $M_\star$. \label{q_fig}}
\end{figure*}

We now examine trends in $q$ with $M_\star$ and SFR. Each panel of Figure~\ref{q_fig} is analogous to the right panel of Figure~\ref{m10_fmr}, but shows the median value of ionization parameter $q$ (instead of metallicity) in bins of $M_\star$ and SFR, where $q$ is calculated using each of the four different D13 grids. Generally, ionization parameter increases with SFR at low $M_\star$, but this trend either flattens or reverses at high $M_\star$. The two grids that depend on [N \textsc{ii}]/[S \textsc{ii}] yield a positive correlation between $M_\star$ and $q$ at high $M_\star$. These four grids produce strikingly different relationships between $q$, SFR, and $M_\star$, and in particular, yield correlations between $q$ and $M_\star$ that have different senses.

It has been hypothesized that the mean ionization parameter should decrease with increasing $M_\star$, since stellar winds with higher metallicities are more opaque, allowing fewer ionizing photons to escape into the ISM \citep{dopita06, kewley13}. Figure~\ref{q_fig} shows that the correlation of ionization parameter with mass derived using some of the D13 grids (including the fiducial grid, top left panel) do not conform to this expectation. The only grid that produces the expected correlation between $q$ and $M_\star$ is the N2O2--O3O2 grid (bottom right panel).

A common feature of ionization parameters derived from all four grids is that $q$ does not depend strongly on SFR at high stellar masses (above $\log (M_\star/M_\odot) \sim 10.5$). The strong correlation with SFR at low masses is likely due to the fact that at low stellar masses, only a small number of star-forming regions are covered by the spectroscopic fiber. More vigorous star formation yields more ionizing radiation and therefore a higher ionization parameter, so the measured value of $q$ is highly dependent on the properties of the few observed star-forming regions in those galaxies. The SFR independence at high $M_\star$ can be explained by the spectra of high mass galaxies covering a much larger number of star-forming regions (up to hundreds), therefore averaging over the conditions in all of those regions and fully sampling the IMF.

\end{appendix}


\end{document}